\documentclass{article}
\usepackage[utf8]{inputenc}
\usepackage{cite}
\usepackage{amsmath,amssymb,amsfonts,amsthm}
\usepackage{algorithmic}
\usepackage{graphicx}
\usepackage{textcomp}
\usepackage{xcolor}
\usepackage{mathtools}
\usepackage{calc}
\usepackage{physics}
\usepackage{hyperref}
\usepackage{cleveref}
\usepackage{multicol}
\usepackage{stmaryrd}
\usepackage{enumerate}
\usepackage[normalem]{ulem}
\usepackage{subcaption}
\usepackage{supertabular}
\usepackage{fullpage}

\def\BibTeX{{\rm B\kern-.05em{\sc i\kern-.025em b}\kern-.08em
		T\kern-.1667em\lower.7ex\hbox{E}\kern-.125emX}}

\usepackage{tikz}
\usepackage{pgfplots}
\usetikzlibrary{decorations.markings,decorations.pathmorphing, decorations.text,positioning}
\usetikzlibrary{backgrounds,folding}
\usetikzlibrary{shapes.geometric, shapes.misc}
\usetikzlibrary{arrows.meta}
\pgfdeclarelayer{edgelayer}
\pgfdeclarelayer{nodelayer}
\pgfsetlayers{background,edgelayer,nodelayer,main}
\tikzstyle{every picture}=[baseline=-0.25em]
\tikzset{every path/.style={draw=black!80, line width=0.6pt}}
\tikzstyle{none}=[inner sep=0mm]
\tikzstyle{black dot}=[inner sep=0.5mm,minimum width=0pt,minimum height=0pt,fill=black,draw=black,shape=circle]
\tikzstyle{dot}=[black dot]
\tikzstyle{white dot}=[inner sep=0.5mm,minimum width=0pt,minimum height=0pt,fill=white,draw=black,shape=circle]
\tikzstyle{white phase dot}=[minimum size=4mm, font={\footnotesize\boldmath}, shape=rectangle, rounded corners=1.5mm, inner sep=0.8mm, outer sep=-2mm, scale=0.8, draw=black, fill=white]
\tikzstyle{box}=[rectangle,fill=white,draw=black, font=\scriptsize, inner sep=2pt]
\tikzstyle{box-no-outline}=[rectangle, inner sep=2pt]
\tikzstyle{fswap}=[inner sep=0.7mm,minimum width=0pt,minimum height=0pt,draw=purple,shape=circle]
\tikzset{
	tickedge/.style={
		decoration={ markings,
			mark=at position .5 with {\draw (0,2pt) -- (0,-2pt);}
		},
		postaction={decorate}
	},
}
\tikzstyle{compact dash}=[dash pattern={on 2pt off 1pt}]

\pgfdeclarelayer{edgelayer}
\pgfdeclarelayer{nodelayer}
\pgfsetlayers{background,edgelayer,nodelayer,main}
\tikzstyle{every loop}=[]

\usepackage{etoolbox}

\newtoggle{extern}
\togglefalse{extern}

\newcommand{\tikzfig}[2][]{
	{\setlength{\fboxsep}{0pt}\colorbox{gray!15}{~#1\strut\input{./figures/#2.tikz}#1}}
}

\def\fig{}

\iftoggle{extern}{
	\usetikzlibrary{external}
	\tikzexternalize[prefix=tikzfigs/]
	
	\renewcommand{\tikzfig}[1]{
		\tikzsetnextfilename{#1}
		
	{\setlength{\fboxsep}{0pt}\colorbox{gray!15}{~\strut\input{./figures/#1.tikz}}}
}
	
}{
	
}

\usetikzlibrary{circuits.ee.IEC}

\newcommand{\groundnobg}
{\iftoggle{extern}{\tikzsetnextfilename{groundnobg}}{}
	{
		\begin{tikzpicture}[circuit ee IEC,yscale=0.9,xscale=0.8]
			\draw[solid,arrows=-] (0,1ex) to (0,0) node[anchor=center,ground,rotate=-90,xshift=.66ex] {};
\end{tikzpicture}}}

\newcommand{\ground}
{\iftoggle{extern}{\tikzsetnextfilename{ground}}{}
	{\setlength{\fboxsep}{0pt}\colorbox{gray!15}{~\strut
			\begin{tikzpicture}[circuit ee IEC,yscale=0.9,xscale=0.8]
				\draw[solid,arrows=-] (0,1ex) to (0,0) node[anchor=center,ground,rotate=-90,xshift=.66ex] {};
			\end{tikzpicture}~}}}

\newcommand{\eq}[2][]{
	#1
	\underset{\substack{#2}}{=}
	#1
}

\newcommand{\interp}[1]{\left\llbracket #1 \right\rrbracket}

\newcommand{\cat}[1]{\mathbf{#1}}
\newcommand{\zwtick}{\cat{ZW}^{\nmid}}

\renewcommand{\vdots}{\rotatebox[origin=c]{90}{...}}

\newtheorem{definition}{Definition}
\newtheorem{proposition}{Proposition}
\newtheorem{theorem}{Theorem}
\newtheorem{lemma}{Lemma}
\newtheorem{corollary}{Corollary}
\newtheorem{example}{Example}

\begin{document}
	
	\title{Complete Graphical Language for Hermiticity-Preserving Superoperators}
	
	\author{\textbf{Titouan Carette},\\
		Centre for Quantum Computer Science, Faculty of Computing, University of Latvia,\\ Raina 19, Riga, Latvia, LV-1586.\vspace{0.5cm}\\ 
		\textbf{Timothée Hoffreumon},\\ Centre for Quantum Information and Communication (QuIC),\\ École polytechnique de Bruxelles,\\ CP 165, Université libre de	Bruxelles, 1050 Brussels, Belgium.\vspace{0.5cm}\\
		\textbf{Emile Larroque},\\ Université Paris-Saclay, ENS Paris-Saclay, Inria, CNRS, LMF,\\ 91190, Gif-sur-Yvette, France.\vspace{0.5cm}\\
		\textbf{Renaud Vilmart},\\ Université Paris-Saclay, ENS Paris-Saclay, Inria, CNRS, LMF,\\ 91190, Gif-sur-Yvette, France.}

	\maketitle
	%
	
	
	\begin{abstract}
		Universal and complete graphical languages have been successfully designed for pure state quantum mechanics, corresponding to linear maps between Hilbert spaces, and mixed states quantum mechanics, corresponding to completely positive superoperators. In this paper, we go one step further and present a universal and complete graphical language for Hermiticity-preserving superoperators. Such a language opens the possibility of diagrammatic compositional investigations of antilinear transformations featured in various physical situations, such as the Choi-Jamio{\l}kowski isomorphism, spin-flip, or entanglement witnesses. Our construction relies on an extension of the ZW-calculus exhibiting a normal form for Hermitian matrices.
	\end{abstract}
	
	
	\section{Introduction}
	
	Experimentally, all one can infer from a process theory is an outcome distribution. In the case of quantum theory, this distribution is given by the Born rule, stating that the probability of measuring state \footnote{We use the usual Dirac notation for complex vectors ($\ket{\Psi}\in\mathbb C^n$ called \emph{ket}) and their dual ($\bra{\Psi}=\ket{\Psi}^\dag$, called \emph{bra}, and where $\dag$ is the \emph{dagger}, that is, the transpose conjugate or Hermitian adjoint). See \cite{nielsen_chuang_2010} for more info.} $\ket{\phi}$ from a prepared state $\ket{\psi}$ is $\left|\braket{\phi}{\psi}\right|^2 $. 
	A symmetry of the theory is a transformation of the states that leave this rule invariant, \textit{i.e.} $T$ is a symmetry of quantum theory if it obeys 
	\begin{equation}\label{eq:def-symmetry}
		\left|\braket{\phi}{\psi}\right|^2 = \left|\braket{T\left(\phi\right)}{T\left(\psi\right)}\right|^2 \:,
	\end{equation} 
	for any two states $\ket{\psi}$ and $\ket{\phi}$. 
	In 1931, Wigner proved that the symmetries of quantum theory should be either unitary or antiunitary (see the appendix to Chapter 20 of \cite{Wigner1931}), meaning that the general form of $T$ in Eq. \eqref{eq:def-symmetry} is 
	\begin{equation}
		\ket{T\left(\psi\right)}\: = \: U\ket{\psi} \quad \text{or} \quad \ket{T\left(\psi\right)}\: = \: U\overline{\ket{\psi}} \: ,
	\end{equation}
	where $U$ is a linear operator respecting $U^\dag U = I = UU^\dag$, and $\overline{\ket{\psi}}$ is the complex conjugation of $\ket{\psi}$.
	
	This abstract point of view turned out to be extremely useful for all aspects of quantum theory. On the one hand, the study of symmetries has been a very fruitful method for simplifying problems encountered at all scales, from solid-state physics to high-energy physics, with notable uses in atomic physics and quantum information sciences. On the other hand, symmetries have been a guiding principle for the construction of theories such as the standard model and, by extension, Yang-Mills theory (see e.g. \cite{Weinberg1995,Weinberg1996}).
	
	So far, physicists and quantum computer scientists have been mainly concerned with unitary transformations, the ubiquitous ingredient in the study of dynamics, and little attention has been devoted to antiunitary transformations.
	There is a clear reason why: unitary transformations represent what can be realised and tested in a (closed) lab. In contrast, one year after his theorem, Wigner showed that antiunitaries are typically involved in the case of a time reversal \cite{Wigner1932} (see, for example, §11.4.2 of \cite{Schwabl2008} for a source in English), something experimentalists cannot achieve (at least within the current theoretical framework of physics). Still, some quantum computer scientists have started investigating the possible computational advantages processes involving time reversal could provide. We mention in particular the complexity-theoretic study of \cite{aaronson2009closed} and the categorical semantics of \cite{pinzani2019categorical}. However, little has been done on the computational power of antiunitaries in general, although some results suggest possible advantages in quantum information \cite{Gisin1999,Massar2000,Cerf2001}, and that they may be simulable \cite{Buzek1999,Dong2019,Regula2021}. Such investigations are actually made difficult by the lack of a clearly defined computational framework because it would require going beyond the usual mathematics underlying quantum circuits. Rigorously put, antiuniatries are associated with positive but not completely positive mappings. Hence, they are outside the framework of quantum computing with open systems, and are assumed non-physical.
	
	Despite this practical limitation and even apart from speculative investigations of computational advantages, these transformations are still of fundamental importance, for example, in the proofs of the CPT and spin statistics theorems \cite{PCT}.
	In particular, there is a long history of the use of antiunitary transformation in the mathematics of quantum information theory to identify non-classical behaviour. 
	The canonical example is the Positive Partial Transpose criterion (PPT, also called Peres-Horodecki \cite{Peres1996,HORODECKI1996}), giving a necessary condition for states to be separable. 
	Transposition is indeed the prototypical example of an antiunitary transformation in the space of density matrices. It is also involved in the definition of Wooters concurrence \cite{Hill1997,Wooters1998}, and it has been argued to play a particular role in the EPR paradox as well as in quantum teleportation (see \cite{Uhlmann2016}).
	
	To better understand the role of antiunitaries in quantum weirdness, it is crucial to have a language that can handle both unitary and antiunitary transformations on the same footing.
	Therefore, this paper aims to introduce a universal and complete graphical language with enough generators to represent both transformations.
	
	Different graphical languages for depicting quantum states and evolutions exist, from the ubiquitous circuits to the more lax and abstract ZX-, ZW- and ZH-calculi. By weakening the requirement of unitarity, the three latter enjoy a richer structure, namely compact closure, making them more convenient to manipulate. All the aforementioned languages now enjoy a \emph{complete} equational theory, meaning that any two (pure) circuits or diagrams that represent the same operator can be turned into one another \cite{Backens2019complete, Clement2022complete, Hadzihasanovic2018complete, Vilmart2019nearminimal}. Using the \emph{discard}-construction \cite{Carette2019completeness}, it is possible to extend the expressiveness of the languages to \emph{mixed} states while at the same time preserving their completeness.
	
	In this paper, we add a ``anti-unitarity-inducing'' generator to one of these languages, explain how it relates to Hermiticity-preserving superoperators, and provide an equational theory, the completeness of which is proven via normal forms, rather than by a characterization of equivalence in the new semantics as it was done for mixed states. The ZW-calculus turns out to be well suited for the matter, as its pure version enjoys a natural normal form that can be leveraged in our case.
	
	All proofs are provided in the appendices.
	
	\section{Preliminaries on Pictorial Quantum Mechanics}
	
	In this section, we recall the necessary definitions and properties of the pictorial approach to quantum computing.
	
	\subsection{Props and Graphical Languages}
	
	We represent quantum processes as boxes with $n$ input and $m$ output wires, each wire corresponding to a qubit:
	
	\begin{center}
		\tikzfig{D}
	\end{center}
	
	Formally those different processes are organised into props.
	
	\begin{definition}[Prop]
		A \textbf{prop} $\cat{P}$ is a collection of sets of processes $\cat{P}[n,m]$ indexed by integers $n,m \in\mathbb{N}$, together with:
		\begin{itemize}
			\item An associative vertical composition operation: $\_\circ\_ : \cat{P}[b,c]\times \cat{P}[a,b] \to \cat{P}[a,c]$ pictured as:
			
			\begin{center}
				$\tikzfig{D2}\circ\tikzfig{D1}=\tikzfig{compo}$.
			\end{center}
			
			\item An associative horizontal composition operation: $\_\otimes\_ : \cat{P}[a,b]\times \cat{P}[c,d] \to \cat{P}[a+c ,b+d]$ pictured as:
			
			\begin{center}
				$\tikzfig{D1}\otimes\tikzfig{D2}=\tikzfig{D1}\tikzfig{D2}$
			\end{center}

			and satisfying $(f\circ g) \otimes (h\circ k)=(f\otimes h) \circ (g\otimes k) $.
			\item Identity processes $id_0 \in \cat{P}[0,0] $ and $id_1 \in \cat{P}[1,1] $, which by horizontal compositions combine into identity processes $id_n \in \cat{P}[n,n] $ pictured as:
			\[id_0:\tikzfig{empty},\qquad id_1:\tikzfig{id},\qquad id_n:\tikzfig{id-n}\]
			%
			%
			and satisfying: $f \otimes id_0 = id_0 \otimes f = id_n \circ f = f\circ id_m = f$ for all $f\in \cat{P}[n,m]$.
			\item A swap $\sigma_{1,1}\in \cat{P}[1+1, 1+1]$ that, combined with identities, provides generalized swaps $\sigma_{a,b} \in \cat{P}[a+b,b+a]$ depicted as $\tikzfig{swap}$ satisfying: $\def\fig{sigma-involutive}{\setlength{\fboxsep}{0pt}\colorbox{gray!15}{~\strut\begin{tikzpicture}
	\begin{pgfonlayer}{nodelayer}
		\node [style=dot] (0)  at (0.0, 0.5) {};
		\node [style=dot] (1)  at (0.0, 0.75) {};
		\node [style=none] (2)  at (0.0, 1.0) {};
		\node [style=dot] (5)  at (0.0, -0.5) {};
		\node [style=dot] (6)  at (0.0, -0.75) {};
		\node [style=none] (7)  at (0.0, -1.0) {};
		\node [style=white phase dot] (8)  at (-0.25, 0.0) {$r$};
		\node [style=white phase dot] (9)  at (0.25, 0.0) {$s$};
	\end{pgfonlayer}
	\begin{pgfonlayer}{edgelayer}
		\draw (0) to (2.center);
		\draw [in=90, out=-135] (0) to (8);
		\draw [in=90, out=-45] (0) to (9);
		\draw (5) to (7.center);
		\draw [in=-90, out=135] (5) to (8);
		\draw [in=-90, out=45] (5) to (9);
	\end{pgfonlayer}
\end{tikzpicture}}}\eq{}{\setlength{\fboxsep}{0pt}\colorbox{gray!15}{~\strut\begin{tikzpicture}
	\begin{pgfonlayer}{nodelayer}
		\node [style=dot] (11)  at (0.0, -0.337) {};
		\node [style=dot] (12)  at (0.0, -0.588) {};
		\node [style=none] (13)  at (0.0, -0.588) {};
		\node [style=dot] (14)  at (0.5, -0.337) {};
		\node [style=dot] (15)  at (0.5, -0.588) {};
		\node [style=none] (16)  at (0.5, -0.838) {};
		\node [style=white dot] (17)  at (0.0, 0.163) {};
		\node [style=white dot] (18)  at (0.5, 0.163) {};
		\node [style=none] (19)  at (-0.5, -0.588) {};
		\node [style=none] (20)  at (-0.5, 0.838) {};
		\node [style=white phase dot] (21)  at (0.0, 0.662) {$r$};
		\node [style=white phase dot] (22)  at (0.5, 0.662) {$s$};
	\end{pgfonlayer}
	\begin{pgfonlayer}{edgelayer}
		\draw (11) to (13.center);
		\draw (11) to (17);
		\draw (11) to (18);
		\draw (14) to (16.center);
		\draw (14) to (17);
		\draw (14) to (18);
		\draw [bend right=90, looseness=1.50] (19.center) to (13.center);
		\draw (20.center) to (19.center);
		\draw (21) to (17);
		\draw (22) to (18);
	\end{pgfonlayer}
\end{tikzpicture}}}$
			and \def\fig{naturality}${\setlength{\fboxsep}{0pt}\colorbox{gray!15}{~\strut}}\eq{}{\setlength{\fboxsep}{0pt}\colorbox{gray!15}{~\strut}}$ for all $D\in \cat{P}[n,m]$.
		\end{itemize}
		
	\end{definition}
	
	We often use the type-theoretic notation $f:n\to m $ for $f\in \cat{P}[n,m]$. 
	Processes with no inputs, i.e. those in $\cat{P}[0,m]$ for some $m$, are usually referred to as states.
	
	In category-theoretic terms, a prop is a strict symmetric monoidal category (SMC) with $(\mathbb{N},+)$ as a monoid of objects. In other words, there is a distinguished object $X$ in the category such that every object has the form $X^{\otimes n}$ for $n \in \mathbb{N}^+$. 
	In categorical quantum information theory, one usually deals with $\cat{FdHilb}$, the category whose processes (or arrows) are linear maps between finite-dimensional Hilbert spaces over complex numbers. 
	However, $\cat{FdHilb}$ is not a prop since its generating objects (wires) can have different dimensions in general. We thus need to fix the dimension of each wire.
	
	This paper will focus on the prop spanned by 2-dimensional complex vectors and its relevant sub-props.  
	We define it as:
	\begin{definition}[$\cat{Hilb}$]
		The prop $\cat{Hilb}$ has for processes $n\to m$ the linear maps $\mathbb{C}^{2^n} \to \mathbb{C}^{2^m}$. That is, the space of $2^m \times 2^n$ matrices with complex entries, denoted $\mathcal{M}_{2^m\times 2^n}(\mathbb{C})$.
	\end{definition}
	$\cat{Hilb}$ is thus the prop whose wires are qubit spaces. For this reason, it is sometimes called $\cat{Qubit}$ in the literature.
	
	A graphical language is then a set of elementary processes (or gates) called generators, together with a set of equations between diagrams formed from the generators called rules.
	
	Categorically, graphical languages are axiomatisations of props. We will often identify a graphical language with the prop it defines, and whose processes are diagrams obtained by combining generators, quotiented by the rewriting rules. A graphical language comes with an interpretation functor defining its semantics; see \cite{baez2017props} for more details. 
	
	\begin{definition}[Graphical language]
		A graphical language is a tuple $(\Sigma, E, \interp{\cdot}: G\to \cat{P})$ where $\Sigma$ is a set of processes, called the \textit{generators}; $E$ is a set of equations relating diagrams made of generators, called the \textit{rules}; $G$ is the prop axiomatised by $\Sigma$ and $E$; and $\cat{P}$ is a prop where we define the semantics. The functor $\interp{\cdot}:G \to \cat{P}$ is called the \textit{interpretation} of the graphical language in $\cat{P}$.
	\end{definition}
	
	A desirable property is that the interpretations of generators span the full semantics.
	\begin{definition}[Universality]
		A graphical language is called \textbf{universal} if $\interp{\cdot}$ is full, meaning that all processes in $\cat{P}$ can be represented by a diagram built using the tensor product and composition from swaps, identities and generators of the language.
	\end{definition}
	
	The equations are required to be sound, meaning that elements of the graphical language which are equivalent up to rewriting rules have the same interpretation.
	The converse property -- that diagrams with the same interpretation are equivalent in the graphical language -- is called completeness.
	
	\begin{definition}[Completeness]
		A graphical language is called \textbf{complete} if $\interp{\cdot}$ is faithful, meaning that the equational theory identifies together all the diagrams that have the same interpretation.
	\end{definition}
	
	Several graphical languages have been shown to be universal and complete for $\cat{Hilb}$: the ZX-calculus, the ZH-calculus and the ZW-calculus.
	
	\subsection{Doubling}
	The maps in $\cat{Hilb}$ allow us to represent pure quantum processes like unitary gates but also un-physical scalars; to represent faithfully quantum mechanics, one must shift to the `doubled' theory.
	
	In this picture, the prop corresponding to $\cat{Hilb}$ is called $\cat{Lin}$; its processes are linear maps between operators, called superoperators, and its states are operators themselves. We will often see a superoperator state $0\to n$ as a matrix in $\mathcal{M}_{2^n \times 2^n }(\mathbb{C})$. 
	
	\begin{definition}[$\cat{Lin}$]
		The prop $\cat{Lin}$ has for processes $ n\to m $ the linear maps $\mathcal{M}_{2^n \times 2^n }(\mathbb{C}) \to \mathcal{M}_{2^m \times 2^m }(\mathbb{C}) $.
	\end{definition}
	The two props are connected by a construction called `doubling'. Colloquially, it amounts to passing from states being `ket' vectors to them being `ket-bra' matrices (i.e. projectors).
	
	\begin{definition}[Doubling]
		The \textbf{doubling} prop functor $\textbf{double}: \cat{Hilb}\to \cat{Lin}$ is defined on all $A\in \mathcal{M}_{2^m \times 2^n }(\mathbb{C})$ as $\textbf{double}(A): \rho \mapsto A\rho A^{\dagger}$. In particular on a state $\ket{\phi}:0\to n $ we get: $\textbf{double}(\ket{\phi})= \ketbra{\phi}{\phi} \in  \mathcal{M}_{2^n \times 2^n }(\mathbb{C})$.
	\end{definition}
	
	The image of $\cat{Hilb}$ in $\cat{Lin}$ under doubling is a sub-prop of $\cat{Lin}$ that we define below.
	
	\begin{definition}[Pure maps]\label{def:pure}
		A linear map $ n \to m $ is said pure if it is in the image of $\textbf{double}$. 
		We denote by $\cat{Pure}$ the prop whose processes $n\to m $ are pure maps.
	\end{definition}
	
	\subsection{Process-State Duality}
	In the language of categories, the props $\cat{Hilb}$ and $\cat{Lin}$ are dagger compact categories \cite{Dodo}. It means that in addition to being SMC, they come equipped with an involution, the dagger $\dag$, as well as a distinguished state, the cap $\tikzfig{cap}: 0\to 2$, and its corresponding dagger dual, the cup $\tikzfig{cup}: 2\to 0$. Pictorially, the dagger reverses the direction of the diagrams while conjugating their coefficients. As for the cup and the cap, they satisfy the `yanking equations':
	\begin{equation}
		\label{eq:snake}
		\def\fig{snake-equations}{\setlength{\fboxsep}{0pt}\colorbox{gray!15}{~\strut}}\eq{}{\setlength{\fboxsep}{0pt}\colorbox{gray!15}{~\strut}}\eq{}{\setlength{\fboxsep}{0pt}\colorbox{gray!15}{~\strut\begin{tikzpicture}
	\begin{pgfonlayer}{nodelayer}
		\node [style=dot] (27)  at (0.25, 0.288) {};
		\node [style=dot] (28)  at (0.25, 0.038) {};
		\node [style=none] (29)  at (0.5, -0.838) {};
		\node [style=none] (33)  at (-0.5, 0.838) {};
		\node [style=white phase dot] (34)  at (0.0, 0.662) {$r$};
		\node [style=white phase dot] (35)  at (0.5, 0.662) {$s$};
		\node [style=white dot] (36)  at (0.25, -0.287) {};
	\end{pgfonlayer}
	\begin{pgfonlayer}{edgelayer}
		\draw (27) to (35);
		\draw (27) to (36);
		\draw (34) to (27);
		\draw [in=90, out=-45] (36) to (29.center);
		\draw [in=-90, out=-180] (36) to (33.center);
	\end{pgfonlayer}
\end{tikzpicture}}}\:,
	\end{equation}
	and they are invariant to the swap:
	\def\fig{swapcupcap}
	\begin{align}
		{\setlength{\fboxsep}{0pt}\colorbox{gray!15}{~\strut}}
		\eq{}{\setlength{\fboxsep}{0pt}\colorbox{gray!15}{~\strut}}
		\text{ and }{\setlength{\fboxsep}{0pt}\colorbox{gray!15}{~\strut}}
		\eq{}{\setlength{\fboxsep}{0pt}\colorbox{gray!15}{~\strut\begin{tikzpicture}
	\begin{pgfonlayer}{nodelayer}
		\node [style=none] (40)  at (0.5, -0.838) {};
		\node [style=none] (41)  at (-0.5, 0.838) {};
		\node [style=white phase dot] (42)  at (0.25, 0.163) {$r{+}s$};
		\node [style=white dot] (44)  at (0.25, -0.287) {};
	\end{pgfonlayer}
	\begin{pgfonlayer}{edgelayer}
		\draw (42) to (44);
		\draw [in=90, out=-45] (44) to (40.center);
		\draw [in=-90, out=-180] (44) to (41.center);
	\end{pgfonlayer}
\end{tikzpicture}}} \:.
	\end{align}
	
	The compactness of these props, i.e. having a cup and cap, induces a process-state (sometimes called map/state or channel/state) duality: any process can be made into a state by appending caps to all of its inputs:
	\begin{equation}
		\cat{P}[n,m] \rightarrow \cat{P}[0,m+n]\;:\;
		\tikzfig{D}\mapsto \tikzfig{process-state-duality} .
	\end{equation}
	
	In the language of quantum information, the cap is a maximally entangled state (unnormalised for equation \eqref{eq:snake} to hold), and the cup is the corresponding postselected measurement (unnormalised again). 
	Since there are several maximally entangled states, it is customary to pick $\sum_{k\in\{0,1\}^n} \ket{k}\otimes \ket{k}$. Therefore, for $\cat{Hilb}$ whose underlying Hilbert spaces are two-dimensional, it is an unnormalised version of the $\ket{\phi^+}$ Bell state. 
	
	The process-state duality of $(\cat{Fd})\cat{Hilb}$ is usually called \textit{vectorisation} in the literature \cite{gilchrist2009vectorization}. In computer science and category theory, this vectorisation process corresponds to the geometry of interaction construction $\cat{Int}$ \cite{Joyal1996traced}. In fact $\cat{Lin}$ can be seen as a full subcategory of $\cat{Int(Hilb)}$ whose objects are of the form $(n,n)$. This approached has been pionerd by the $\cat{CPM}$ construction \cite{Selinger2006CPM}. Notice that $\cat{Int(Hilb)}$ is equivalent to $\cat{Hilb}$ as $\cat{Hilb}$ is compact closed, however this equivalence is not induced by the doubling functor.
	We will be primarily interested in the process-state duality for $\cat{Lin}$. Since linear maps between density operators are easier to handle when they are turned into density operators themselves, process-state duality occurs much more often in the literature for $\cat{Lin}$ than for $\cat{Hilb}$, where it is common practice to omit the normalization of the cap. 
	This specific instance of process-state duality is usually referred to as the \textit{Choi-Jamio{\l}kowski isomorphism}.
	\begin{definition}[Choi-Jamio{\l}kowski isomorphism \cite{Jamiolkowski1972,Choi1975}]\label{def:CJ}
		Let $\mathcal{F}$ be a linear map between square matrices of dimensions $n^2$ and $m^2$. 
		The \textbf{Choi-Jamio{\l}kowski isomorphism} is a bijective isomorphism between $\mathcal{F}$ and a matrix $F$ given by
		\begin{equation}\label{eq:CJ}
			\begin{aligned}
				\left(\mathcal{M}_{n\times n}(\mathbb{C}) \rightarrow \mathcal{M}_{m\times m}(\mathbb{C})\right) \rightarrow \mathcal{M}_{nm \times nm}(\mathbb{C}):\\
				\mathcal{F} \mapsto F := \left(\mathcal{I} \otimes \mathcal{F}\right)\left(\sum_{k,l=0}^{n-1} \ketbra{k}{l}\otimes \ketbra{k}{l}\right)\:, 
			\end{aligned}
		\end{equation}
		where $\mathcal{I}$ is the identity map. 
		$F$ is called the \textbf{Choi matrix} of $\mathcal{F}$. 
		
		The \textbf{reverse direction} of the isomorphism is the image in $\mathcal{M}_{m\times m}(\mathbb{C})$ of the action of $\mathcal{F}$ on a matrix $\rho$:
		\begin{equation}\label{eq:CJ^-1}
			\mathcal{F}(\rho) = \mathrm{Tr}_{\mathcal{M}_{2^n\times 2^n}(\mathbb{C})}\left[F \left(\rho^T \otimes 1_{\mathcal{M}_{2^m\times 2^m}(\mathbb{C})} \right)\right] \:.
		\end{equation}
		In the above, $^T$ is the transposition  with respect to a fixed basis, and $\mathrm{Tr}_{\mathcal{M}_{2^n\times 2^n}(\mathbb{C})}$ is the \emph{partial trace} over subsystem $\mathcal{M}_{2^n\times 2^n}(\mathbb{C})$.
	\end{definition}	
	
	\subsection{Completely Positive Maps}
	$\cat{Pure}$ only represents closed quantum dynamics between pure states. 
	In order to be more general, we have to consider another sub-prop of $\cat{Lin}$ containing mixed-state quantum theory. That is, the prop generated by Completely Positive (CP) maps.
	\begin{definition}[Complete Positivity]\label{def:CP}
		A superoperator $\mathcal{F}$ is said to be \textbf{Completely-Positive} (CP) if, for all input positive operators $\rho$, it produces a positive operator, even if applied locally:
		\begin{equation}
			\forall\rho,\quad \rho \geq 0 \quad \Rightarrow \quad \left(\mathcal{I}\otimes\mathcal{F}\right)(\rho) \geq 0 \:,
		\end{equation}
		where $\mathcal{I}$ is the identity superoperator.
		
		We denote by $\cat{CP}$ the prop whose processes $n\to m $ are the CP superoperators in $\mathcal{M}_{2^n \times 2^{n}}(\mathbb{C}) \to \mathcal{M}_{2^m \times 2^{m}}(\mathbb{C})$.
	\end{definition}
	
	Through process-state duality, CP maps have a nice characterisation in terms of matrices:
	\begin{theorem}[Choi \cite{Choi1975}]\label{theo:Choi}
		The Choi-Jamio{\l}kowski isomorphism maps CP maps to positive semi-definite matrices.
	\end{theorem}
	
	Moreover, the structure of CP maps as an extension of pure maps is well understood: Purification -- a standard result (see e.g.~\cite{nielsen_chuang_2010}) -- states these can be obtained by composition of pure maps and partial traces. A canonical way to extend a graphical language for pure to CP maps is then to introduce a generator that will perform the partial trace. 
	In \cite{Carette2019completeness,Coecke2012environment,Coecke2008}, this generator is represented as $\ground:1\to 0$.
	
	\begin{proposition}[\!\!\cite{Coecke2008bis}]\label{prop:ground}
		Adding the discard map, i.e.~partial trace superoperator, as a generator to a universal set of generators for $\cat{Pure}$ provides a universal set of generators for completely positive maps.
	\end{proposition}
	\begin{proposition}[\!\!\cite{Carette2019completeness}]\label{prop:isometry}
		Any universal and complete graphical language for pure maps extends into a universal and complete graphical language for CP maps by adding the discard generator $\ground$ and the equation:
		\[\tikzfig{discard-isometry} = \ground\]	
		for each isometry $V$, that is, for each map $V$ satisfying $V^{\dagger}V=I$.
	\end{proposition}
	
	Proposition \ref{prop:isometry} was used in \cite{Carette2019completeness} to extend the ZX-, ZH-, and ZW-calculi, which are universal and complete for pure maps, to universal and complete graphical languages for CP maps. It is noteworthy that this can be done with a finite number of (parameterised) equations since a finite set of generators for isometries is known. 
	
	
	\section{Depicting Hermiticity-Preserving Superoperators}
	
	\subsection{Antilinearity and the Transpose Map}
	
	As the initial motivation for this work is the graphical representation of antiunitary maps, we begin by recollecting some facts about these maps.
	\begin{definition}\label{def:antilinear}
		A map $\theta:\mathbb{C}^n \to \mathbb{C}^m $ is called \emph{antilinear} (or \emph{conjugate-linear}) if it satisfies
		\begin{equation}\label{eq:def-Antilinear}
			\theta\left(c_1\ket{\psi} + c_2 \ket{\phi}\right) \:=\: \overline{c_1}\;\theta\left(\ket{\psi}\right) \:+\: \overline{c_2}\;\theta\left(\ket{\phi}\right) \:,
		\end{equation}%
		for all vectors $\ket{\psi}, \ket{\phi} \in \mathbb{C}^n$, and for all complex numbers $c_1, c_2 \in \mathbb{C}$.
	\end{definition}
	If the space has an inner product $\braket{\cdot}{\cdot}: \mathbb{C}^n \to \mathbb{C}$, the Hermitian adjoint of an antilinear map can be defined. Let $\theta$ be an antilinear map from $\mathbb{C}^n$ to $\mathbb{C}^m$, then its adjoint is the map $\theta^\dag: \mathbb{C}^m \to \mathbb{C}^n$ which obeys
	\begin{equation}\label{eq:def-HermitianAdjoint}
		\braket{y}{\theta(x)} \:=\: \braket{x}{\theta^\dag(y)} \quad \forall x \in \mathbb{C}^n\:,\: \forall y \in \mathbb{C}^m \:.
	\end{equation}
	A special class of antilinear operators are the antiunitaries. 
	\begin{definition}\label{def:antiunitary}
		An antilinear operator $A$ is an \emph{antiunitary} if it is \emph{normal}, \textit{i.e.} it commutes with its adjoint $AA^\dag = A^\dag A$, and if it conjugates the scalar product,
		\begin{equation}\label{eq:def-antiunitary}
			\braket{A(y)}{A(x)} = \overline{\braket{y}{x}} = \braket{x}{y} \:.
		\end{equation}
	\end{definition}
	
	Mathematically, antiunitaries are `one coset away' from unitaries, in the sense that they can always be written as \cite{Wigner1993}
	\begin{equation}\label{eq:A=UK}
		A = U \circ K \:,
	\end{equation}
	where $U$ is a unitary and $K$ is an antilinear involution, meaning that it obeys $K(cId) = \overline c K$ for $c\in \mathbb{C}$ and $K^2=Id$. Without loss of generality, the same $K$ can be used for all $A$. 
	Thus, constructing a graphical language for antiunitary maps amounts to depicting the map $K$ and its interaction with the $\mathbb{C}$-linear maps.
	In the case of pure state (single wire) quantum mechanics, the standard choice is to take $K$ as the operation of complex conjugation,
	\begin{equation}
		K \ket{x} = \overline{\ket{x}} \:,
	\end{equation}
	so that $K\left(c_1 \ket{0} + c_2 \ket{1}\right) = \overline{c_1} \ket{0} + \overline{c_2} \ket{1}$ for example. 
	
	This choice induces the use of the transposition as the involution for the mixed state (i.e. doubled) representation. The doubling of a pure state can be written as: $\ket{x} \mapsto \ket{x} \otimes \ket{x}^\dag = \ket{x} \otimes \bra{x} \cong \dyad{x}$. Thus, the doubling of an antiunitary action is: $A\ket{x} \mapsto A\ket{x} \otimes \left(A\ket{x}\right)^\dag = U\overline{\ket{x}} \otimes \left(U\overline{\ket{x}}\right)^\dag = U\overline{\ket{x}} \otimes \overline{\bra{x}} U^\dag \cong  U\overline{\dyad{x}}U^\dag$. And since $\dyad{x}$ is self-adjoint this yields
	\begin{equation}
		A\ket{x} \mapsto U (\dyad{x})^T U^\dag \:.
	\end{equation}
	Since the transposition is a linear map on the space of density matrices, this construction linearly extends from the pure processes to all processes.
	
	Using the result of Wigner, unitaries and antiunitaries are symmetries of density matrices,
	\begin{equation}
		\mathcal{U}(\rho) \:=\: U \rho U^\dag \quad \text{or} \quad \mathcal{A}(\rho) \: = \: U \rho^T U^\dag \:.
	\end{equation}
	Surprisingly, going to the doubled picture does not induce any new symmetries beside these two (see App. A and B of \cite{Chiribella2021} for a proof).
	
	When deriving the mixed state representation of antiunitaries, we implicitly assumed that they behave properly under the tensor product.
	Nonetheless, this is not true in general. 
	Consider the case where we want to apply a complex conjugation on one part in a bipartite state. On the one hand:
	\begin{equation}
		\begin{aligned}
			(K \otimes I)e^{i\theta}\ket{xy} & = (K \otimes I)(e^{i\theta}\ket{x})\otimes \ket{y}\\[0.5em]
			&= \overline{e^{i\theta}\ket{x}}\otimes \ket{y}
			= e^{-i\theta} \overline{\ket{x}}\otimes \ket{y}\:.
		\end{aligned}
	\end{equation}
	On the other hand:
	\begin{equation}
		\begin{aligned}
			(K \otimes I)e^{i\theta}\ket{xy} &= (K \otimes I)\ket x\otimes (e^{i\theta}\ket{y})\\[0.5em]
			&= \overline{\ket{x}}\otimes (e^{i\theta}\ket{y}) = e^{i\theta} \overline{\ket{x}}\otimes \ket{y}\:,
		\end{aligned}
	\end{equation}
	a contradiction! In reality, there is no mathematically consistent direct product of a linear and antilinear operator within the category of complex linear spaces \cite{Uhlmann2016}. However, the equality holds up to a phase, a norm one complex number, suggesting that the difficulty disappears when moving to mixed-state quantum mechanics and completely positive maps.
	As we have seen above, in mixed-state quantum mechanics, antiuniatries are represented by a transposition followed by a unitary. 
	As this representation is linear on the space of density matrices, there is no longer a \textit{mathematical} problem of mixing unitaries and antiunitaries; a map like
	\begin{equation}
		\left(\mathcal{A} \otimes \mathcal{I}\right)\left( \rho \otimes \sigma\right) = U \rho^T U^\dag \otimes \sigma 
	\end{equation}
	is well-defined. Therefore, we can establish a suitable process theory and start looking for a graphical representation of the antiunitary maps. Moreover, we already know that antiunitary maps decompose into $\mathcal{A} = \mathcal{U} \circ \mathcal{T}$, where $\mathcal{U}$ is a unitary adjoint action, and $\mathcal{T}$ is the transposition map. Consequently, the graphical study of antiunitaries reduces to adding the transposition as a new operation in mixed-states quantum mechanics. 
	
	There is nonetheless a \textit{physical} problem that remains: the transpose map is a positive (P) but not completely positive (CP) map. For this reason, it is not a valid quantum evolution; it does not map density matrices to density matrices when applied locally. It can be seen, for example, when applying a transpose on one part of a Bell state: the resulting matrix is no longer positive semi-definite; hence it cannot be a valid quantum state. Thus, adding the transposition requires one to pay the price of leaving the category of CP maps for a larger category of superoperators.
	
	\subsection{Hermiticity-Preserving Superoperators}
	
	Luckily, the sub-prop of $\cat{Lin}$ freely spanned by tensor and composition of CP maps with $\mathcal{T}$ admits a nice description as Hermiticity-preserving superoperators.
	
	\begin{definition}[Hermiticity-Preserving superoperators]
		A superoperator $\mathcal{H}$ is said to be \textbf{Hermiticity-Preserving} (HP) if for all operator $\rho$:
		\begin{equation}
			\rho = \rho^\dagger \quad \Rightarrow \quad \mathcal{H}(\rho)=\mathcal{H}(\rho)^{\dagger} \:.
		\end{equation}
		
		We denote by $\cat{HP}$ the prop whose arrows $n\to m $ are the HP superoperators in $\mathcal{M}_{2^n \times 2^{n}}(\mathbb{C}) \to \mathcal{M}_{2^m \times 2^{m}}(\mathbb{C})$.
		
	\end{definition}
	
\begin{figure}[!h]
		\centering
		\begin{tikzpicture}
	\begin{pgfonlayer}{nodelayer}
		\node [style=none] (0) at (4.75, 0) {$\cat{HP}$};
		\node [style=none] (1) at (6.5, 0) {$\cat{Lin}$};
		\node [style=none] (2) at (1.5, 0) {$\cat{Pure}$};
		\node [style=none] (3) at (3, 0) {$\cat{CP}$};
		\node [style=none] (4) at (1.5, -0.25) {};
		\node [style=none] (5) at (2.25, 0) {};
		\node [style=none] (6) at (1.5, 0.25) {};
		\node [style=none] (7) at (0.75, 0) {};
		\node [style=none] (9) at (2.25, -0.5) {};
		\node [style=none] (10) at (4, 0) {};
		\node [style=none] (11) at (2.25, 0.5) {};
		\node [style=none] (12) at (0.5, 0) {};
		\node [style=none] (19) at (3, -0.75) {};
		\node [style=none] (20) at (5.75, 0) {};
		\node [style=none] (21) at (3, 0.75) {};
		\node [style=none] (22) at (0.25, 0) {};
		\node [style=none] (29) at (3.75, -1) {};
		\node [style=none] (30) at (7.5, 0) {};
		\node [style=none] (31) at (3.75, 1) {};
		\node [style=none] (32) at (0, 0) {};
	\end{pgfonlayer}
	\begin{pgfonlayer}{edgelayer}
		\draw [in=-90, out=0, looseness=0.75] (4.center) to (5.center);
		\draw [in=0, out=90, looseness=0.75] (5.center) to (6.center);
		\draw [in=90, out=180, looseness=0.75] (6.center) to (7.center);
		\draw [in=180, out=-90, looseness=0.75] (7.center) to (4.center);
		\draw [in=-90, out=0, looseness=0.75] (9.center) to (10.center);
		\draw [in=0, out=90, looseness=0.75] (10.center) to (11.center);
		\draw [in=90, out=180, looseness=0.75] (11.center) to (12.center);
		\draw [in=180, out=-90, looseness=0.75] (12.center) to (9.center);
		\draw [in=-90, out=0, looseness=0.75] (19.center) to (20.center);
		\draw [in=0, out=90, looseness=0.75] (20.center) to (21.center);
		\draw [in=90, out=180, looseness=0.75] (21.center) to (22.center);
		\draw [in=180, out=-90, looseness=0.75] (22.center) to (19.center);
		\draw [in=-90, out=0, looseness=0.75] (29.center) to (30.center);
		\draw [in=0, out=90, looseness=0.75] (30.center) to (31.center);
		\draw [in=90, out=180, looseness=0.75] (31.center) to (32.center);
		\draw [in=180, out=-90, looseness=0.75] (32.center) to (29.center);
	\end{pgfonlayer}
\end{tikzpicture}
		\caption{Inclusion of the different props.\label{fig:venn}}
		
	\end{figure}
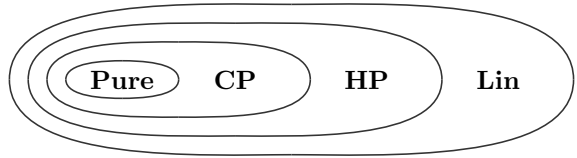

	Contrary to positive superoperators, HP superoperators are stable by tensor product. Hence there is no need to consider a notion of ``completely HP" superoperator. We have $\cat{Pure} \subset \cat{CP} \subset \cat{HP} \subset \cat{Lin}$, as represented in Figure \ref{fig:venn}.
	
	Moreover, $\cat{HP}$ has the following property:
	
	\begin{theorem}[De Pillis \cite{depillis1967}, Proposition 1.2]
		The Choi-Jamio{\l}kowski isomorphism maps HP maps to Hermitian matrices.
	\end{theorem}
	
	In category-theoretical terms, this mapping corresponds to the fact that $\cat{HP}$ is a compact closed subcategory of $\cat{Lin}$, with the same objects but morphisms restricted to HP maps. 
	Thus, to each map $n\to m$ in $\cat{HP}$, we can associate a unique state $0\to n+m$. This duality allows to show results on HP maps while manipulating only Hermitian matrices. 
	
	We can sum up the process-state duality between the different props as follows:
	
	\begin{center}
		\begin{tabular}{|c|c|c|}
			\hline
			Prop & Processes & States\\
			\hline
			$\cat{Pure}$ & pure & rank-$1$ positive Hermitian matrices \\
			\hline
			$\cat{CP}$ & CP & positive Hermitian matrices \\
			\hline
			$\cat{HP}$ & HP & Hermitian matrices \\
			\hline
			$\cat{Lin}$ & linear & matrices\\
			\hline
		\end{tabular} 
	\end{center}
	
	One may have expected the positive maps to form a sub-prop in-between CP and HP. However, they do not since P maps are not stable by tensor composition. Note for completeness that the corresponding states are the \textit{positive on pure tensor} matrices \cite{Klay1987}.

	\section{$\zwtick$-Calculus}
	
	\label{sec:zwtick-calculus}
	
	We choose the ZW-Calculus \cite{Coecke2010compositional,Hadzihasanovic2015diagrammatic} as the starting graphical language for our extension, as it features a natural normal form that will help us get completeness. For those who already the language, our extension consists in adding a $1\to1$ generator -- which we may consider as an edge type as shown in the following -- depicted as $\tikzfig{tickedge}$.
	
	One ZW-calculus is defined for each subring of $\mathbb C$ \cite{Hadzihasanovic2017algebra} stable by conjugation. This ring defines the domain in which some parameters will live. This generalisation allows us to represent different fragments of quantum computations, each time with a complete equational theory.
	For simplicity, we fix a subring $R$ right away, such that $\frac12\in R$, and take the liberty not to recall it every time. We then define $\cat{Hilb}_R$ as the subcategory of $\cat{Hilb}$ where maps have coefficients in $R$. We similarly define\footnote{Those are well defined, and correspond to superoperators whose matrix entries are elements of $R$} $\cat{Lin}_R$ and $\cat{HP}_R$.
	
	\subsection{Diagrams}
	
	The qubit $\zwtick$-Calculus is a $\dagger$-compact prop as defined above, with the following set of generators in string-diagram representation:
	\begin{itemize}
		\item $\tikzfig{Z-spider}:n\to m$
		\hfill$\left(r\in R,~\text{by convention: }\def\fig{Z-spider-1}
		\scalebox{0.8}{${\setlength{\fboxsep}{0pt}\colorbox{gray!15}{~\strut}}$}:=\scalebox{0.8}{${\setlength{\fboxsep}{0pt}\colorbox{gray!15}{~\strut}}$}\right)$
	\end{itemize}
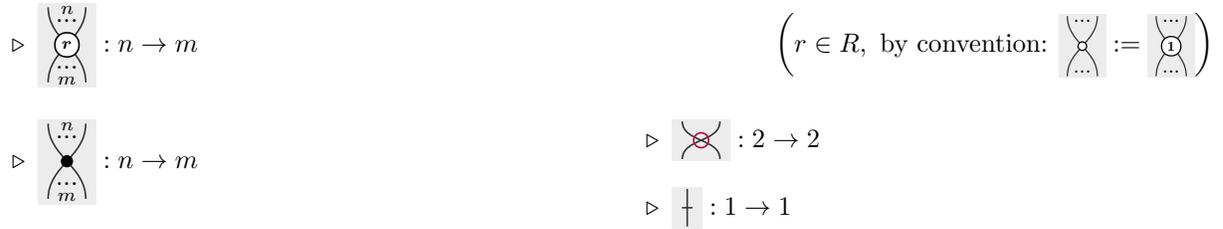
	\begin{multicols}{2}
		\begin{itemize}
			\setlength\itemsep{0.5em}
			\item $\tikzfig{W-spider}:n\to m$
			\item $\tikzfig{fswap}:2\to2$
			\item $\tikzfig{tickedge}:1\to1$
		\end{itemize}
	\end{multicols}
	
	The dagger $\dagger$ acts on the generators as follows:
	\begin{multicols}{2}
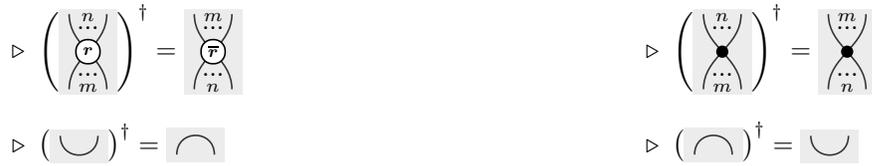
		\begin{itemize}
			\setlength{\itemsep}{0.5em}
			\item $\left(\tikzfig{Z-spider}\right)^\dagger = \tikzfig{Z-spider-dagger}$ 
			\item $\left(\tikzfig{cup}\right)^\dagger = \tikzfig{cap}$
			\item $\left(\tikzfig{W-spider}\right)^\dagger = \tikzfig{W-spider-m-n}$
			\item $\left(\tikzfig{cap}\right)^\dagger = \tikzfig{cup}$
		\end{itemize}
	\end{multicols}
	and leave the other generators unchanged.
	
	A property of these generators is that diagrams built from them can be deformed at will.
	More formally the tick, the white nodes and black nodes are \emph{flexsymmetric} \cite{DBLP:phd/hal/Carette21}, meaning that for any permutations $\sigma$ encoded by swaps we have:
	\def\fig{flex}
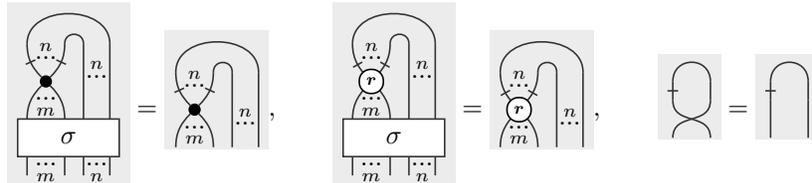
	\begin{align*}
		{\setlength{\fboxsep}{0pt}\colorbox{gray!15}{~\strut}}
		\eq{} {\setlength{\fboxsep}{0pt}\colorbox{gray!15}{~\strut}},\qquad
		{\setlength{\fboxsep}{0pt}\colorbox{gray!15}{~\strut}}
		\eq{}{\setlength{\fboxsep}{0pt}\colorbox{gray!15}{~\strut}},\qquad {\setlength{\fboxsep}{0pt}\colorbox{gray!15}{~\strut\begin{tikzpicture}
	\begin{pgfonlayer}{nodelayer}
		\node [style=white phase dot] (46)  at (0.0, 0.0) {$r{+}s$};
		\node [style=none] (47)  at (0.0, 0.5) {};
		\node [style=none] (48)  at (0.0, -0.5) {};
	\end{pgfonlayer}
	\begin{pgfonlayer}{edgelayer}
		\draw (48.center) to (47.center);
	\end{pgfonlayer}
\end{tikzpicture}}}
		\eq{}{\setlength{\fboxsep}{0pt}\colorbox{gray!15}{~\strut\input{./figures/\fig/\fig_05.tikz}}}
	\end{align*}
	
	This last equation implies in particular:
	\def\fig{tick-snake}
	\[
	{\setlength{\fboxsep}{0pt}\colorbox{gray!15}{~\strut}}
	\eq{}{\setlength{\fboxsep}{0pt}\colorbox{gray!15}{~\strut}}
	\eq{}{\setlength{\fboxsep}{0pt}\colorbox{gray!15}{~\strut}}
	\]
	Notice, however, that one should understand $\tikzfig{fswap}$ as a kind of swap as it is not flexsymmetric. We cannot freely change the order of its inputs/outputs. The axioms governing this generator will be made clear in the upcoming equational theory.
	
	In the following, we explain how to understand a $\zwtick$-diagram as a quantum operator. Equations introduced so far are sound with respect to this interpretation.
	
	\subsection{Semantics}
	
	The vanilla $\cat{ZW}$-Calculus comes with a standard interpretation which allows us to understand $\cat{ZW}$-diagrams as pure quantum operators over the ring $R$: $\interp{.}:\cat{ZW} \to \cat{Hilb}_R$.
	
	The standard interpretation $\interp{.}$ can be defined as a monoidal functor -- both sequential and parallel compositions are preserved --, acting as follows on the generators:
	\begin{itemize}
		\setlength\itemsep{0.5em}
		\item $\interp{\tikzfig{Z-spider}} = \ketbra{0^m}{0^n} + r\ketbra{1^m}{1^n}$
		\item $\displaystyle\interp{\tikzfig{W-spider}} = \sum_{\substack{x,y\in\{0,1\}^{n+ m}\\\abs{x\cdot y}=1}} \ketbra{y}{x}$
		\item $\displaystyle\interp{\tikzfig{fswap}} = \sum_{i,j\in\{0,1\}} (-1)^{ij} \ketbra{ji}{ij}$
		\item $\interp{\tikzfig{cap}} =\interp{\tikzfig{cup}}^\dagger = \ket{00} + \ket{11}$
		\item $\displaystyle\interp{\tikzfig{swap}} = \sum_{i,j\in\{0,1\}} \ketbra{ji}{ij}$
	\end{itemize}
	where $x\cdot y$ is the concatenation of the two bitstrings, and $\abs{x\cdot y}$ is the Hamming weight -- the number of non-0 symbols -- in that concatenation. For instance, $\interp{\tikzfig{W-0-3}} = \ket{001}+\ket{010}+\ket{100}$.
	
	It is possible to leverage the existence of this standard interpretation to define semantics for $\zwtick$-diagrams. To do so, we need to map $\zwtick$-diagrams to $\cat{ZW}$-diagrams in a meaningful way.
	
	First, we will define a monoidal functor to usual $\cat{ZW}$-diagrams. Although this functor is very straightforward and easy to work with, it does not provide us with the interpretation we want for $\zwtick$-diagrams, i.e.~as Hermiticity-preserving maps, in particular, because the usual semantics of $\cat{ZW}$-diagrams is into $\cat{Hilb}_R$ and not $\cat{Lin}_R$.
	
	To fix this issue, we will define another functor into a particular construction over $\cat{ZW}$-diagrams fit for describing superoperators.
	
	
	\subsubsection{Unzipping}
	\label{sec:unzip}
	We call this first functor $\operatorname{Unzip}$ to distinguish it from the usual \emph{doubling} used to include pure maps into superoperators.
	\begin{definition}
		We define the monoidal functor $\operatorname{Unzip}:\zwtick\to\cat{ZW}$ which maps any $n\to m$ diagram to a $2n\to 2m$ diagram via:\\
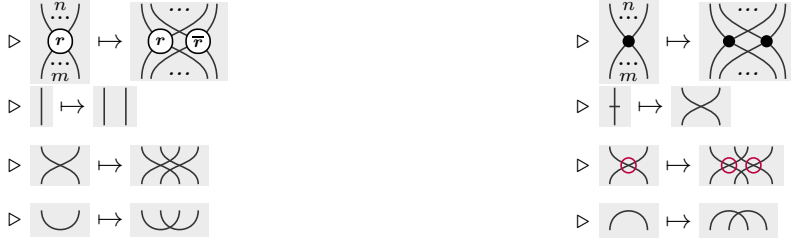
		\begin{minipage}{0.45\columnwidth}
			$\triangleright$ $\tikzfig{Z-spider}\mapsto\tikzfig{doubling-Z-spider}$
		\end{minipage}
		\begin{minipage}{0.45\columnwidth}
			$\triangleright$ $\tikzfig{W-spider}\mapsto\tikzfig{doubling-W-spider}$
		\end{minipage}\\
		\medskip
		\begin{minipage}{0.45\columnwidth}
			$\triangleright$ $\tikzfig{id}\mapsto\tikzfig{id}\tikzfig{id}$
		\end{minipage}
		\begin{minipage}{0.45\columnwidth}
			$\triangleright$ $\tikzfig{tickedge}\mapsto\tikzfig{swap}$
		\end{minipage}\\
		\medskip
		\begin{minipage}{0.45\columnwidth}
			$\triangleright$ $\tikzfig{swap}\mapsto\tikzfig{doubling-swap}$
		\end{minipage}
		\begin{minipage}{0.45\columnwidth}
			$\triangleright$ $\tikzfig{fswap}\mapsto\tikzfig{doubling-fswap}$
		\end{minipage}\\
		\medskip
		\begin{minipage}{0.45\columnwidth}
			$\triangleright$ $\tikzfig{cup}\mapsto\tikzfig{doubling-cup}$
		\end{minipage}
		\begin{minipage}{0.45\columnwidth}
			$\triangleright$ $\tikzfig{cap}\mapsto\tikzfig{doubling-cap}$
		\end{minipage}
		%
	\end{definition}
	As explained previously, it is possible to give the semantic of $\zwtick$-diagrams as $\interp{\operatorname{Unzip}(\cdot)}$, and this is actually enough to define semantical equivalence between diagrams.
	
	\subsubsection{Superoperators}
	
	Here we first wrap $\cat{ZW}$-diagrams into a category that allows us to define superoperators. It turns out that a subcategory of the so-called ``Int''-construction (see e.g.~\cite{Abramsky2002GoI}) is well suited.
	\begin{definition}
		We define the category $\cat{Lin(ZW)}$ as the prop whose arrows $n\to m$ are $n+m \to n+m$ $\cat{ZW}$-diagrams, the identity (on $1$) is \def\fig{tick-tick}${\setlength{\fboxsep}{0pt}\colorbox{gray!15}{~\strut}}$, and the compositions are defined as follows:
		\begin{itemize}
			\setlength{\itemsep}{0.5em}
			\item \def\fig{Int-compo}${\setlength{\fboxsep}{0pt}\colorbox{gray!15}{~\strut}}\circ{\setlength{\fboxsep}{0pt}\colorbox{gray!15}{~\strut}}
			\eq{}{\setlength{\fboxsep}{0pt}\colorbox{gray!15}{~\strut}}$
			\item \def\fig{Int-tensor}${\setlength{\fboxsep}{0pt}\colorbox{gray!15}{~\strut}}\otimes{\setlength{\fboxsep}{0pt}\colorbox{gray!15}{~\strut}}
			\eq{}{\setlength{\fboxsep}{0pt}\colorbox{gray!15}{~\strut}}$
		\end{itemize}
	\end{definition}
	
	Notice that $\cat{Lin(ZW)}$ corresponds to the full subcategory of $\cat{Int(ZW)}$ whose object are of the form $(n,n)$. The $\centerdot$ symbols in the definition above have no semantical meaning, they are merely visual cues to avoid specifying the domain and codomain of each diagram (which would otherwise be necessary, as we have to distinguish between e.g.~$1\to1$ from $0\to 2$ morphisms). This construction forms a monoidal category \cite{Joyal1996traced}. 
	
	Intuitively, time flows from left to right in $\cat{Lin(ZW)}$-diagrams, and $\centerdot$ separates the inputs, on its left, from the outputs on its right.
	It is possible to turn $\cat{Lin(ZW)}[n, m]$-diagrams into usual $\cat{ZW}[n+m , n+m]$-diagrams through functor $\iota$, which, graphically, simply removes the $\centerdot$ symbols.
	
	We may define an interpretation $\interp{\cdot}_{\cat{Lin}}$ of such diagrams, this time as superoperators, i.e.~$\interp{\cdot}_{\cat{Lin}}:\cat{Lin(ZW)}\to\cat{Lin}$, as:
	\[\def\fig{Int-compo}\interp{{\setlength{\fboxsep}{0pt}\colorbox{gray!15}{~\strut}}}_{\cat{Lin}} := \rho \mapsto \interp{\tikzfig{interp-int-zw}}\]
	where $\rho\in\mathcal M_{2^n\times2^n}(R)$ and $D_{\rho}$ is any $\cat{ZW}$-diagram such that $\interp{D_\rho}=\rho$. Such a diagram always exists by universality of $\cat{ZW}$-diagrams \cite{Hadzihasanovic2017algebra}, and its choice does not change the result of the interpretation.
	
	We may now define how to understand $\zwtick$-diagrams as $\cat{Lin(ZW)}$-diagrams:	
	\begin{definition}
		We define the monoidal functor $\operatorname{HP}:\zwtick\to \cat{Lin(ZW)}$ that acts on generators as follows:\\
		$\tikzfig{tickedge}\mapsto \hspace*{0.6em} \begin{array}{c}
			\mathclap{\tikzfig{cup}}\!\raisebox{0.4em}{$\centerdot$}\\
			\mathclap{\tikzfig{cap}}\!\raisebox{0em}{$\centerdot$}
		\end{array}$\hfill and \hfill
		$\tikzfig{f} \mapsto \tikzfig{HP-f-pure}$ if $f$ is $\tikzfig{tickedge}$-free
	\end{definition}
	We may illustrate the use of $\centerdot$ by emphasising that:
	\begin{align*}
		\operatorname{HP}\left(\tikzfig{cap}\right) = \begin{array}{c@{}c}
			\raisebox{0.3em}{$\centerdot$}&\tikzfig{cup}\\
			\centerdot&\tikzfig{cap}
		\end{array}
		\hspace*{4em}
		\operatorname{HP}\left(\tikzfig{tickedge}\right) =~~~ \begin{array}{c}
			\mathclap{\tikzfig{cup}}\!\raisebox{0.4em}{$\centerdot$}\\
			\mathclap{\tikzfig{cap}}\!\raisebox{0em}{$\centerdot$}
		\end{array}
	\end{align*}
	\begin{align*}
		\operatorname{HP}\left(\tikzfig{cup}\right) = \begin{array}{c@{}c}
			\tikzfig{cup}&\raisebox{0.3em}{$\centerdot$}\\
			\tikzfig{cap}&\centerdot
		\end{array}
	\end{align*}

	The map's name $\operatorname{HP}$ stands for ``Hermiticity-preserving'', since, as we will see in the following, states of $\zwtick$ represent Hermitian operators, and morphisms represent superoperators that preserve Hermiticity.
	
	We can show that the two semantics above are isomorphic. There exists an invertible map $\operatorname{\Psi}$ such that the following diagram commutes:
	\begin{equation}
		\label{eq:semantics-isomorphism}
		\begin{tikzpicture}
	\begin{pgfonlayer}{nodelayer}
		\node [style=box-no-outline] (0) at (-4.75, 1) {$\zwtick$};
		\node [style=box-no-outline] (1) at (-4.75, -1) {$\operatorname{HP}(\cat{\zwtick})$};
		\node [style=box-no-outline] (2) at (-2.25, 1) {$\operatorname{Unzip}(\zwtick)$};
		\node [style=none, font={\scriptsize}] (3) at (-5, 0) {$\operatorname{HP}$};
		\node [style=none, font={\scriptsize}] (4) at (-3.875, 1.125) {$\operatorname{Unzip}$};
		\node [style=box-no-outline] (8) at (1.25, -1) {$\cat{Lin}_R$};
		\node [style=box-no-outline] (9) at (1.25, 1) {$\cat{Hilb}_R$};
		\node [style=none] (11) at (0.5, 1.25) {$\interp{.}$};
		\node [style=none] (12) at (-0.25, -1.25) {$\interp{.}_{\cat{Lin}}$};
		\node [style=box-no-outline] (13) at (-0.25, 1) {$\cat{ZW}$};
		\node [style=box-no-outline] (14) at (-2, -1) {$\cat{Lin(ZW)}$};
		\node [style=none] (15) at (1, 0) {$\iota$};
		\node [style=none] (16) at (-3.75, 0.25) {$\operatorname{\Psi}$};
	\end{pgfonlayer}
	\begin{pgfonlayer}{edgelayer}
		\draw [style={arrows={->[]}}] (0) to (2);
		\draw [style={arrows={->[]}}] (0) to (1);
		\draw [style={arrows={->[]}}] (8) to (9);
		\draw [style={arrows={Hooks[right]->[]}}] (2) to (13);
		\draw [style={arrows={->[]}}] (13) to (9);
		\draw [style={arrows={Hooks[right]->[]}}] (1) to (14);
		\draw [style={arrows={->[]}}] (14) to (8);
		\draw [style={arrows={->[]}}] (2) to (1);
	\end{pgfonlayer}
\end{tikzpicture}
	\end{equation}
	We define $\operatorname{\Psi}$ on all $\operatorname{Unzip}(\zwtick)$ $f:2n\to2m$, and $\operatorname{\Psi}^{-1}$ as:
	\[\operatorname{\Psi}\left(\tikzfig{f-2n-2m}\right):= \tikzfig{double-to-HP}\]
	\[\operatorname{\Psi}^{-1}\left(\tikzfig{f-HP}\right):=\tikzfig{HP-to-double}\]
	
	
	\begin{proposition}
		\label{prop:semantics-are-isomorphic}
		The above functors $\operatorname{\Psi}$ and $\operatorname{\Psi}^{-1}$ are inverses of one another.
	\end{proposition}
	
	It is now possible to combine the standard interpretation $\interp{.}$ of $\cat{ZW}$-diagrams with either $\operatorname{Unzip}$ or $\operatorname{HP}$ to define a standard interpretation $\interp{.}^\nmid$ for $\zwtick$ diagrams. 
	Regarding the upcoming universality of the language, it is more natural to use $\operatorname{HP}$. Hence we choose $\interp{.}^\nmid := \interp{\operatorname{HP}(.)}_{\cat{Lin}}$.
	
	As we will see in the following, all the equations of $\dagger$-compactness preserve the semantics: if $D_1 = D_2$ then $\interp{D_1}^\nmid=\interp{D_2}^\nmid$. The reciprocal is the problem of \emph{completeness}: capturing all diagrammatic transformations that keep the semantics unchanged. In the matter, the equations of $\dagger$-compactness are not enough, we need to add axioms to the equational theory.
	
	\subsection{Equational Theory}
	
	The vanilla $\cat{ZW}$-Calculus already has a complete equational theory for $\cat{Hilb}_R$. It is reminded\footnote{The equational theory provided here is slightly different from the one provided in e.g.~\cite{Hadzihasanovic2018complete}. The proof that the latter can be recovered from the equational theory provided here is made in \Cref{sec:pure-zw}.} in \Cref{fig:ZW_rules}.
	
	\begin{figure*}
		\centering
		\begin{subfigure}{1\textwidth}
			\centering
			\begin{minipage}{0.3\textwidth}
				\begin{align*}
					\def\fig{Z-spider-rule}{\setlength{\fboxsep}{0pt}\colorbox{gray!15}{~\strut\input{./figures/\fig/\fig_00.tikz}}}\eq{}{\setlength{\fboxsep}{0pt}\colorbox{gray!15}{~\strut\input{./figures/\fig/\fig_01.tikz}}}\label{ax:Z-spider}\tag{zs}
				\end{align*}
			\end{minipage}\hspace*{1em}
			\begin{minipage}{0.2\textwidth}
				\begin{align*}
					\def\fig{Z-binary-rule}{\setlength{\fboxsep}{0pt}\colorbox{gray!15}{~\strut\input{./figures/\fig/\fig_00.tikz}}}\eq{}{\setlength{\fboxsep}{0pt}\colorbox{gray!15}{~\strut\input{./figures/\fig/\fig_01.tikz}}}\label{ax:Z-id}\tag{id}
				\end{align*}
			\end{minipage}\hspace*{1em}
			\begin{minipage}{0.25\textwidth}
				\begin{align*}
					\def\fig{f-loop-rule}{\setlength{\fboxsep}{0pt}\colorbox{gray!15}{~\strut\input{./figures/\fig/\fig_00.tikz}}}\eq{}{\setlength{\fboxsep}{0pt}\colorbox{gray!15}{~\strut\input{./figures/\fig/\fig_01.tikz}}}\label{ax:f-loop}\tag{fl}
				\end{align*}
			\end{minipage}

			\begin{minipage}{0.3\textwidth}
				\begin{align*}
					\def\fig{W-spider-rule}{\setlength{\fboxsep}{0pt}\colorbox{gray!15}{~\strut\input{./figures/\fig/\fig_00.tikz}}}\eq{}{\setlength{\fboxsep}{0pt}\colorbox{gray!15}{~\strut\input{./figures/\fig/\fig_01.tikz}}}\label{ax:W-spider}\tag{ws}
				\end{align*}
			\end{minipage}\hspace*{1em}
			\begin{minipage}{0.2\textwidth}
				\begin{align*}
					\def\fig{W-binary-involutive}{\setlength{\fboxsep}{0pt}\colorbox{gray!15}{~\strut\input{./figures/\fig/\fig_00.tikz}}}\eq{}{\setlength{\fboxsep}{0pt}\colorbox{gray!15}{~\strut\input{./figures/\fig/\fig_01.tikz}}}\label{ax:W-inv}\tag{in}
				\end{align*}
			\end{minipage}\hspace*{1em}
			\begin{minipage}{0.25\textwidth}
				\begin{align*}
					\def\fig{fswap-removal}{\setlength{\fboxsep}{0pt}\colorbox{gray!15}{~\strut\input{./figures/\fig/\fig_00.tikz}}}\eq{}{\setlength{\fboxsep}{0pt}\colorbox{gray!15}{~\strut\input{./figures/\fig/\fig_01.tikz}}}\label{ax:fswap-removal}\tag{rm}
				\end{align*}
			\end{minipage}

			\begin{minipage}{0.2\textwidth}
				\begin{align*}
					\def\fig{W-binary-diffusion}{\setlength{\fboxsep}{0pt}\colorbox{gray!15}{~\strut\input{./figures/\fig/\fig_00.tikz}}}\eq{}{\setlength{\fboxsep}{0pt}\colorbox{gray!15}{~\strut\input{./figures/\fig/\fig_01.tikz}}}\label{ax:W-binary-diffusion}\tag{di}
				\end{align*}
			\end{minipage}\hspace*{1em}
			\begin{minipage}{0.2\textwidth}
				\begin{align*}
					\def\fig{Z-W-bialgebra}{\setlength{\fboxsep}{0pt}\colorbox{gray!15}{~\strut\input{./figures/\fig/\fig_00.tikz}}}\eq{}{\setlength{\fboxsep}{0pt}\colorbox{gray!15}{~\strut\input{./figures/\fig/\fig_01.tikz}}}\label{ax:Z-W-bialgebra}\tag{b}
				\end{align*}
			\end{minipage}\hspace*{1em}
			\begin{minipage}{0.2\textwidth}
				\begin{align*}
					\def\fig{Z-W-Hopf}{\setlength{\fboxsep}{0pt}\colorbox{gray!15}{~\strut\input{./figures/\fig/\fig_00.tikz}}}\eq{}{\setlength{\fboxsep}{0pt}\colorbox{gray!15}{~\strut\input{./figures/\fig/\fig_01.tikz}}}\label{ax:Z-W-Hopf}\tag{ho}
				\end{align*}
			\end{minipage}\hspace*{1em}
			\begin{minipage}{0.2\textwidth}
				\begin{align*}
					\def\fig{sum-rule}{\setlength{\fboxsep}{0pt}\colorbox{gray!15}{~\strut\input{./figures/\fig/\fig_00.tikz}}}\eq{}{\setlength{\fboxsep}{0pt}\colorbox{gray!15}{~\strut\input{./figures/\fig/\fig_01.tikz}}}\label{ax:sum}\tag{ad}
				\end{align*}
			\end{minipage}

			\begin{minipage}{0.25\textwidth}
				\begin{align*}
					\def\fig{W-bialgebra}{\setlength{\fboxsep}{0pt}\colorbox{gray!15}{~\strut\input{./figures/\fig/\fig_00.tikz}}}\eq{}{\setlength{\fboxsep}{0pt}\colorbox{gray!15}{~\strut\input{./figures/\fig/\fig_01.tikz}}}\label{ax:W-bialgebra}\tag{bw}
				\end{align*}
			\end{minipage}\hspace*{1em}
			\begin{minipage}{0.25\textwidth}
				\begin{align*}
					\def\fig{fswaps-through-W}{\setlength{\fboxsep}{0pt}\colorbox{gray!15}{~\strut\input{./figures/\fig/\fig_00.tikz}}}\eq{}{\setlength{\fboxsep}{0pt}\colorbox{gray!15}{~\strut\input{./figures/\fig/\fig_01.tikz}}}\label{ax:fswaps-through-W}\tag{fw}
				\end{align*}
			\end{minipage}\hspace*{1em}
			\begin{minipage}{0.25\textwidth}
				\begin{align*}
					\def\fig{fswap-through-Z}{\setlength{\fboxsep}{0pt}\colorbox{gray!15}{~\strut\input{./figures/\fig/\fig_00.tikz}}}\eq{}{\setlength{\fboxsep}{0pt}\colorbox{gray!15}{~\strut\input{./figures/\fig/\fig_01.tikz}}}\label{ax:fswap-through-Z}\tag{fz}
				\end{align*}
			\end{minipage}

			\begin{minipage}{0.2\textwidth}
				\begin{align*}
					\def\fig{fswap-involution}{\setlength{\fboxsep}{0pt}\colorbox{gray!15}{~\strut\input{./figures/\fig/\fig_00.tikz}}}\eq{}{\setlength{\fboxsep}{0pt}\colorbox{gray!15}{~\strut\input{./figures/\fig/\fig_01.tikz}}}\label{ax:fswap-involution}\tag{fi}
				\end{align*}
			\end{minipage}\hspace*{1em}
			\begin{minipage}{0.22\textwidth}
				\begin{align*}
					\def\fig{fswap-YB}{\setlength{\fboxsep}{0pt}\colorbox{gray!15}{~\strut\input{./figures/\fig/\fig_00.tikz}}}\eq{}{\setlength{\fboxsep}{0pt}\colorbox{gray!15}{~\strut\input{./figures/\fig/\fig_01.tikz}}}\label{ax:fswap-YB}\tag{yb}
				\end{align*}
			\end{minipage}\hspace*{1em}
			\begin{minipage}{0.2\textwidth}
				\begin{align*}
					\def\fig{fswap-rotated}{\setlength{\fboxsep}{0pt}\colorbox{gray!15}{~\strut\input{./figures/\fig/\fig_00.tikz}}}\eq{}{\setlength{\fboxsep}{0pt}\colorbox{gray!15}{~\strut\input{./figures/\fig/\fig_01.tikz}}}\label{ax:fswap-rotated}\tag{fr}
				\end{align*}
			\end{minipage}\hspace*{1em}
			\begin{minipage}{0.2\textwidth}
				\begin{align*}
					\def\fig{fswap-swapped}{\setlength{\fboxsep}{0pt}\colorbox{gray!15}{~\strut\input{./figures/\fig/\fig_00.tikz}}}\eq{}{\setlength{\fboxsep}{0pt}\colorbox{gray!15}{~\strut\input{./figures/\fig/\fig_01.tikz}}}\label{ax:fswap-swapped}\tag{fs}
				\end{align*}
			\end{minipage}
			\caption{Set of rules for the $\cat{ZW}$-Calculus, with $r,s\in R$. In \eqref{ax:Z-W-bialgebra}, either $n=m=0$ or $n>0$.}
			\label{fig:ZW_rules}
		\end{subfigure}
		\hfill
		\begin{subfigure}{0.44\textwidth}
			\centering
			\begin{minipage}{0.45\columnwidth}
				\begin{align*}
					\def\fig{tick-through-Z-spider}{\setlength{\fboxsep}{0pt}\colorbox{gray!15}{~\strut\input{./figures/\fig/\fig_00.tikz}}}\eq{}{\setlength{\fboxsep}{0pt}\colorbox{gray!15}{~\strut\input{./figures/\fig/\fig_01.tikz}}}\label{ax:tick-through-Z-spider}\tag{nz}
				\end{align*}
			\end{minipage}\hspace*{1em}
			\begin{minipage}{0.45\columnwidth}
				\begin{align*}
					\def\fig{tick-through-W-spider}{\setlength{\fboxsep}{0pt}\colorbox{gray!15}{~\strut\input{./figures/\fig/\fig_00.tikz}}}\eq{}{\setlength{\fboxsep}{0pt}\colorbox{gray!15}{~\strut\input{./figures/\fig/\fig_01.tikz}}}\label{ax:tick-through-W-spider}\tag{nw}
				\end{align*}
			\end{minipage}

			\begin{minipage}{0.45\columnwidth}
				\begin{align*}
					\def\fig{tick-through-fswap}{\setlength{\fboxsep}{0pt}\colorbox{gray!15}{~\strut\input{./figures/\fig/\fig_00.tikz}}}\eq{}{\setlength{\fboxsep}{0pt}\colorbox{gray!15}{~\strut\input{./figures/\fig/\fig_01.tikz}}}\label{ax:tick-through-fswap}\tag{nf}
				\end{align*}
			\end{minipage}\hspace*{1em}
			\caption{Naturality axioms for $\tikzfig{tickedge}$.}
			\label{fig:tick-naturality}
		\end{subfigure}
		\hfill
		\begin{subfigure}{0.54\textwidth}
			\centering
			\begin{minipage}{0.48\columnwidth}
				\begin{align*}
					\def\fig{Axiom-NF-Z}{\setlength{\fboxsep}{0pt}\colorbox{gray!15}{~\strut\input{./figures/\fig/\fig_00.tikz}}}\eq{}{\setlength{\fboxsep}{0pt}\colorbox{gray!15}{~\strut\input{./figures/\fig/\fig_01.tikz}}}\label{ax:NF-Z}\tag{zt}
				\end{align*}
			\end{minipage}\hspace*{1em}
			\begin{minipage}{0.45\columnwidth}
				\begin{align*}
					\def\fig{Axiom-grounds-aux}{\setlength{\fboxsep}{0pt}\colorbox{gray!15}{~\strut\input{./figures/\fig/\fig_00.tikz}}}\eq{}{\setlength{\fboxsep}{0pt}\colorbox{gray!15}{~\strut\input{./figures/\fig/\fig_01.tikz}}}\label{ax:tick-loop}\tag{tl}
				\end{align*}
			\end{minipage}

			\begin{minipage}{0.35\columnwidth}
				\begin{align*}
					\def\fig{Axiom-tensor-1}{\setlength{\fboxsep}{0pt}\colorbox{gray!15}{~\strut\input{./figures/\fig/\fig_00.tikz}}}\eq{}{\setlength{\fboxsep}{0pt}\colorbox{gray!15}{~\strut\input{./figures/\fig/\fig_01.tikz}}}\label{ax:nf-tensor-1}\tag{th}
				\end{align*}
			\end{minipage}\hspace*{1em}
			\begin{minipage}{0.56\columnwidth}
				\begin{align*}
					\def\fig{Axiom-tensor-2}{\setlength{\fboxsep}{0pt}\colorbox{gray!15}{~\strut\input{./figures/\fig/\fig_00.tikz}}}\eq{}{\setlength{\fboxsep}{0pt}\colorbox{gray!15}{~\strut\input{./figures/\fig/\fig_01.tikz}}}\label{ax:nf-tensor-2}\tag{td}
				\end{align*}
			\end{minipage}
			\caption{Additional axioms.}
			\label{fig:more-axioms}
		\end{subfigure}
		
		\caption{All $\zwtick$ axioms.}
		\label{fig:all-axioms}
	\end{figure*}
	%
	%
	Each of these equations remains sound in $\zwtick$, but they are not enough for completeness. For instance, the following equation: \def\fig{tick-tick}${\setlength{\fboxsep}{0pt}\colorbox{gray!15}{~\strut}} = \tikzfig{id}$ which is sound since:
	\begin{align*}
		\operatorname{HP}\left({\setlength{\fboxsep}{0pt}\colorbox{gray!15}{~\strut}}\right)
		={\setlength{\fboxsep}{0pt}\colorbox{gray!15}{~\strut}}
		={\setlength{\fboxsep}{0pt}\colorbox{gray!15}{~\strut}}
		=\operatorname{HP}\left(~\tikzfig{id}~\right)
	\end{align*}
	or equivalently:
	\begin{align*}
		\operatorname{Unzip}\left({\setlength{\fboxsep}{0pt}\colorbox{gray!15}{~\strut}}\right)
		=\def\fig{sigma-involutive}
		{\setlength{\fboxsep}{0pt}\colorbox{gray!15}{~\strut}}
		\eq{}{\setlength{\fboxsep}{0pt}\colorbox{gray!15}{~\strut}}
		=\operatorname{Unzip}\left(~\tikzfig{id}~\right)
	\end{align*}
	cannot be inferred from an equational theory that contains no mention of $\tikzfig{tickedge}$. We thus need to add axioms that are specific to the new generator. A first series is straightforward to obtain under the idea that the tick $\tikzfig{tickedge}$ is natural between identity and conjugation. These are gathered in \Cref{fig:tick-naturality}.
	
	
	Those rules are enough to derive the previous equation:
	\def\fig{tick-involution}
	\begin{align*}
		\label{eq:tick-involution}
		{\setlength{\fboxsep}{0pt}\colorbox{gray!15}{~\strut}}
		\eq{}{\setlength{\fboxsep}{0pt}\colorbox{gray!15}{~\strut}}
		\eq{\eqref{ax:Z-id}}{\setlength{\fboxsep}{0pt}\colorbox{gray!15}{~\strut}}
		\eq{\eqref{ax:tick-through-Z-spider}}{\setlength{\fboxsep}{0pt}\colorbox{gray!15}{~\strut}}
		\eq{\eqref{ax:Z-id}}{\setlength{\fboxsep}{0pt}\colorbox{gray!15}{~\strut}}
		\tag{ti}
	\end{align*}
	
	However, it seems to be not enough to reach completeness. Driven by the upcoming proof through normal forms, we add a few new axioms, presented in \Cref{fig:more-axioms}.
	%
	%
	Formally, when a series of rewrites from Figures \ref{fig:ZW_rules}, \ref{fig:tick-naturality} and \ref{fig:more-axioms} are used to transform $D_1$ into $D_2$ we should write $\zwtick\vdash D_1=D_2$. Since there is only one axiomatisation in this paper, we take the liberty to not specify ``$\zwtick\vdash$''. Any equality between diagrams from now on should be understood as an equality permitted by the axiomatisation. Notice that since the whole axiomatisation of $\cat{ZW}$ is contained in that of $\zwtick$, we have $\cat{ZW}\vdash D_1=D_2 \implies \zwtick\vdash D_1=D_2$. Moreover, by completeness of $\cat{ZW}$, any sound equation between $\tikzfig{tickedge}$-free diagrams is necessarily provable in $\zwtick$.
	
	\begin{proposition}[Soundness]
		\label{prop:soundness}
		For any two $\zwtick$-diagrams $D_1$ and $D_2$:
		$\quad\displaystyle\zwtick\vdash D_1=D_2 \implies \interp{D_1}^\nmid=\interp{D_2}^\nmid$
	\end{proposition}

	%
	%
	%
	%
	%
	%
	%

	\section{Universality and Normal Form}
	
	\label{sec:universality-NF}
	
	We show in this section that diagrams of $\zwtick$ capture exactly Hermiticity-preserving operators over $R$. We can first show this claim for states:
	
	\begin{proposition}
		\label{prop:states-are-hermitian}
		The states of $\zwtick$ represent Hermitian operators, i.e.: 
		$\quad\displaystyle\forall f\in \zwtick[0,n],~~\iota\left(\interp{f}^\nmid\right) = \iota\left(\interp{f}^\nmid\right)^\dagger$.
	\end{proposition}
	
	We can then lift this result to arbitrary morphisms of $\zwtick$:
	
	\begin{corollary}
		Diagrams of $\zwtick$ represent Hermiticity-preserving superoperators. Indeed, diagrams of $\zwtick$ map states of $\zwtick$ to states of $\zwtick$, thereby mapping any Hermitian operator to another Hermitian operator.
	\end{corollary}
	In other words the restriction of $\interp{\cdot}_{\cat{Lin}}$ to $\operatorname{HP}(\zwtick)$ maps into $\cat{HP}_R$. Accordingly, we consider the codomain of $\interp{\cdot}^\nmid$ to be $\cat{HP}_R$.
	
	It remains to show the reciprocal: that a $\zwtick$-diagram can represent any morphism of $\cat{HP}_R$. To simplify the proof, we notice that the process-state duality can be adapted here to make $n\to m$ superoperators and $0\to n+m$ superoperators isomorphic. It allows us to focus on states only. We then give a map that will build a $\zwtick$-diagram from any Hermitian operator and show that the obtained diagram does indeed represent the operator:

	\begin{definition}[Normal Form]
		Let $\mathcal N$ be the map that associates every state $f\in\cat{HP}_R[0,n]$ with a $\zwtick[0,n]$-diagram in the following way:
		\[f=\sum_{i=1}^m \lambda_i\ketbra{\vec x_i}{\vec y_i} \mapsto \tikzfig{NF}\]
		where $\ket{\vec x_i} = \bra{\vec x_i}^\dagger = \ket{x_{i_1},...,x_{i_n}}$, $x_{i_j}\in\{0,1\}$ (similarly for $\vec y$), and such that, between the white node with parameter $\lambda_i/2$ and the $k$-th output, there is:
		\begin{itemize}
			\item an edge $\tikzfig{id}$ if $x_{i_k}=1$
			\item a ticked edge $\tikzfig{tickedge}$ if $y_{i_k}=1$
		\end{itemize}
	\end{definition}
	Notice that there can be both a plain edge and a ticked edge between a white node and an output.
	We call this form \emph{normal form}, as it will be used later in the proof of completeness. As shown in \Cref{prop:NF-preserves-semantics}, the map $\mathcal N$ creates a diagram whose semantics is the Hermitian operator.
	
	\begin{example}
		\label{ex:NF-1-qb}
		An arbitrary $1$-qubit Hermitian operator will be mapped as follows:
		\begin{align*}
			\mathcal N \left(\begin{pmatrix}a&\overline b\\b&c\end{pmatrix}\right) = \tikzfig{NF-example}
		\end{align*}
	\end{example}
	
	Although the normal form defined above behaves well for the proof of completeness, it seems not to use the Hermiticity of the morphism. 
	If we, temporarily, allowed $f$ to be any square matrix, $\mathcal N(f)$ would actually build a diagram representing $\frac12(f+f^\dagger)$ which is always Hermitian, and is equal to $f$ iff $f$ is Hermitian. A Hermitian matrix is completely defined by its (upper or lower) triangular coefficients. A diagram that more closely resembles this form can be obtained by merging together nodes that correspond to symmetric coefficients. To do so, we can use \eqref{ax:tick-through-Z-spider} on one of the two nodes, which will make both have exactly the same neighbourhood, and then apply:
	
	\noindent
	\begin{minipage}[t]{0.49\columnwidth}
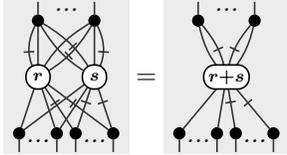
		\begin{lemma}
			\label{lem:sum-branch-NF}
			\def\fig{lemma-sum-branch-NF}
			\begin{align*}
				{\setlength{\fboxsep}{0pt}\colorbox{gray!15}{~\strut}}
				\eq{}{\setlength{\fboxsep}{0pt}\colorbox{gray!15}{~\strut}}
			\end{align*}
		\end{lemma}
	\end{minipage}
	
	\bigskip
	A diagram that has the structure of normal form except that all its symmetric coefficients are merged, is called in \emph{reduced normal form}. For instance, the diagram of \Cref{ex:NF-1-qb} rewritten this way gives:
	\begin{center}
		\tikzfig{NF-reduced-example}
	\end{center}

	The map $\mathcal N$ creates a $\zwtick$-state with the semantics of the starting Hermitian operator, i.e.:
	\begin{proposition}
		\label{prop:NF-preserves-semantics}
		For all $f\in \cat{HP}_R[0, n]$, $\interp{\mathcal N(f)}^\nmid = f$.
	\end{proposition}
	
	\begin{corollary}[Universality]
		For any $f\in \cat{HP}_R[0, n]$, there exists $D\in\zwtick$, such that:$\quad\displaystyle\interp{D}^\nmid = f$
	\end{corollary}
	
	\section{Capturing Completely Positive Maps}\label{sec:CPM}
	
	In the last section, we proved that adding the tick generator to the ZW-calculus provides a sound and universal graphical language whose interpretation yields the set of Hermiticity-Preserving maps. Considering that the set of $\cat{CP}_R$ maps (the intersection of $\cat{CP}$ and $\cat{HP}_R$ maps) is a subset of the $\cat{HP}_R$ maps, the discard generator should be contained within $\zwtick$. In this section, we show how to represent specifically the $\cat{CP}_R$ maps and how the axioms for the naturality of $\tikzfig{tickedge}$ are essentially enough for completeness in that fragment.
	
	Discard should have the following images in $\operatorname{unzip}$ and $\operatorname{HP}$:
	\[\operatorname{unzip}\left(\ground\right) = \tikzfig{cup} \hspace*{4em}
	\operatorname{HP}\left(\ground\right) = \def\fig{ground-with-tick}{\setlength{\fboxsep}{0pt}\colorbox{gray!15}{~\strut}}\:.\]
	This generator can be represented in $\zwtick$ as:
	\def\fig{ground-with-tick}
	\begin{align*}
		\operatorname{HP}\left({\setlength{\fboxsep}{0pt}\colorbox{gray!15}{~\!\!\!\!\strut\!\!\!\!}}\right)
		\eq{}{\setlength{\fboxsep}{0pt}\colorbox{gray!15}{~\strut}}
		\eq{}{\setlength{\fboxsep}{0pt}\colorbox{gray!15}{~\strut}}
		=\operatorname{HP}	\left(\ground\right)\:.
	\end{align*}
	Reference \cite{Carette2019completeness} also provides a simple set of rules concerning $\ground$ that makes the extended language complete for CP maps. When replacing $\ground$ by ${\setlength{\fboxsep}{0pt}\colorbox{gray!15}{~\!\!\!\!\strut\!\!\!\!}}$, these equations become:
	\begin{equation}
		\label{eq:ground-ax}
		\begin{array}{c}
			\def\fig{CPM-axiom-phase}{\setlength{\fboxsep}{0pt}\colorbox{gray!15}{~\strut}}\eq{}{\setlength{\fboxsep}{0pt}\colorbox{gray!15}{~\strut}}\hspace*{3em}
			\def\fig{CPM-axiom-scalar}{\setlength{\fboxsep}{0pt}\colorbox{gray!15}{~\!\!\!\!\strut\!\!\!\!}}\eq{}{\setlength{\fboxsep}{0pt}\colorbox{gray!15}{~\strut}}\hspace*{3em}
			\def\fig{CPM-axiom-Z}{\setlength{\fboxsep}{0pt}\colorbox{gray!15}{~\!\!\strut\!\!}}\eq{}{\setlength{\fboxsep}{0pt}\colorbox{gray!15}{~\!\!\!\!\strut\!\!\!\!}}\\[3em]
			\def\fig{CPM-axiom-fswap}{\setlength{\fboxsep}{0pt}\colorbox{gray!15}{~\!\!\!\!\strut\!\!\!\!}}\eq{}{\setlength{\fboxsep}{0pt}\colorbox{gray!15}{~\!\!\!\!\strut\!\!\!\!}}\hspace*{4em}
			\def\fig{CPM-axiom-H}{\setlength{\fboxsep}{0pt}\colorbox{gray!15}{~\strut}}\eq{}{\setlength{\fboxsep}{0pt}\colorbox{gray!15}{~\!\!\!\!\strut\!\!\!\!}}
		\end{array}
	\end{equation}
	
	\begin{proposition}
		\label{prop:ground-derivable}
		If half phases are allowed in $R$ (i.e.~$e^{i\alpha}\in R\iff e^{i\frac\alpha2}\in R$), the rules of \Cref{fig:ZW_rules} and \Cref{fig:tick-naturality} are enough to derive the equations \eqref{eq:ground-ax}. If we do not have half-angles but $e^{i\frac\pi4}\in R$, then adding the axiom $\def\fig{Axiom-grounds-aux}{\setlength{\fboxsep}{0pt}\colorbox{gray!15}{~\strut}}\eq{}{\setlength{\fboxsep}{0pt}\colorbox{gray!15}{~\!\!\!\!\strut\!\!\!\!}}$ is enough. 
	\end{proposition}
	
	\section{Completeness}
	
	\label{sec:completeness}
	
	We show here the central result of the paper: the axiomatisation given in Figures~\ref{fig:ZW_rules},~\ref{fig:tick-naturality} and \ref{fig:more-axioms} entirely captures semantical equivalence. That is to say, any two diagrams with the same interpretation can be turned into one another only by local application of the diagram transformations given in this axiomatisation.
	
	To show this result, we reuse the normal form defined in \Cref{sec:universality-NF}. The normal form will constitute the canonical representative of an equivalence class. If we show that any diagram can be put in normal form, we show as a consequence that any two diagrams with the same semantics can be rewritten into one another. We shall prove that all the generators can be put in normal form, and that all compositions of diagrams in normal form can be put in normal form.
	
	First, we show a pretty handy transformation on normal forms, adapted from \cite{Hadzihasanovic2017algebra}:
	\begin{lemma}[Negation]
		\label{lem:NF-negation}
		Given a diagram in (reduced) normal form, the diagram obtained by adding (or removing) \tikzfig{W-1-1} to one of the outputs can be rewritten in (reduced) normal form by independently complementing the $\tikzfig{id}$ and $\tikzfig{tickedge}$ connections of that output to the white vertices, in the following sense:
		\def\fig{NF-negation-2}
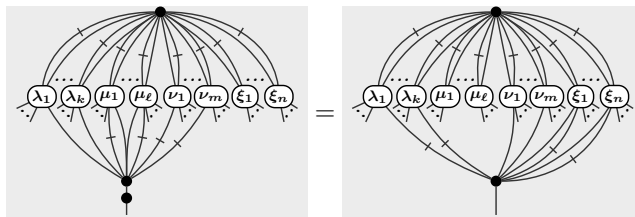
		\begin{align*}
			\scalebox{0.9}{${\setlength{\fboxsep}{0pt}\colorbox{gray!15}{~\strut}}$}
			\eq{}\scalebox{0.9}{${\setlength{\fboxsep}{0pt}\colorbox{gray!15}{~\strut}}$}
		\end{align*}
	\end{lemma}
	
	This lemma proves useful in particular in the upcoming proof of tensor product of normal forms: 
	
	\begin{proposition}
		\label{prop:NF-tensor}
		The tensor product of two diagrams in normal form can be put in normal form.
	\end{proposition}
	
	For showing the result on compositions of normal form, as well as for generators, the following two lemmas prove useful:
	
	\begin{lemma}
		\label{lem:NF-bra-0}
		Applying either $\tikzfig{bra-0}$ or $\tikzfig{bra-1}$ on an output of a diagram in normal form gives a diagram that can be put in normal form.
	\end{lemma}
	
	\begin{lemma}
		\label{lem:NF-W-2-1}
		Applying $\tikzfig{W-2-1}$ to a pair of outputs of a diagram in normal form gives a diagram that can be put in normal form.
	\end{lemma}
	
	We can now prove the following:
	\begin{proposition}
		\label{prop:NF-compo}
		The sequential composition of diagrams in normal form (which, through the process-state duality, amounts to several applications of $\tikzfig{cup}$ to pairs of outputs) gives a diagram that can be put in normal form.
	\end{proposition}
	
	\begin{proof}[Proof of \Cref{prop:NF-compo}]
		One can easily show that $\tikzfig{cup}\eq{}\tikzfig{cup-decomposition}$. Then, using Lemmas \ref{lem:NF-negation}, \ref{lem:NF-W-2-1} and \ref{lem:NF-bra-0}, one can turn the diagram obtained by application of $\tikzfig{cup}$ to a pair of outputs into normal form.
	\end{proof}
	
	Compositions of normal forms can therefore be put in normal form. It now remains to show that the generators of $\zwtick$-diagrams can themselves be put in normal form:
	\begin{proposition}
		\label{prop:NF-generators}
		The generators of the $\zwtick$-Calculus can be put in normal form.
	\end{proposition}
	
	The main result of the paper follows:
	\begin{theorem}
		\label{thm:completeness}
		The axiomatisation in Figures \ref{fig:ZW_rules}, \ref{fig:tick-naturality} and \ref{fig:more-axioms} is complete for Hermiticity-preserving operators, i.e.:
		\[\forall D_1,D_2\in\zwtick,~~\interp{D_1}^\nmid=\interp{D_2}^\nmid \iff \zwtick\vdash D_1=D_2\]
	\end{theorem}
	
	\begin{proof}
		Since all generators can be put in normal form, and all compositions of diagrams in normal form can be put in normal form, any diagram can be put in normal form. Since the rewrite rules are sound, the diagram in normal form has the same semantics as the initial one. By the uniqueness of the normal form, any two diagrams with the same semantics can be put in the same normal form, showing that the two can be turned into one another.
	\end{proof}
	
	\section{Applications}
	
	We conclude by providing some illustrations of the possibilities offered by our graphical language. These are brief sketches of applications whose thorough development could be the subject of further investigations.
	
	\subsection{Scalar Product and Dagger}
	
	In categorical quantum mechanics, the scalar product is often computed using the cup. However, the corresponding bilinear form is strictly linear in both components and thus does not match the usual Hermitian scalar product of the Hilbert space, which has an antilinear component. 
	This is fine as long as we consider processes in the real-linear subspace. It amounts to identifying the underlying Hilbert space with its dual.
	
	The problem is that, by doing so, the cap and the maximally entangled states get confused into a single equation. Nevertheless, if observed carefully, the cap `connects' the Hilbert space with its dual (which should be represented by a complex conjugated copy of itself since the inner product is sesquilinear). In contrast, the maximally entangled state `connects' the Hilbert space with a copy of itself (i.e. it is a bipartite state).
	
	The two are different on non-real vectors. Consider, for example, the +1 eigenstate of the $Y$ matrix, $\ket{i}:=\frac{1}{\sqrt{2}}(\ket{0}+i\ket{1})$. In single wire, the cap of two such elements is different from their inner product, 
	\begin{equation}
		\left(\bra{i} \otimes \bra{i}\right) \circ \left( \sum_{k=0}^1 \frac{1}{\sqrt{2}}\:\ket{k}\otimes \ket{k}\right) = 0 \neq \braket{i} =  1 \:.
	\end{equation} 
	In the above, the $\ket{i}$ and the cap were expressed in their usual vector form in the computational basis, $\bra{i} \otimes \bra{i} = \frac{1}{2}\left(\begin{smallmatrix} 1 & -i & -i & -1 \end{smallmatrix}\right)$, and $\left( \sum_k \frac{1}{\sqrt{2}}\:\ket{k}\otimes \ket{k}\right) = \frac{1}{\sqrt{2}} \left(\begin{smallmatrix}  1\\ 0 \\ 0 \\ 1  \end{smallmatrix}\right)$. 
	Diagrammatically:
	\def\fig{scalar1}
	\begin{align*}
		{\setlength{\fboxsep}{0pt}\colorbox{gray!15}{~\strut}}
		\eq{}{\setlength{\fboxsep}{0pt}\colorbox{gray!15}{~\strut}}
		\eq{}{\setlength{\fboxsep}{0pt}\colorbox{gray!15}{~\strut}}
		\eq{}0
		\neq{\setlength{\fboxsep}{0pt}\colorbox{gray!15}{~\strut}}
		\eq{}1
	\end{align*}
	
	In doubled, this becomes:
	\begin{equation}
		\begin{aligned}
			\Tr{\left(\dyad{i}\otimes \dyad{i}\right)^\dag \cdot \phi^+ } = \Tr{\dyad{i}^\dag\cdot \dyad{i}^T} = 0\\
			\neq \Tr{\dyad{i}^\dag\cdot \dyad{i}}=1\:.
		\end{aligned}
	\end{equation}
	The usual workaround for this problem, is to label direct and dual Hilbert spaces using arrows, resulting in `Hairy Spiders' \cite[§8.6.3]{Dodo}, which were pioneered in \cite{Coecke2008,Coecke2011}. 
	These can be awkward to work with, but luckily, the tick provides a way to circumvent the $Y$ basis issue without using hairy spiders.
	In the doubled theory, the tick has a natural `arrow-less' formulation, allowing us to define a new kind of cup which indeed recovers the scalar product:
	\def\fig{scalar2}
	\begin{align*}
		\interp{{\setlength{\fboxsep}{0pt}\colorbox{gray!15}{~\strut}}}
		\eq{}|\braket{x}{y}|^2
	\end{align*}
	
	In the same way that the cap and cup can be used to transpose linear transforms, we can use the tick to represent $\dagger$ of diagrams internally as:
	\def\fig{internaldagger}
	\begin{align*}
		{\setlength{\fboxsep}{0pt}\colorbox{gray!15}{~\strut}}
		\eq{}{\setlength{\fboxsep}{0pt}\colorbox{gray!15}{~\strut}}
	\end{align*}
	
	It can be checked  that it satisfies the defining property of the adjoint i.e. $\braket{D^{\dagger} x}{y}=\braket{x}{Dy}$, in the doubled picture:
	\def\fig{internaldaggerproof2}
	\begin{align*}
		{\setlength{\fboxsep}{0pt}\colorbox{gray!15}{~\strut}}
		\eq{}{\setlength{\fboxsep}{0pt}\colorbox{gray!15}{~\strut}}
		\eq{}{\setlength{\fboxsep}{0pt}\colorbox{gray!15}{~\strut}}
	\end{align*}
	
	Hence, the definition of a diagram being unitary becomes: 
	\def\fig{unitarydef}
	\begin{align*}
		{\setlength{\fboxsep}{0pt}\colorbox{gray!15}{~\strut}}
		\eq{}{\setlength{\fboxsep}{0pt}\colorbox{gray!15}{~\strut}}
		\eq{}{\setlength{\fboxsep}{0pt}\colorbox{gray!15}{~\strut}}
	\end{align*}
	
	\subsection{Proper Choi-Jamio{\l}kowski}
	
	The reader may have noticed the transposition in the definition of the reverse direction of the Choi-Jamio{\l}kowski (CJ) isomorphism, eq. \eqref{eq:CJ^-1}. It is again a symptom of identifying a Hilbert space with its dual. In other words, when one uses the cap in process-state duality. The cap indeed provides a linear identification of spaces that should be anti-isomorphic. Therefore, the transpose appears to `repay' the forgotten antilinearity.
	
	It has been argued elsewhere (see e.g.~\cite{Jiang2013,Leifer2013,Uhlmann2016,Hoffreumon2021,Frembs2022}) that process-state duality should be conducted using an antilinear map, with the most straightforward solution being to add a transposition on one part of the dual state.
	The advantage is that there is no longer a transposition lurking in the computations. In terms of diagrams, there is no risk of omitting a complex conjugation.
	
	The fix proposed here to represent the CJ isomorphism properly is the same as above: replacing the cap (resp.~cup) with a cap (resp.~cup) with a tick.
	\def\fig{TrueCJ}
	\begin{align*}
		{\setlength{\fboxsep}{0pt}\colorbox{gray!15}{~\strut}}\mapsto {\setlength{\fboxsep}{0pt}\colorbox{gray!15}{~\strut}} \qquad {\setlength{\fboxsep}{0pt}\colorbox{gray!15}{~\strut}} \mapsto {\setlength{\fboxsep}{0pt}\colorbox{gray!15}{~\strut}}
	\end{align*}
	
	The difference between the two is flagrant when considering the map state duality when a symmetry group act on the Hilbert space via a unitary representation. Equivariant maps are then defined as:
	\def\fig{equivdef}${\setlength{\fboxsep}{0pt}\colorbox{gray!15}{~\strut}}= {\setlength{\fboxsep}{0pt}\colorbox{gray!15}{~\strut}}$
	
	Using the antilinear isomorphism we can see that equivariant maps corresponds to invariant states, which is not true when using the linear isomorphism.
	
	\subsection{PPT Criterion}
	
	One of the main applications of antilinearity in quantum information theory is the design of entanglement witnesses. A state is said separable if it can be written as a convex sum of pure tensor states, and is said entangled otherwise. Detecting entanglement is a difficult task. A necessary condition can be obtained using antilinearity; it is called the PPT criterion.
	
	\begin{lemma} [PPT criterion \cite{Peres1996,HORODECKI1996}]
		If a bipartite state $\rho $ is separable, then it satisfies:
		
		\begin{center}
			$\tikzfig{pptcrit} \in \cat{CP}$
		\end{center}
		
	\end{lemma}
	
	We will rephrase it graphically. First, we start by giving a diagrammatical characterisation of separability: 
	
	\begin{lemma}
		A bipartite state $\rho $ is separable if and only if there are CP maps $A$ and $B$ such that:
		
		\begin{center}
			$\interp{\tikzfig{pptsep}}=\rho$
		\end{center}
	\end{lemma}
	
	\begin{proof}
		First, starting with a diagram of the right form and computing the semantics, we get $\sum_k A(\ketbra{k}{k}) \otimes B(\ketbra{k}{k}) $ which is separable. Now for the converse, given a separable state $\rho = \sum_k A_k \otimes B_k $ we only have to find superoperators $A$ and $B$ such that $A(\ketbra{k}{k})=A_k $ and $B(\ketbra{k}{k})=B_k $. Given any family of CP maps $A_k $, we can construct such map $A$ as follows. We purify all $A_k $ with an auxiliary system of the same size (assuming maximal Kraus rank in general), giving pure maps $A'_k$. Then we control all those pure maps into a pure map $A'$ satisfying $A'(\ketbra{k}{k})=A'_k$. Finally, we trace out the control input and get the desired map $A$.
	\end{proof}
	
	From this characterisation, we can directly show that the PPT criterion is a necessary condition:
	
	\def\fig{pptproof}
	\begin{align*}
		{\setlength{\fboxsep}{0pt}\colorbox{gray!15}{~\strut}}= {\setlength{\fboxsep}{0pt}\colorbox{gray!15}{~\strut}}= {\setlength{\fboxsep}{0pt}\colorbox{gray!15}{~\strut}} \in \cat{CP}
	\end{align*}
	as $A$, $\overline B$ and $\ground$ are all CP maps.
	
	\subsection{Spin Flip}
	
	We end this list of illustrations with an example of a new kind of transformation one can represent using ticks: the spin-flip. This transformation acts on a qubit state $\rho$ as $\rho \mapsto Y\rho^T Y$,
	and can be represented diagrammatically as: $\tikzfig{spinflip}$.

	Recall that a 1-qubit state can be parametrised by a vector in the Bloch ball, given by the tuple $\vec r = (r_x,r_y,r_z)$, in the following way:
	\begin{align*}
		\rho &= \frac{I+\vec r\cdot \vec \sigma}2 = \frac{I+r_xX+r_yY+r_zZ}2\\
		&=\frac12\begin{pmatrix}
			1+r_z & r_x-ir_y \\ r_x+ir_y & 1-r_z
		\end{pmatrix}
	\end{align*}
	The spin-flip then amounts to a central symmetry of the Bloch sphere, matching each point to its antipodes. In a sense, this is the true (logical) negation of a qubit in the Bloch sphere.
	
	The above morphism is thus supposed to map $\rho$ defined by $\vec r$ to $\rho'$ defined by $-\vec r$. This can be checked using e.g.~the normal form of $\rho$ (ignoring overall scalar of $\frac14$):
	%
	
	\def\fig{spin-check-arbitrary}
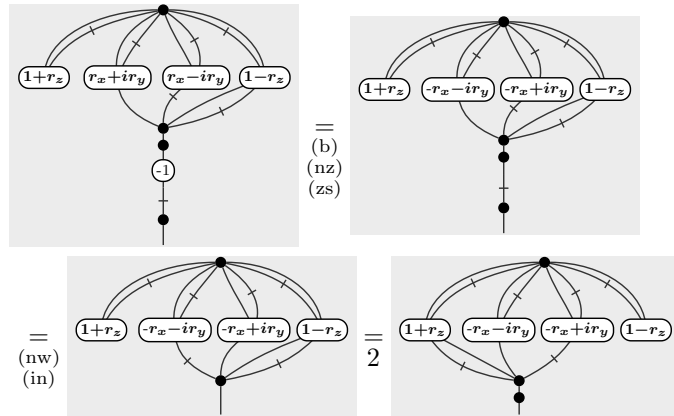
	\begin{align*}
		&\scalebox{0.9}{${\setlength{\fboxsep}{0pt}\colorbox{gray!15}{~\strut}}$}
		\eq{\eqref{ax:Z-W-bialgebra}\\\eqref{ax:tick-through-Z-spider}\\\eqref{ax:Z-spider}}\scalebox{0.9}{${\setlength{\fboxsep}{0pt}\colorbox{gray!15}{~\strut}}$}\\
		&\eq{\eqref{ax:tick-through-W-spider}\\\eqref{ax:W-inv}}\scalebox{0.9}{${\setlength{\fboxsep}{0pt}\colorbox{gray!15}{~\strut}}$}
		\eq{\ref{lem:NF-negation}}\scalebox{0.9}{${\setlength{\fboxsep}{0pt}\colorbox{gray!15}{~\strut}}$}
	\end{align*}
	
	Spin-flipping offers an example of the peculiar role played by antiunitaries in quantum information. In \cite{Gisin1999} it was shown that more information could be stored in a pair of antiparallel spin states rather than two parallel ones. This purely quantum phenomenon can be tracked to the fact that, in the former case, the states are related by an antiunitary transformation, the spin-flip. This corroborates the fact that two local operations enriched with an antiunitary generator (in this case, the preparation of the two spin states, followed by the spin-flip of one of them) allows to do more than without the antiunitary (in this case, to succeed a protocol with better efficiency).
	
	\section{Discussion}
	
	We showed that adding a generator for partial transpose, the tick, is sufficient to extend universal and complete graphical languages for pure quantum operations to one for Hermiticity-preserving operations. In addition, we provided a normal form for diagrams using the ZW-calculus, providing a completeness result. We also extended the doubling construction to include the tick under the name of the $\operatorname{Unzip}$ functor and provided a connection to the $\cat{Int}$ functor.
	
	A consequence of our results is a partial answer to the question of completeness for Clifford+T CP maps, left open in \cite{Carette2019completeness}. Indeed, when taking $R:=\mathbb Z[\frac12, e^{i\frac\pi4}]$ (which obviously contains $\frac12$) we precisely end up in the Clifford+T fragment \cite{Jeandel2018complete}. This means we can now show diagrammatic equality between two Clifford+T CPM diagrams. However, we may have to use the full power of the present axiomatisation and, in the process, have diagrams that are not locally CPMs but merely Hermiticity-preserving operators.
	
	Contrary to \cite{Carette2019completeness} we have completeness for the Clifford+T fragment and unrestricted quantum computation (when taking $R:=\mathbb C$). But, not for the Clifford fragment, which the $\zwtick$ diagram cannot represent, no matter what ring $R$ we take (since the W-state $\tikzfig{W-0-3}$ itself is outside of the fragment). We leave getting an complete equational theory for Clifford HP maps as an open question.
	
	An important caveat, compared to other more established axiomatisations, is that there does not appear to be a good interpretation of the axioms of \Cref{fig:more-axioms}. Finding such interpretations, or simplifying the current axioms with ones that do have natural meanings, is left as an open question as well. Another unanswered question in this work is the necessity of the axioms, and consequently the minimality of the equational theory.
	
	As stated above, we managed to provide the completeness result using normal forms --a different method than \cite{Carette2019completeness}. The reason is that the positivity of a matrix is challenging to express algebraically. This is why the proof of completeness for $\cat{CP}$ uses purification instead of the normal form. Hermiticity is, however, a simpler property, allowing direct proof of completeness using normal forms.
	The restatement of the proof in terms of purification is left for future works generalising the notion of purification for Hermiticity-preserving operators, paving the way toward a categorical characterisation of $\cat{HP}$ via a universal property as done for $\cat{CP}$ in \cite{huot2019quantum}. Such uniform characterisation would allow us to extend completeness to other graphical languages like ZX- and ZH-calculi more naturally than by direct translation of the present axioms, and could even prove useful for similar issues in quantum circuits, which are not compact close.
	
	\section*{Acknowledgements}
	
	We thank Marc de Visme for fruitful discussions about the paper. T.H.~is grateful to the QuaCS group at Inria Saclay for their hospitality during a stay during which part of this research was conducted. This stay was made possible through the support of the ID\#61466 grant from the John Templeton Foundation, as part of the “The Quantum Information Structure of Spacetime (QISS)” Project (qiss.fr). The opinions expressed in this publication are those of the authors and do not necessarily reflect the views of the John Templeton Foundation. T. H. benefits from the support of the French Community of Belgium within the framework of the financing of a FRIA grant. T.H. would like to thank Alexandra Elbakyan for providing access to the scientific literature. 
	R.V.~acknowledges support from the PEPR integrated project EPiQ ANR-22-PETQ-0007 part of Plan France 2030, the ANR projects TaQC ANR-22-CE47-0012 and HQI ANR-22-PNCQ-0002, as well as the European project HPCQS. T.C. was supported by the ERDF project 1.1.1.5/18/A/020 “Quantum algorithms: from complexity theory to experiment”.
	
	\bibliography{biblio}

\begin{thebibliography}{10}

\bibitem{aaronson2009closed}
Scott Aaronson and John Watrous.
\newblock Closed timelike curves make quantum and classical computing
  equivalent.
\newblock {\em Proceedings of the Royal Society A: Mathematical, Physical and
  Engineering Sciences}, 465(2102):631--647, 2009.

\bibitem{Abramsky2002GoI}
Samson Abramsky, Esfandiar Haghverdi, and Philip~J. Scott.
\newblock Geometry of interaction and linear combinatory algebras.
\newblock {\em Mathematical Structures in Computer Science}, 12:625 -- 665,
  2002.

\bibitem{Backens2019complete}
Miriam Backens and Aleks Kissinger.
\newblock {ZH}: A complete graphical calculus for quantum computations
  involving classical non-linearity.
\newblock In Peter Selinger and Giulio Chiribella, editors, {\em {\textrm
  Proceedings of the 15th International Conference on} Quantum Physics and
  Logic, {\textrm Halifax, Canada, 3-7th June 2018}}, volume 287 of {\em
  Electronic Proceedings in Theoretical Computer Science}, pages 23--42, 2019.

\bibitem{baez2017props}
John~C Baez, Brandon Coya, and Franciscus Rebro.
\newblock Props in network theory.
\newblock {\em arXiv preprint arXiv:1707.08321}, 2017.

\bibitem{Buzek1999}
V.~Bužek, M.~Hillery, and R.~F. Werner.
\newblock Optimal manipulations with qubits: Universal-not gate.
\newblock {\em Physical Review A}, 60(4):R2626–R2629, Oct 1999.

\bibitem{DBLP:phd/hal/Carette21}
Titouan Carette.
\newblock {\em Wielding the ZX-calculus, Flexsymmetry, Mixed States, and
  Scalable Notations. (Manier le ZX-calcul, flexsym{\'{e}}trie, syst{\`{e}}mes
  ouverts et limandes)}.
\newblock PhD thesis, University of Lorraine, Nancy, France, 2021.

\bibitem{Carette2019completeness}
Titouan Carette, Emmanuel Jeandel, Simon Perdrix, and Renaud Vilmart.
\newblock {Completeness of Graphical Languages for Mixed States Quantum
  Mechanics}.
\newblock In Christel Baier, Ioannis Chatzigiannakis, Paola Flocchini, and
  Stefano Leonardi, editors, {\em 46th International Colloquium on Automata,
  Languages, and Programming (ICALP 2019)}, volume 132 of {\em Leibniz
  International Proceedings in Informatics (LIPIcs)}, pages 108:1--108:15,
  Dagstuhl, Germany, 2019. Schloss Dagstuhl--Leibniz-Zentrum fuer Informatik.

\bibitem{Cerf2001}
N.~J. Cerf and S.~Iblisdir.
\newblock Quantum cloning machines with phase-conjugate input modes.
\newblock {\em Phys. Rev. Lett.}, 87:247903, Nov 2001.

\bibitem{Chiribella2021}
Giulio Chiribella, Erik Aurell, and Karol {\.{Z}}yczkowski.
\newblock Symmetries of quantum evolutions.
\newblock {\em Physical Review Research}, 3(3), jul 2021.

\bibitem{Choi1975}
Man-Duen Choi.
\newblock {Positive Linear Maps on Complex Matrices}.
\newblock {\em Linear Algebra and its Applications}, 10(3):285--290, 1975.

\bibitem{Clement2022complete}
Alexandre Clément, Nicolas Heurtel, Shane Mansfield, Simon Perdrix, and Benoit
  Valiron.
\newblock A complete equational theory for quantum circuits, 2022.

\bibitem{Coecke2008bis}
Bob Coecke.
\newblock Axiomatic description of mixed states from selinger's
  cpm-construction.
\newblock {\em Electronic Notes in Theoretical Computer Science}, 210:3--13,
  2008.
\newblock Proceedings of the 4th International Workshop on Quantum Programming
  Languages (QPL 2006).

\bibitem{Coecke2011}
Bob Coecke and Ross Duncan.
\newblock Interacting quantum observables: categorical algebra and
  diagrammatics.
\newblock {\em New Journal of Physics}, 13(4):043016, Apr 2011.

\bibitem{Coecke2010compositional}
Bob Coecke and Aleks Kissinger.
\newblock The compositional structure of multipartite quantum entanglement.
\newblock In {\em Automata, Languages and Programming}, pages 297--308.
  Springer Berlin Heidelberg, 2010.

\bibitem{Dodo}
Bob Coecke and Aleks Kissinger.
\newblock {\em Picturing Quantum Processes: A First Course in Quantum Theory
  and Diagrammatic Reasoning}.
\newblock Cambridge University Press, Cambridge, 2017.

\bibitem{Coecke2012environment}
Bob Coecke and Simon Perdrix.
\newblock {Environment and Classical Channels in Categorical Quantum
  Mechanics}.
\newblock {\em {Logical Methods in Computer Science}}, {Volume 8, Issue 4}, Nov
  2012.

\bibitem{Coecke2008}
Bob Coecke, Simon Perdrix, and {\'E}ric~Oliver Paquette.
\newblock Bases in diagrammatic quantum protocols.
\newblock {\em Electronic Notes in Theoretical Computer Science},
  218:131–152, Oct 2008.

\bibitem{depillis1967}
John de~Pillis.
\newblock Linear transformations which preserve hermitian and positive
  semidefinite operators.
\newblock {\em Pacific J. Math.}, 23(1):129--137, 1967.

\bibitem{Dong2019}
Qingxiuxiong Dong, Marco~Túlio Quintino, Akihito Soeda, and Mio Murao.
\newblock Implementing positive maps with multiple copies of an input state.
\newblock {\em Physical Review A}, 99(5), May 2019.

\bibitem{Frembs2022}
Markus Frembs and Eric~G. Cavalcanti.
\newblock Variations on the choi-jamiolkowski isomorphism, 2022.

\bibitem{gilchrist2009vectorization}
Alexei Gilchrist, Daniel~R. Terno, and Christopher~J. Wood.
\newblock Vectorization of quantum operations and its use, 2009.

\bibitem{Gisin1999}
Nicolas Gisin and Sandu Popescu.
\newblock Spin flips and quantum information for antiparallel spins.
\newblock {\em Physical Review Letters}, 83:432--435, Jul 1999.

\bibitem{Hadzihasanovic2015diagrammatic}
Amar Hadzihasanovic.
\newblock A diagrammatic axiomatisation for qubit entanglement.
\newblock In {\em 2015 30th Annual ACM/IEEE Symposium on Logic in Computer
  Science}, pages 573--584, Jul 2015.

\bibitem{Hadzihasanovic2017algebra}
Amar Hadzihasanovic.
\newblock {\em The Algebra of Entanglement and the Geometry of Composition}.
\newblock PhD thesis, University of Oxford, 2017.

\bibitem{Hadzihasanovic2018complete}
Amar Hadzihasanovic, Kang~Feng Ng, and Quanlong Wang.
\newblock Two complete axiomatisations of pure-state qubit quantum computing.
\newblock In {\em Proceedings of the 33rd Annual ACM/IEEE Symposium on Logic in
  Computer Science}, LICS '18, pages 502--511, New York, NY, USA, 2018. ACM.

\bibitem{Hill1997}
Sam~A. Hill and William~K. Wootters.
\newblock Entanglement of a pair of quantum bits.
\newblock {\em Physical Review Letters}, 78:5022--5025, Jun 1997.

\bibitem{Hoffreumon2021}
Timothée Hoffreumon and Ognyan Oreshkov.
\newblock The multi-round process matrix.
\newblock {\em Quantum}, 5:384, Jan 2021.

\bibitem{HORODECKI1996}
Michał Horodecki, Paweł Horodecki, and Ryszard Horodecki.
\newblock Separability of mixed states: necessary and sufficient conditions.
\newblock {\em Physics Letters A}, 223(1):1--8, 1996.

\bibitem{huot2019quantum}
Mathieu Huot and Sam Staton.
\newblock Quantum channels as a categorical completion.
\newblock In {\em 2019 34th Annual ACM/IEEE Symposium on Logic in Computer
  Science (LICS)}, pages 1--13. IEEE, 2019.

\bibitem{Jamiolkowski1972}
Andrzej Jamio{\l}kowski.
\newblock {Linear transformations which preserve trace and positive
  semidefiniteness of operators}.
\newblock {\em Reports on Mathematical Physics}, 3(4):275--278, dec 1972.

\bibitem{Jeandel2018complete}
Emmanuel Jeandel, Simon Perdrix, and Renaud Vilmart.
\newblock A complete axiomatisation of the {ZX}-calculus for {C}lifford+{T}
  quantum mechanics.
\newblock In {\em Proceedings of the 33rd Annual ACM/IEEE Symposium on Logic in
  Computer Science}, LICS '18, pages 559--568, New York, NY, USA, 2018. ACM.

\bibitem{Jiang2013}
Min Jiang, Shunlong Luo, and Shuangshuang Fu.
\newblock Channel-state duality.
\newblock {\em Phys. Rev. A}, 87:022310, Feb 2013.

\bibitem{Joyal1996traced}
André Joyal, Ross Street, and Dominic Verity.
\newblock Traced monoidal categories.
\newblock {\em Mathematical Proceedings of the Cambridge Philosophical
  Society}, 119(3):447–468, 1996.

\bibitem{Klay1987}
M.~Kl{\"a}y, C.~Randall, and D.~Foulis.
\newblock Tensor products and probability weights.
\newblock {\em International Journal of Theoretical Physics}, 26(3):199--219,
  1987.

\bibitem{Leifer2013}
M.~S. Leifer and Robert~W. Spekkens.
\newblock Towards a formulation of quantum theory as a causally neutral theory
  of bayesian inference.
\newblock {\em Physical Review A}, 88(5), Nov 2013.

\bibitem{Massar2000}
S.~Massar.
\newblock Collective versus local measurements on two parallel or antiparallel
  spins.
\newblock {\em Phys. Rev. A}, 62:040101, Sep 2000.

\bibitem{nielsen_chuang_2010}
Michael~A. Nielsen and Isaac~L. Chuang.
\newblock {\em Quantum Computation and Quantum Information: 10th Anniversary
  Edition}.
\newblock Cambridge University Press, 2010.

\bibitem{Peres1996}
Asher Peres.
\newblock Separability criterion for density matrices.
\newblock {\em Physical Review Letters}, 77:1413--1415, Aug 1996.

\bibitem{pinzani2019categorical}
Nicola Pinzani, Stefano Gogioso, and Bob Coecke.
\newblock Categorical semantics for time travel.
\newblock In {\em 2019 34th Annual ACM/IEEE Symposium on Logic in Computer
  Science (LICS)}, pages 1--20. IEEE, 2019.

\bibitem{Regula2021}
Bartosz Regula, Ryuji Takagi, and Mile Gu.
\newblock Operational applications of the diamond norm and related measures in
  quantifying the non-physicality of quantum maps.
\newblock {\em Quantum}, 5:522, Aug 2021.

\bibitem{Schwabl2008}
Franz Schwabl.
\newblock {\em Advanced Quantum Mechanics}.
\newblock Springer-Verlag, Berlin Heidelberg, fourth edition, 2008.

\bibitem{Selinger2006CPM}
Peter Selinger.
\newblock Dagger compact closed categories and completely positive maps.
\newblock {\em Electronic Notes in Theoretical Computer Science}, 170:139--163,
  Mar 2007.

\bibitem{PCT}
Raymond~F. Streater and Arthur~S. Wightman.
\newblock {\em {PCT}, Spin and Statistics, and All That}.
\newblock Princeton Landmarks in Physics. Princeton University Press,
  Princeton, New Jersey, 2001.

\bibitem{Uhlmann2016}
Armin Uhlmann.
\newblock Anti- (conjugate) linearity.
\newblock {\em Science China Physics, Mechanics \& Astronomy}, 59(3), jan 2016.

\bibitem{vandewetering2020zxcalculus}
John van~de Wetering.
\newblock {ZX}-calculus for the working quantum computer scientist, 2020.

\bibitem{Vilmart2019nearminimal}
Renaud {Vilmart}.
\newblock A near-minimal axiomatisation of zx-calculus for pure qubit quantum
  mechanics.
\newblock In {\em 2019 34th Annual ACM/IEEE Symposium on Logic in Computer
  Science (LICS)}, pages 1--10, June 2019.

\bibitem{Weinberg1995}
Steven Weinberg.
\newblock {\em The Quantum Theory of Fields}, volume~1.
\newblock Cambridge University Press, 1995.

\bibitem{Weinberg1996}
Steven Weinberg.
\newblock {\em The Quantum Theory of Fields}, volume~2.
\newblock Cambridge University Press, 1996.

\bibitem{Wigner1932}
Eugene~P. Wigner.
\newblock Ueber die operation der zeitumkehr in der quantenmechanik.
\newblock {\em Nachrichten von der Gesellschaft der Wissenschaften zu
  Göttingen, Mathematisch-Physikalische Klasse}, 1932:546--559, 1932.

\bibitem{Wigner1931}
Eugene~P. Wigner.
\newblock {\em Group Theory and its application to the quantum mechanics of
  atomic spectra}, volume~5 of {\em Pure and Applied Physics}.
\newblock Academic Press, 1959.

\bibitem{Wigner1993}
Eugene~P. Wigner.
\newblock Normal form of antiunitary operators.
\newblock In Arthur~S. Wightman, editor, {\em The Collected Works of Eugene
  Paul Wigner: Part A: The Scientific Papers}, pages 551--555. Springer Berlin
  Heidelberg, Berlin, Heidelberg, 1993.

\bibitem{Wooters1998}
William~K. Wootters.
\newblock Entanglement of formation of an arbitrary state of two qubits.
\newblock {\em Physical Review Letters}, 80:2245--2248, Mar 1998.

\end{thebibliography}
	\bibliographystyle{plain}
	
	\appendix
	
	\subsection{Proofs of \Cref{sec:zwtick-calculus}}
	
	\begin{proof}[Proof of \Cref{prop:semantics-are-isomorphic}]
		That $\operatorname{\Psi}$ and $\operatorname{\Psi}^{-1}$ are inverses of one another is direct, one merely undoes the permutation of inputs/outputs from the other one. For the rest, we may want to prove that $\operatorname{\Psi}\circ\operatorname{unzip} = \operatorname{HP}$. We show this by induction on the structure of the diagrams:
		\begin{itemize}
			\item When $D = D_1\otimes D_2$:
			\def\fig{double-to-HP-tensor}
			\begin{align*}
				\operatorname{\Psi}&\left(\operatorname{unzip}\left(\tikzfig{D1}\tikzfig{D2}\right)\right)\\
				&\eq{}\operatorname{\Psi}\left({\setlength{\fboxsep}{0pt}\colorbox{gray!15}{~\strut}}\right)\\
				&\eq{}{\setlength{\fboxsep}{0pt}\colorbox{gray!15}{~\strut}}\\
				&\eq{}{\setlength{\fboxsep}{0pt}\colorbox{gray!15}{~\strut}}
				\eq{}\operatorname{HP}\left(\tikzfig{D1}\tikzfig{D2}\right)	
			\end{align*}
			\item When $D = D_2\circ D_1$:
			\def\fig{double-to-HP-compo}
			\begin{align*}
				\operatorname{\Psi}&\left(\operatorname{unzip}\left(\tikzfig{compo}\right)\right)
				\eq{}\operatorname{\Psi}\left({\setlength{\fboxsep}{0pt}\colorbox{gray!15}{~\strut}}\right)\\
				&\eq{}{\setlength{\fboxsep}{0pt}\colorbox{gray!15}{~\strut}}\\
				&\eq{}{\setlength{\fboxsep}{0pt}\colorbox{gray!15}{~\strut}}\\
				&\eq{}{\setlength{\fboxsep}{0pt}\colorbox{gray!15}{~\strut}}
				\eq{}\operatorname{HP}\left(\tikzfig{compo}\right)	
			\end{align*}
			\item For $\tikzfig{tickedge}$-free generator $g$:
			\def\fig{double-to-HP-pure-generator}
			\begin{align*}
				\operatorname{\Psi}&\left(\operatorname{unzip}\left({\setlength{\fboxsep}{0pt}\colorbox{gray!15}{~\strut}}\right)\right)
				\eq{}\operatorname{\Psi}\left({\setlength{\fboxsep}{0pt}\colorbox{gray!15}{~\strut}}\right)\\
				&\eq{}{\setlength{\fboxsep}{0pt}\colorbox{gray!15}{~\strut}}
				\eq{}{\setlength{\fboxsep}{0pt}\colorbox{gray!15}{~\strut}}
				\eq{}\operatorname{HP}\left({\setlength{\fboxsep}{0pt}\colorbox{gray!15}{~\strut}}\right)	
			\end{align*}
			where $\overline g$ stands for the ``conjugate'' of the generator $g$: it is $g$ itself for all generators except for the Z-spider, in which case the parameter $r$ becomes $\overline r$. Exchanging inputs and outputs performs the transpose, hence $g^\dagger$ is obtained by composition of inputs-outputs swapping and conjugation.
			\item Finally, for $\tikzfig{tickedge}$:
			\begin{align*}
				\operatorname{\Psi}\left(\operatorname{unzip}\left(\tikzfig{tickedge}\right)\right)
				&\eq{}\operatorname{\Psi}\left(\tikzfig{swap}\right)
				\eq{}\tikzfig{double-to-HP-tick}
				\eq{}\begin{array}{@{}c@{}}\tikzfig{cup}\\\tikzfig{cap}\end{array}\\
				&\eq{}\operatorname{HP}\left(\tikzfig{tickedge}\right)	
			\end{align*}
		\end{itemize}
	\end{proof}
	
	\begin{proof}[Proof of \Cref{prop:soundness}]
		The axioms of \Cref{fig:ZW_rules} are obviously sound, as they are in $\cat{ZW}$ (through $\operatorname{unzip}$ for instance, we simply end up with two copies of the same diagram). It is also pretty straightforward to see that the axioms of \ref{fig:tick-naturality} are sound using $\operatorname{unzip}$. We can show the one with the Z-spider for example:
		\def\fig{soundness-tick-through-Z}
		\begin{align*}
			{\setlength{\fboxsep}{0pt}\colorbox{gray!15}{~\strut}}
			\mapsto{\setlength{\fboxsep}{0pt}\colorbox{gray!15}{~\strut}}
			\eq{}{\setlength{\fboxsep}{0pt}\colorbox{gray!15}{~\strut}}
			\eq{}{\setlength{\fboxsep}{0pt}\colorbox{gray!15}{~\strut}}
			\mapsfrom{\setlength{\fboxsep}{0pt}\colorbox{gray!15}{~\strut}}
		\end{align*}
		The axioms from \Cref{fig:more-axioms} are the ones that really need a proof:
		\begin{itemize}
			\setlength{\itemsep}{1em}
			\item \def\fig{soundness-Axiom-grounds}$
			\begin{aligned}[t]
				{\setlength{\fboxsep}{0pt}\colorbox{gray!15}{~\strut}}
				\mapsto{\setlength{\fboxsep}{0pt}\colorbox{gray!15}{~\strut}}
				\eq{\eqref{ax:Z-spider}}{\setlength{\fboxsep}{0pt}\colorbox{gray!15}{~\strut}}
				\eq{\eqref{ax:Z-W-bialgebra}\\\eqref{ax:W-inv}}{\setlength{\fboxsep}{0pt}\colorbox{gray!15}{~\strut}}
				\eq{\eqref{ax:sum}}{\setlength{\fboxsep}{0pt}\colorbox{gray!15}{~\strut}}\\
				\eq{\eqref{ax:Z-W-bialgebra}\\\eqref{ax:Z-spider}\\\eqref{ax:Z-id}}{\setlength{\fboxsep}{0pt}\colorbox{gray!15}{~\strut\input{./figures/\fig/\fig_05.tikz}}}
				\eq{\eqref{ax:Z-spider}\\\eqref{ax:Z-id}}{\setlength{\fboxsep}{0pt}\colorbox{gray!15}{~\!\!\!\!\strut\input{./figures/\fig/\fig_06.tikz}\!\!\!\!}}
				\mapsfrom{\setlength{\fboxsep}{0pt}\colorbox{gray!15}{~\!\!\!\!\strut\input{./figures/\fig/\fig_07.tikz}\!\!\!\!}}
			\end{aligned}$
			\item \def\fig{soundness-Axiom-NF-Z}
			$\begin{aligned}[t]
				\interp{\operatorname{Unzip}\left({\setlength{\fboxsep}{0pt}\colorbox{gray!15}{~\strut}}\right)}
				\eq{}\interp{{\setlength{\fboxsep}{0pt}\colorbox{gray!15}{~\strut}}}
				\eq{\eqref{ax:Z-spider}\\\eqref{ax:Z-id}}{\setlength{\fboxsep}{0pt}\colorbox{gray!15}{~\strut}}\\
				\eq{\eqref{ax:W-spider}\\\eqref{ax:Z-W-bialgebra}}\interp{{\setlength{\fboxsep}{0pt}\colorbox{gray!15}{~\strut}}}
				\eq{}\interp{{\setlength{\fboxsep}{0pt}\colorbox{gray!15}{~\strut}}}
				+\interp{{\setlength{\fboxsep}{0pt}\colorbox{gray!15}{~\strut\input{./figures/\fig/\fig_05.tikz}}}}\\
				\eq{\eqref{ax:W-inv}\\\eqref{ax:W-bialgebra}\\\eqref{ax:W-spider}}\interp{{\setlength{\fboxsep}{0pt}\colorbox{gray!15}{~\strut\input{./figures/\fig/\fig_06.tikz}}}}
				+\interp{{\setlength{\fboxsep}{0pt}\colorbox{gray!15}{~\strut\input{./figures/\fig/\fig_07.tikz}}}}
				\eq{}\interp{{\setlength{\fboxsep}{0pt}\colorbox{gray!15}{~\strut\input{./figures/\fig/\fig_08.tikz}}}}
				\eq{}\interp{\operatorname{Unzip}\left({\setlength{\fboxsep}{0pt}\colorbox{gray!15}{~\strut\input{./figures/\fig/\fig_09.tikz}}}\right)}
			\end{aligned}$\\
			Here, to ease the computation, we used the following, easily checked, equalities:
			\def\fig{Z-decomp}
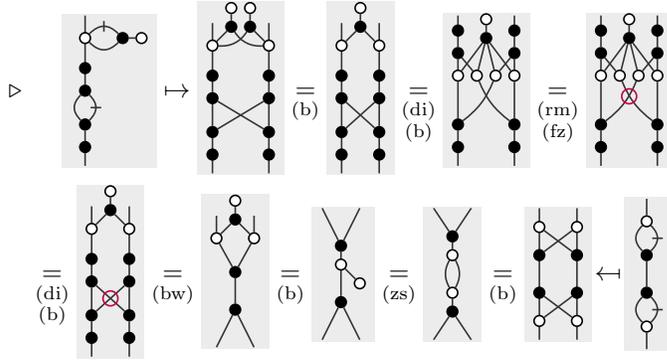
			\begin{align*}
				\interp{{\setlength{\fboxsep}{0pt}\colorbox{gray!15}{~\strut}}}
				&\eq{}\interp{{\setlength{\fboxsep}{0pt}\colorbox{gray!15}{~\strut}}}
				+\interp{{\setlength{\fboxsep}{0pt}\colorbox{gray!15}{~\strut}}}\\
				\def\fig{soundness-Axiom-NF-Z}
				\interp{{\setlength{\fboxsep}{0pt}\colorbox{gray!15}{~\strut\input{./figures/\fig/\fig_08.tikz}}}}
				&\def\fig{soundness-Axiom-NF-Z}
				\eq{}\interp{{\setlength{\fboxsep}{0pt}\colorbox{gray!15}{~\strut\input{./figures/\fig/\fig_06.tikz}}}}
				+\interp{{\setlength{\fboxsep}{0pt}\colorbox{gray!15}{~\strut\input{./figures/\fig/\fig_07.tikz}}}}
			\end{align*}
			\item \def\fig{soundness-Axiom-tensor-1}
			$\begin{aligned}[t]
				{\setlength{\fboxsep}{0pt}\colorbox{gray!15}{~\strut}}
				\mapsto{\setlength{\fboxsep}{0pt}\colorbox{gray!15}{~\strut}}
				\eq{\eqref{ax:Z-W-bialgebra}}{\setlength{\fboxsep}{0pt}\colorbox{gray!15}{~\strut}}
				\eq{\eqref{ax:W-binary-diffusion}\\\eqref{ax:Z-W-bialgebra}}{\setlength{\fboxsep}{0pt}\colorbox{gray!15}{~\strut}}
				\eq{\eqref{ax:fswap-removal}\\\eqref{ax:fswap-through-Z}}{\setlength{\fboxsep}{0pt}\colorbox{gray!15}{~\strut}}\\
				\eq{\eqref{ax:W-binary-diffusion}\\\eqref{ax:Z-W-bialgebra}}{\setlength{\fboxsep}{0pt}\colorbox{gray!15}{~\strut\input{./figures/\fig/\fig_05.tikz}}}
				\eq{\eqref{ax:W-bialgebra}}{\setlength{\fboxsep}{0pt}\colorbox{gray!15}{~\strut\input{./figures/\fig/\fig_06.tikz}}}
				\eq{\eqref{ax:Z-W-bialgebra}}{\setlength{\fboxsep}{0pt}\colorbox{gray!15}{~\strut\input{./figures/\fig/\fig_07.tikz}}}
				\eq{\eqref{ax:Z-spider}}{\setlength{\fboxsep}{0pt}\colorbox{gray!15}{~\strut\input{./figures/\fig/\fig_08.tikz}}}
				\eq{\eqref{ax:Z-W-bialgebra}}{\setlength{\fboxsep}{0pt}\colorbox{gray!15}{~\strut\input{./figures/\fig/\fig_09.tikz}}}
				\!\mapsfrom\!{\setlength{\fboxsep}{0pt}\colorbox{gray!15}{~\strut\input{./figures/\fig/\fig_10.tikz}}}
			\end{aligned}$
			\item 
			To prove the soundness of the last axiom, we are going to show the equality for every element in a basis that spans the space of some inputs. More precisely, we have to show the equality between two $3\to2$ diagrams. After unzipping, these become $6\to 4$ diagrams. We are going to show the equality for all classical states spanning the space of the two leftmost and the two rightmost inputs:
			\begin{itemize}
				\item {$\mathbf{\ket{00**00}:}$} Since the two leftmost (resp.~the two rightmost) input states are equal, we can reason with the diagram before unzipping:
				\def\fig{soundness-Axiom-tensor-2-2-00}
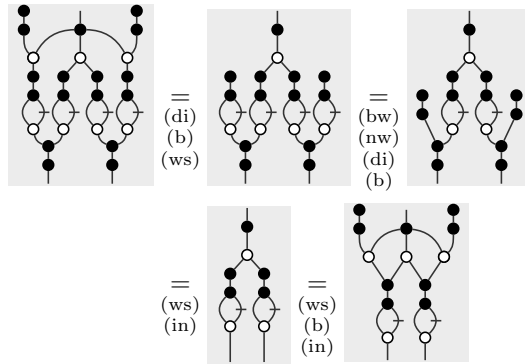
				\begin{align*}
					{\setlength{\fboxsep}{0pt}\colorbox{gray!15}{~\strut}}
					&\eq{\eqref{ax:W-binary-diffusion}\\\eqref{ax:Z-W-bialgebra}\\\eqref{ax:W-spider}}{\setlength{\fboxsep}{0pt}\colorbox{gray!15}{~\strut}}
					\eq{\eqref{ax:W-bialgebra}\\\eqref{ax:tick-through-W-spider}\\\eqref{ax:W-binary-diffusion}\\\eqref{ax:Z-W-bialgebra}}{\setlength{\fboxsep}{0pt}\colorbox{gray!15}{~\strut}}\\
					&\eq{\eqref{ax:W-spider}\\\eqref{ax:W-inv}}{\setlength{\fboxsep}{0pt}\colorbox{gray!15}{~\strut}}
					\def\fig{soundness-Axiom-tensor-2-1-00}
					\eq{\eqref{ax:W-spider}\\\eqref{ax:Z-W-bialgebra}\\\eqref{ax:W-inv}}{\setlength{\fboxsep}{0pt}\colorbox{gray!15}{~\strut}}
				\end{align*}
				\item {$\mathbf{\ket{11****}:}$} Notice that by symmetry of the diagrams, the derivation for $\ket{****11}$ is completely similar. Again, we can reason here with the diagrams before unzipping:
				\begin{align*}
					&\def\fig{soundness-Axiom-tensor-2-2-11}
					{\setlength{\fboxsep}{0pt}\colorbox{gray!15}{~\strut}}
					\eq{\eqref{ax:Z-W-bialgebra}}{\setlength{\fboxsep}{0pt}\colorbox{gray!15}{~\strut}}
					\eq{\eqref{ax:W-spider}\\\eqref{ax:W-bialgebra}}{\setlength{\fboxsep}{0pt}\colorbox{gray!15}{~\strut}}\\
					&\def\fig{soundness-Axiom-tensor-2-2-11}
					\eq{\eqref{ax:W-inv}\\\eqref{ax:Z-W-bialgebra}}{\setlength{\fboxsep}{0pt}\colorbox{gray!15}{~\strut}}
					\def\fig{soundness-Axiom-tensor-2-2-11}
					\eq{\eqref{ax:W-bialgebra}\\\eqref{ax:tick-through-W-spider}\\\eqref{ax:Z-W-bialgebra}}{\setlength{\fboxsep}{0pt}\colorbox{gray!15}{~\strut}}
					\eq{\eqref{ax:W-spider}\\\eqref{ax:W-inv}}{\setlength{\fboxsep}{0pt}\colorbox{gray!15}{~\strut\input{./figures/\fig/\fig_05.tikz}}}\\
					&\def\fig{soundness-Axiom-tensor-2-1-11}
					\eq{\eqref{ax:W-bialgebra}\\\eqref{ax:tick-through-W-spider}\\\eqref{ax:Z-W-bialgebra}}{\setlength{\fboxsep}{0pt}\colorbox{gray!15}{~\strut}}
					\eq{\eqref{ax:W-inv}\\\eqref{ax:Z-W-bialgebra}}{\setlength{\fboxsep}{0pt}\colorbox{gray!15}{~\strut}}
					\eq{\eqref{ax:W-spider}\\\eqref{ax:W-bialgebra}}{\setlength{\fboxsep}{0pt}\colorbox{gray!15}{~\strut}}\\
					&\def\fig{soundness-Axiom-tensor-2-1-11}
					\eq{\eqref{ax:Z-W-bialgebra}}{\setlength{\fboxsep}{0pt}\colorbox{gray!15}{~\strut}}
				\end{align*}
				\item {$\mathbf{\ket{10****}:}$} In that case, we are going to show that the obtained diagram is the zero map in each case:
				\def\fig{soundness-Axiom-tensor-2-1-10}
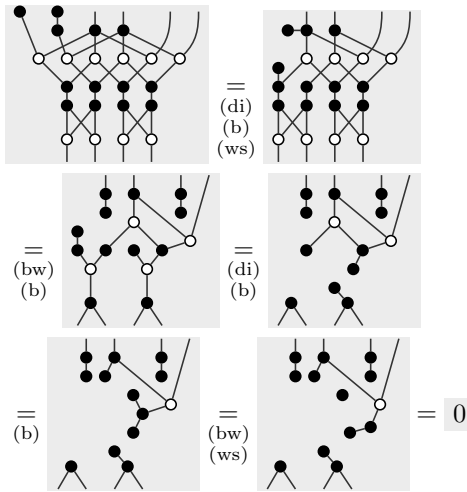
				\begin{align*}
					&{\setlength{\fboxsep}{0pt}\colorbox{gray!15}{~\strut}}
					\eq{\eqref{ax:W-binary-diffusion}\\\eqref{ax:Z-W-bialgebra}\\\eqref{ax:W-spider}}{\setlength{\fboxsep}{0pt}\colorbox{gray!15}{~\strut}}\\
					&\eq{\eqref{ax:W-bialgebra}\\\eqref{ax:Z-W-bialgebra}}{\setlength{\fboxsep}{0pt}\colorbox{gray!15}{~\strut}}
					\eq{\eqref{ax:W-binary-diffusion}\\\eqref{ax:Z-W-bialgebra}}{\setlength{\fboxsep}{0pt}\colorbox{gray!15}{~\strut}}\\
					&\eq{\eqref{ax:Z-W-bialgebra}}{\setlength{\fboxsep}{0pt}\colorbox{gray!15}{~\strut}}
					\eq{\eqref{ax:W-bialgebra}\\\eqref{ax:W-spider}}{\setlength{\fboxsep}{0pt}\colorbox{gray!15}{~\strut\input{./figures/\fig/\fig_05.tikz}}}
					\eq{}{\setlength{\fboxsep}{0pt}\colorbox{gray!15}{~\strut\input{./figures/\fig/\fig_06.tikz}}}
				\end{align*}
				where we used the fact that $\interp{\tikzfig{zero}}=0$. Then:
				\def\fig{soundness-Axiom-tensor-2-2-10}
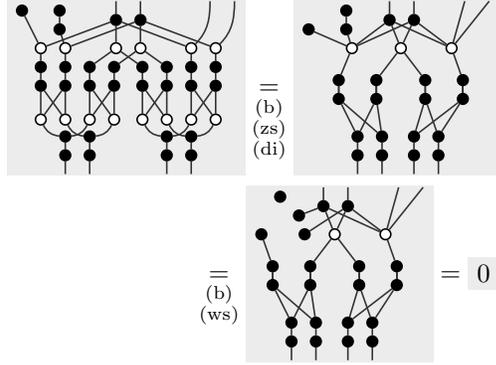
				\begin{align*}
					{\setlength{\fboxsep}{0pt}\colorbox{gray!15}{~\strut}}
					\eq{\eqref{ax:Z-W-bialgebra}\\\eqref{ax:Z-spider}\\\eqref{ax:W-binary-diffusion}}{\setlength{\fboxsep}{0pt}\colorbox{gray!15}{~\strut}}\\
					\eq{\eqref{ax:Z-W-bialgebra}\\\eqref{ax:W-spider}}{\setlength{\fboxsep}{0pt}\colorbox{gray!15}{~\strut}}
					\eq{}{\setlength{\fboxsep}{0pt}\colorbox{gray!15}{~\strut}}
				\end{align*}
				Again, by symmetry, the similar equalities are obtained for inputs $\ket{01****}$, $\ket{****01}$ and $\ket{****10}$.
			\end{itemize}
			Finally, all the ways to apply a classical state to the two leftmost and the two rightmost inputs end up with the same diagrams on both side. By linearity, this implies that the two diagrams have the same interpretation.
		\end{itemize}
	\end{proof}
	
	The following lemmas are fairly direct and are used extensively in the derivations to come:\\
	\noindent
	\begin{minipage}[t]{0.49\columnwidth}
		\begin{lemma}
			\def\fig{lemma-obvious}
			\label{lem:obvious}
			\begin{align*}
				{\setlength{\fboxsep}{0pt}\colorbox{gray!15}{~\strut}}
				\eq{}{\setlength{\fboxsep}{0pt}\colorbox{gray!15}{~\strut}}
				\eq{}{\setlength{\fboxsep}{0pt}\colorbox{gray!15}{~\strut}}
			\end{align*}
		\end{lemma}
	\end{minipage}
	\hfill
	\begin{minipage}[t]{0.49\columnwidth}
		\begin{lemma}
			\label{lem:obvious-2}
			\def\fig{lemma-obvious-2}
			\begin{align*}
				{\setlength{\fboxsep}{0pt}\colorbox{gray!15}{~\strut}}
				\eq{}{\setlength{\fboxsep}{0pt}\colorbox{gray!15}{~\strut}}
			\end{align*}
		\end{lemma}
	\end{minipage}
	
	\bigskip
	\noindent
	\begin{minipage}[t]{0.49\columnwidth}
		\begin{lemma}
			\label{lem:untitled}
			\def\fig{lemma-untitled}
			\begin{align*}
				{\setlength{\fboxsep}{0pt}\colorbox{gray!15}{~\strut}}
				\eq{}{\setlength{\fboxsep}{0pt}\colorbox{gray!15}{~\strut\input{./figures/\fig/\fig_05.tikz}}}
			\end{align*}
		\end{lemma}
	\end{minipage}
	\hfill
	\begin{minipage}[t]{0.49\columnwidth}
		\begin{lemma}
			\label{lem:E3-through-Z}
			\def\fig{E3-through-Z}
			\begin{align*}
				{\setlength{\fboxsep}{0pt}\colorbox{gray!15}{~\strut}}
				\eq{}{\setlength{\fboxsep}{0pt}\colorbox{gray!15}{~\strut}}
			\end{align*}
		\end{lemma}
	\end{minipage}
	
	\bigskip
	\noindent
	\begin{minipage}[t]{0.49\columnwidth}
		\begin{lemma}
			\label{lem:ground-removed-by-E3}
			\def\fig{ground-removed-by-E3}
			\begin{align*}
				{\setlength{\fboxsep}{0pt}\colorbox{gray!15}{~\strut}}
				\eq{}{\setlength{\fboxsep}{0pt}\colorbox{gray!15}{~\strut}}
			\end{align*}
		\end{lemma}
	\end{minipage}
	\hfill
	\begin{minipage}[t]{0.49\columnwidth}
		\begin{lemma}
			\label{lem:E3-through-W}
			\def\fig{E3-through-W}
			\begin{align*}
				{\setlength{\fboxsep}{0pt}\colorbox{gray!15}{~\strut}}
				\eq{}{\setlength{\fboxsep}{0pt}\colorbox{gray!15}{~\strut}}
			\end{align*}
		\end{lemma}
	\end{minipage}

	\begin{proof}[Proof of \Cref{lem:obvious}]
		\def\fig{lemma-obvious}
		\begin{align*}
			{\setlength{\fboxsep}{0pt}\colorbox{gray!15}{~\strut}}
			\eq{\eqref{ax:tick-through-W-spider}}{\setlength{\fboxsep}{0pt}\colorbox{gray!15}{~\strut}}
			\eq{\eqref{eq:tick-involution}}{\setlength{\fboxsep}{0pt}\colorbox{gray!15}{~\strut}}
			\eq{\eqref{ax:tick-through-Z-spider}}{\setlength{\fboxsep}{0pt}\colorbox{gray!15}{~\strut}}
		\end{align*}
	\end{proof}
	
	\begin{proof}[Proof of \Cref{lem:obvious-2}]
		\def\fig{lemma-obvious-2}
		\begin{align*}
			{\setlength{\fboxsep}{0pt}\colorbox{gray!15}{~\strut}}
			\eq{\eqref{ax:Z-spider}}{\setlength{\fboxsep}{0pt}\colorbox{gray!15}{~\strut}}
			\eq{\eqref{ax:tick-through-Z-spider}}{\setlength{\fboxsep}{0pt}\colorbox{gray!15}{~\strut}}
			\eq{\eqref{eq:tick-involution}}{\setlength{\fboxsep}{0pt}\colorbox{gray!15}{~\strut}}
			\eq{\eqref{ax:Z-spider}}{\setlength{\fboxsep}{0pt}\colorbox{gray!15}{~\strut}}
		\end{align*}
	\end{proof}
	
	\begin{proof}[Proof of \Cref{lem:untitled}]
		\def\fig{lemma-untitled}
		\begin{align*}
			{\setlength{\fboxsep}{0pt}\colorbox{gray!15}{~\strut}}
			&\eq{\eqref{ax:Z-spider}}{\setlength{\fboxsep}{0pt}\colorbox{gray!15}{~\strut}}
			\eq{\eqref{ax:Z-W-bialgebra}\\\eqref{ax:Z-spider}}{\setlength{\fboxsep}{0pt}\colorbox{gray!15}{~\strut}}
			\eq{\eqref{ax:tick-through-W-spider}}{\setlength{\fboxsep}{0pt}\colorbox{gray!15}{~\strut}}\\
			&\eq{\ref{lem:obvious-2}}{\setlength{\fboxsep}{0pt}\colorbox{gray!15}{~\strut}}
			\eq{\eqref{ax:Z-W-bialgebra}\\\eqref{ax:Z-spider}}{\setlength{\fboxsep}{0pt}\colorbox{gray!15}{~\strut\input{./figures/\fig/\fig_05.tikz}}}
		\end{align*}
	\end{proof}
	
	\begin{proof}[Proof of \Cref{lem:E3-through-Z}]
		\def\fig{E3-through-Z}
		\begin{align*}
			{\setlength{\fboxsep}{0pt}\colorbox{gray!15}{~\strut}}
			\eq{\eqref{ax:Z-W-bialgebra}}{\setlength{\fboxsep}{0pt}\colorbox{gray!15}{~\strut}}
			\eq{\eqref{ax:tick-through-Z-spider}\\\eqref{ax:Z-spider}}{\setlength{\fboxsep}{0pt}\colorbox{gray!15}{~\strut}}
		\end{align*}
	\end{proof}
	
	\begin{proof}[Proof of \Cref{lem:E3-through-W}]
		\def\fig{E3-through-W}
		\begin{align*}
			{\setlength{\fboxsep}{0pt}\colorbox{gray!15}{~\strut}}
			\eq{\eqref{ax:Z-W-bialgebra}}{\setlength{\fboxsep}{0pt}\colorbox{gray!15}{~\strut}}
			\eq{\eqref{ax:tick-through-W-spider}\\\eqref{ax:W-spider}}{\setlength{\fboxsep}{0pt}\colorbox{gray!15}{~\strut}}
		\end{align*}
	\end{proof}
	
	\begin{proof}[Proof of \Cref{lem:ground-removed-by-E3}]
		\def\fig{ground-removed-by-E3}
		\begin{align*}
			{\setlength{\fboxsep}{0pt}\colorbox{gray!15}{~\strut}}
			\eq{\eqref{ax:Z-spider}}{\setlength{\fboxsep}{0pt}\colorbox{gray!15}{~\strut}}
			\eq{\ref{lem:E3-through-Z}}{\setlength{\fboxsep}{0pt}\colorbox{gray!15}{~\strut}}
			\eq{\eqref{ax:tick-through-W-spider}}{\setlength{\fboxsep}{0pt}\colorbox{gray!15}{~\strut}}
			\eq{\ref{lem:E3-through-Z}\\\eqref{ax:Z-spider}\\\eqref{ax:Z-id}}{\setlength{\fboxsep}{0pt}\colorbox{gray!15}{~\strut}}
		\end{align*}
	\end{proof}
	
	\subsection{Proofs of \Cref{sec:universality-NF}}
	
	\begin{proof}[Proof of \Cref{prop:states-are-hermitian}]
		Let $f$ be a state in $\zwtick$. Suppose is has $p$ occurrences of \tikzfig{tickedge} (we may try to reduce this number using the equational theory, but this is not a concern). It is then possible to push them out using elementary diagram deformations: 
		\def\fig{states-are-hermitian}
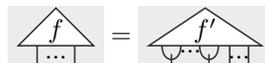
		\begin{align*}
			{\setlength{\fboxsep}{0pt}\colorbox{gray!15}{~\strut}}
			\eq{}{\setlength{\fboxsep}{0pt}\colorbox{gray!15}{~\strut}}
		\end{align*}
		We actually do not have to go through the whole interpretation to show that this is mapped to a Hermitian operator; we only have to look at what $\operatorname{HP}$ maps it to. First, we can show that the cup with a tick on it is mapped to the following:
		\def\fig{cup-ticked}
		\begin{align*}
			{\setlength{\fboxsep}{0pt}\colorbox{gray!15}{~\strut}}
			&\eq{}{\setlength{\fboxsep}{0pt}\colorbox{gray!15}{~\strut}}
			=\tikzfig{cup}\circ\left(\tikzfig{tickedge}\otimes\tikzfig{id}\right)\\
			&\mapsto{\setlength{\fboxsep}{0pt}\colorbox{gray!15}{~\strut}}
			\eq{}{\setlength{\fboxsep}{0pt}\colorbox{gray!15}{~\strut\input{./figures/\fig/\fig_05.tikz}}}
		\end{align*}
		It is direct to check that an $n$-fold tensor of identities is mapped as follows:
		\[{\setlength{\fboxsep}{0pt}\colorbox{gray!15}{\!\!\!\!\!\!\strut\tikzfig{id}\!\!...\!\!\!\!\!\!\tikzfig{id}\!\!}}
		\mapsto \tikzfig{HP-id}\]
		From the previous two mappings, we deduce that their tensor product is mapped to:
		\def\fig{HP-cuptick-id}
		\[{\setlength{\fboxsep}{0pt}\colorbox{gray!15}{~\strut}}\mapsto{\setlength{\fboxsep}{0pt}\colorbox{gray!15}{~\strut}}\]
		Finally, we get that:
		\def\fig{states-are-hermitian}
		\begin{align*}
			&{\setlength{\fboxsep}{0pt}\colorbox{gray!15}{~\strut}}\\
			&\mapsto\scalebox{0.9}{{\setlength{\fboxsep}{0pt}\colorbox{gray!15}{~\strut}}}\\
			&\quad\eq{}{\setlength{\fboxsep}{0pt}\colorbox{gray!15}{~\strut}}
		\end{align*}
		since $f'$ is \tikzfig{tickedge}-free. Now, if we cast this $0\to n$ superoperator into the corresponding ($n\to n$) operator through $\iota$, we can easily check that the latter is indeed Hermitian in $\cat{ZW}$ and hence $\interp{f}^\nmid$ is Hermitian as taking the dagger of this map gives exactly the same map.
	\end{proof}
	
	\begin{proof}[Proof of \Cref{lem:sum-branch-NF}]
		\def\fig{lemma-sum-branch-NF}
		\begin{align*}
			{\setlength{\fboxsep}{0pt}\colorbox{gray!15}{~\strut}}
			\eq{\eqref{ax:W-spider}\\\eqref{ax:tick-through-W-spider}}{\setlength{\fboxsep}{0pt}\colorbox{gray!15}{~\strut}}
			\eq{\eqref{ax:Z-W-bialgebra}\\\eqref{ax:W-spider}\\\eqref{ax:tick-through-W-spider}}{\setlength{\fboxsep}{0pt}\colorbox{gray!15}{~\strut}}
			\eq{\eqref{ax:sum}\\\eqref{ax:Z-spider}}{\setlength{\fboxsep}{0pt}\colorbox{gray!15}{~\strut}}
		\end{align*}
	\end{proof}
	
	\begin{proof}[Proof of \Cref{prop:NF-preserves-semantics}]
		First of all, let us consider that the diagram is already in reduced normal form:
		\[ \mathcal N \left( \sum_{\substack{i=1}}^{m} \left(\lambda_i'\ketbra{\vec x_i}{\vec y_i}+\overline\lambda_i'\ketbra{\vec y_i}{\vec x_i}\right) \right) = \def\fig{NF-decomp}{\setlength{\fboxsep}{0pt}\colorbox{gray!15}{~\strut}}\]
		We may prove the result by induction on $m$.
		\begin{itemize}
			\item \textbf{Case $\mathbf{m=0}$:} In this case, the diagram is reduced to: $\tikzfig{NF-0}$, which indeed represents the $0$ map, as $\interp{\tikzfig{zero}}^\nmid=0$.
			\item \textbf{Case $\mathbf{m=1}$:} This case is not necessary, but it will be reused in the following. The diagram in this case is reduced to: \def\fig{NF-decomp}{\setlength{\fboxsep}{0pt}\colorbox{gray!15}{~\strut}}. By reordering the outputs, we can write the diagram as follows:
			\def\fig{NF-1}
			{\setlength{\fboxsep}{0pt}\colorbox{gray!15}{~\strut}} with $\lambda = \lambda_1'$. It is then mapped through $\operatorname{HP}$ to:
			\begin{align*}
				&{\setlength{\fboxsep}{0pt}\colorbox{gray!15}{~\strut}}
				\eq{\eqref{ax:W-spider}\\\eqref{ax:Z-W-bialgebra}}{\setlength{\fboxsep}{0pt}\colorbox{gray!15}{~\strut}}\\
				&\eq{\eqref{ax:W-binary-diffusion}\\\eqref{ax:Z-W-bialgebra}\\\eqref{ax:W-bialgebra}\\\eqref{ax:W-spider}}{\setlength{\fboxsep}{0pt}\colorbox{gray!15}{~\strut}}\\
				&~=~\overline\lambda~{\setlength{\fboxsep}{0pt}\colorbox{gray!15}{~\strut}}
				~+~\lambda~{\setlength{\fboxsep}{0pt}\colorbox{gray!15}{~\strut\input{./figures/\fig/\fig_05.tikz}}}
			\end{align*}
			by decomposing the Z-spiders, and whose interpretation is 
			\[\lambda \ketbra{0^a1^b0^c1^d}{0^a0^b1^c1^d} + \overline\lambda\ketbra{0^a0^b1^c1^d}{0^a1^b0^c1^d}\]
			Undoing the permutation we did to group outputs together does put the non-zero coefficients to their designated place.
			\item \textbf{Case $\mathbf{m}$:}
			We will use the following identity:
			\def\fig{E3-decomposition}
			\begin{align*}
				\interp{{\setlength{\fboxsep}{0pt}\colorbox{gray!15}{~\strut}}}^\nmid
				=\interp{{\setlength{\fboxsep}{0pt}\colorbox{gray!15}{~\strut}}}^\nmid
				+\interp{{\setlength{\fboxsep}{0pt}\colorbox{gray!15}{~\strut}}}^\nmid
			\end{align*}
			which can be proven as:
			\def\fig{E3-decomposition-proof}
			\begin{align*}
				&\interp{\operatorname{HP}\left({\setlength{\fboxsep}{0pt}\colorbox{gray!15}{~\strut}}\right)}
				\eq{}\interp{{\setlength{\fboxsep}{0pt}\colorbox{gray!15}{~\strut}}}
				\eq{\eqref{ax:Z-spider}\\\eqref{ax:W-spider}\\\eqref{ax:Z-W-bialgebra}\\\eqref{ax:W-binary-diffusion}}\interp{{\setlength{\fboxsep}{0pt}\colorbox{gray!15}{~\strut}}}\\
				&\!\!\eq{}\!\interp{{\setlength{\fboxsep}{0pt}\colorbox{gray!15}{~\strut}}}
				\!\!+\!\!\interp{{\setlength{\fboxsep}{0pt}\colorbox{gray!15}{~\strut}}}
				\eq{\eqref{ax:W-bialgebra}\\\eqref{ax:Z-W-bialgebra}\\\eqref{ax:W-spider}}\interp{{\setlength{\fboxsep}{0pt}\colorbox{gray!15}{~\strut\input{./figures/\fig/\fig_05.tikz}}}}
				\!\!+\!\!\interp{{\setlength{\fboxsep}{0pt}\colorbox{gray!15}{~\strut\input{./figures/\fig/\fig_06.tikz}}}}
			\end{align*}
			and ${\setlength{\fboxsep}{0pt}\colorbox{gray!15}{~\strut\input{./figures/\fig/\fig_07.tikz}}}~~\overset{\operatorname{HP}}\mapsto~~{\setlength{\fboxsep}{0pt}\colorbox{gray!15}{~\strut\input{./figures/\fig/\fig_05.tikz}}}$ and ${\setlength{\fboxsep}{0pt}\colorbox{gray!15}{~\strut\input{./figures/\fig/\fig_08.tikz}}}~~\overset{\operatorname{HP}}\mapsto~~{\setlength{\fboxsep}{0pt}\colorbox{gray!15}{~\strut\input{./figures/\fig/\fig_06.tikz}}}$.
			If we apply this to the diagram in normal form, on the first coefficient, we get:
			\def\fig{NF-decomp}
			\begin{align*}
				&\interp{{\setlength{\fboxsep}{0pt}\colorbox{gray!15}{~\strut}}}^\nmid\\
				&\eq{}\!\interp{{\setlength{\fboxsep}{0pt}\colorbox{gray!15}{~\strut}}}^\nmid\!\!\!
				+\!\interp{{\setlength{\fboxsep}{0pt}\colorbox{gray!15}{~\strut}}}^\nmid\\
				&\eq{\eqref{ax:Z-W-bialgebra}\\\eqref{ax:W-inv}\\\eqref{ax:W-bialgebra}\\\eqref{ax:tick-through-W-spider}\\\eqref{ax:W-spider}}\interp{{\setlength{\fboxsep}{0pt}\colorbox{gray!15}{~\strut}}}^\nmid
				+\interp{{\setlength{\fboxsep}{0pt}\colorbox{gray!15}{~\strut}}}^\nmid\\
			\end{align*}
			Notice that the left term is the case $m=1$, and that right term is a diagram in normal form with $m-1$ white nodes. Hence, we may apply the induction hypothesis on it to get:
			\begin{align*}
				&\interp{{\setlength{\fboxsep}{0pt}\colorbox{gray!15}{~\strut}}}^\nmid\\
				&\qquad=\lambda_1'\ketbra{\vec x_1}{\vec y_1}+\overline\lambda_1'\ketbra{\vec y_1}{\vec x_1}\\
				&\qquad\qquad+ \sum_{\substack{i=2}}^{m} \left(\lambda_i'\ketbra{\vec x_i}{\vec y_i}+\overline\lambda_i'\ketbra{\vec y_i}{\vec x_i}\right)
			\end{align*}
			This finishes the proof.
		\end{itemize}
	\end{proof}
	
	\subsection{Proofs of \Cref{sec:CPM}}
	
	\begin{lemma}
		\label{lem:pure-zw-lemmas}
		If $R$ contains all half angles, then $e^{i\frac\pi4},e^{-i\frac\pi4}\in R$. If $e^{i\frac\pi4}\in R$, then $\sqrt2,\frac1{\sqrt2}\in R$.\\
		The following equations are sound\footnote{These equations are direct translations of ZX \cite{vandewetering2020zxcalculus} axioms or known lemmas, when taking the black $K_3$ gadget as ZX's $\pi$-X-spider, and \def\fig{ZW-pure-H-decomp}${\setlength{\fboxsep}{0pt}\colorbox{gray!15}{~\strut}}$ as un unnormalised Hadamard gate.} and $\tikzfig{tickedge}$-free, whence they are derivable in $\cat{ZW}$ by completeness, and a fortiori also in $\zwtick$:\\
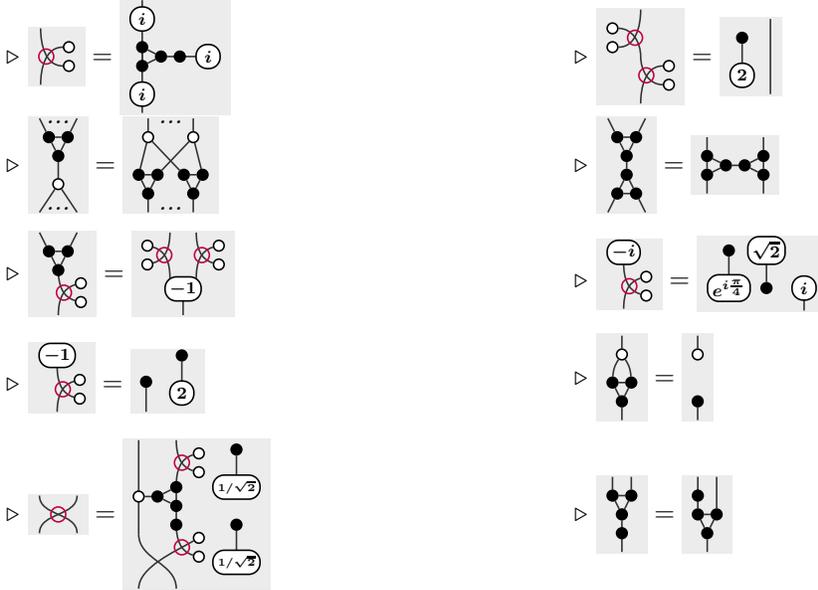
		\begin{minipage}{0.45\columnwidth}
			$\triangleright$ \def\fig{ZW-pure-H-decomp}${\setlength{\fboxsep}{0pt}\colorbox{gray!15}{~\strut}}\eq{}{\setlength{\fboxsep}{0pt}\colorbox{gray!15}{~\strut}}$
		\end{minipage}
		\begin{minipage}{0.45\columnwidth}
			$\triangleright$ \def\fig{ZW-pure-H-involution}${\setlength{\fboxsep}{0pt}\colorbox{gray!15}{~\strut}}\eq{}{\setlength{\fboxsep}{0pt}\colorbox{gray!15}{~\strut}}$
		\end{minipage}\\
		\medskip
		\begin{minipage}{0.45\columnwidth}
			$\triangleright$ \def\fig{ZW-pure-Z-X-bialgebra}${\setlength{\fboxsep}{0pt}\colorbox{gray!15}{~\strut}}\eq{}{\setlength{\fboxsep}{0pt}\colorbox{gray!15}{~\strut}}$
		\end{minipage}
		\begin{minipage}{0.45\columnwidth}
			$\triangleright$ \def\fig{ZW-pure-X-Frobenius}${\setlength{\fboxsep}{0pt}\colorbox{gray!15}{~\strut}}\eq{}{\setlength{\fboxsep}{0pt}\colorbox{gray!15}{~\strut}}$
		\end{minipage}\\
		\medskip
		\begin{minipage}{0.45\columnwidth}
			$\triangleright$ \def\fig{ZW-pure-colour-change}${\setlength{\fboxsep}{0pt}\colorbox{gray!15}{~\strut}}\eq{}{\setlength{\fboxsep}{0pt}\colorbox{gray!15}{~\strut}}$
		\end{minipage}
		\begin{minipage}{0.45\columnwidth}
			$\triangleright$ \def\fig{ZW-pure-i-minus-i}${\setlength{\fboxsep}{0pt}\colorbox{gray!15}{~\strut}}\eq{}{\setlength{\fboxsep}{0pt}\colorbox{gray!15}{~\strut}}$
		\end{minipage}\\
		\medskip
		\begin{minipage}{0.45\columnwidth}
			$\triangleright$ \def\fig{ZW-pure-H-ket1}${\setlength{\fboxsep}{0pt}\colorbox{gray!15}{~\strut}}\eq{}{\setlength{\fboxsep}{0pt}\colorbox{gray!15}{~\strut}}$
		\end{minipage}
		\begin{minipage}{0.45\columnwidth}
			$\triangleright$ \def\fig{ZW-pure-Hopf}${\setlength{\fboxsep}{0pt}\colorbox{gray!15}{~\strut}}\eq{}{\setlength{\fboxsep}{0pt}\colorbox{gray!15}{~\strut}}$
		\end{minipage}\\
		\medskip
		\begin{minipage}{0.45\columnwidth}
			$\triangleright$ \def\fig{ZW-pure-fswap-decomp}${\setlength{\fboxsep}{0pt}\colorbox{gray!15}{~\strut}}\eq{}{\setlength{\fboxsep}{0pt}\colorbox{gray!15}{~\strut}}$
		\end{minipage}
		\begin{minipage}{0.45\columnwidth}
			$\triangleright$ \def\fig{ZW-pure-X-assoc}${\setlength{\fboxsep}{0pt}\colorbox{gray!15}{~\strut}}\eq{}{\setlength{\fboxsep}{0pt}\colorbox{gray!15}{~\strut}}$
		\end{minipage}
	\end{lemma}
	
	\begin{proof}[Proof of \Cref{prop:ground-derivable}]
		If $R$ contains all half angles, $-1=e^{i\pi}\in R$, $e^{i\frac\pi2}\in R$ so $e^{i\frac\pi4}\in R$, hence $e^{-i\frac\pi4}=(e^{i\frac\pi4})^7\in R$. Then $\sqrt2 = e^{i\frac\pi4}+e^{-i\frac\pi4}\in R$ and $\frac1{\sqrt2}=\frac{\sqrt2}2\in R$. In the following, we denote $\omega:=e^{i\frac\pi4}$.
		\begin{itemize}
			\setlength\itemsep{0.5em}
			\item \def\fig{CPM-axiom-scalar-derivation}
			$\begin{aligned}[t]
				{\setlength{\fboxsep}{0pt}\colorbox{gray!15}{~\strut}}
				\eq{\eqref{ax:W-inv}\\\eqref{ax:tick-through-W-spider}\\\eqref{ax:W-binary-diffusion}\\\eqref{ax:Z-W-bialgebra}}{\setlength{\fboxsep}{0pt}\colorbox{gray!15}{~\strut}}
				\eq{\eqref{ax:tick-through-W-spider}}{\setlength{\fboxsep}{0pt}\colorbox{gray!15}{~\strut}}
				\eq{\eqref{ax:W-bialgebra}}{\setlength{\fboxsep}{0pt}\colorbox{gray!15}{~\strut}}
			\end{aligned}$
			\item If $R$ contains all half-angles:
			\def\fig{CPM-axiom-Z-derivation-half-angles}
			\begin{align*}
				{\setlength{\fboxsep}{0pt}\colorbox{gray!15}{~\strut}}
				&\eq{\eqref{ax:Z-spider}}{\setlength{\fboxsep}{0pt}\colorbox{gray!15}{~\strut}}
				\eq{\eqref{ax:tick-through-Z-spider}}{\setlength{\fboxsep}{0pt}\colorbox{gray!15}{~\strut}}\\
				&\eq{\eqref{ax:Z-spider}}{\setlength{\fboxsep}{0pt}\colorbox{gray!15}{~\strut}}
				\eq{\eqref{ax:Z-id}}{\setlength{\fboxsep}{0pt}\colorbox{gray!15}{~\strut}}
			\end{align*}
			else:
			\def\fig{CPM-axiom-Z-derivation-no-half-angles}
			\begin{align*}
				{\setlength{\fboxsep}{0pt}\colorbox{gray!15}{~\strut}}
				&\eq{\eqref{ax:W-inv}\\\eqref{ax:tick-through-W-spider}\\\eqref{ax:W-binary-diffusion}}{\setlength{\fboxsep}{0pt}\colorbox{gray!15}{~\strut}}
				\eq{\eqref{ax:tick-loop}}{\setlength{\fboxsep}{0pt}\colorbox{gray!15}{~\strut}}
				\eq{\eqref{ax:Z-W-bialgebra}}{\setlength{\fboxsep}{0pt}\colorbox{gray!15}{~\strut}}\\
				&\eq{\eqref{ax:tick-through-Z-spider}}{\setlength{\fboxsep}{0pt}\colorbox{gray!15}{~\strut}}
				\eq{\eqref{ax:Z-spider}}{\setlength{\fboxsep}{0pt}\colorbox{gray!15}{~\strut\input{./figures/\fig/\fig_05.tikz}}}
				\eq{\eqref{ax:tick-loop}}{\setlength{\fboxsep}{0pt}\colorbox{gray!15}{~\strut\input{./figures/\fig/\fig_06.tikz}}}
				\eq{\eqref{ax:W-inv}\\\eqref{ax:tick-through-W-spider}\\\eqref{ax:W-binary-diffusion}}{\setlength{\fboxsep}{0pt}\colorbox{gray!15}{~\strut\input{./figures/\fig/\fig_07.tikz}}}
			\end{align*}
			\item \def\fig{CPM-axiom-phase-derivation}
			$\begin{aligned}[t]
				{\setlength{\fboxsep}{0pt}\colorbox{gray!15}{~\strut}}
				&\eq{}{\setlength{\fboxsep}{0pt}\colorbox{gray!15}{~\strut}}
				\eq{\eqref{ax:Z-W-bialgebra}}{\setlength{\fboxsep}{0pt}\colorbox{gray!15}{~\strut}}
				\eq{\eqref{ax:Z-spider}}{\setlength{\fboxsep}{0pt}\colorbox{gray!15}{~\strut}}\\
				&\eq{}{\setlength{\fboxsep}{0pt}\colorbox{gray!15}{~\strut}}
				\eq{}{\setlength{\fboxsep}{0pt}\colorbox{gray!15}{~\strut\input{./figures/\fig/\fig_05.tikz}}}
			\end{aligned}$
			\item First:\hspace*{-1em}
			\def\fig{CPM-red-i-aux}
			$\begin{aligned}[t]
				{\setlength{\fboxsep}{0pt}\colorbox{gray!15}{~\strut}}
				\eq{\eqref{ax:Z-spider}}{\setlength{\fboxsep}{0pt}\colorbox{gray!15}{~\strut}}
				\eq{\ref{lem:pure-zw-lemmas}\\\eqref{ax:tick-through-Z-spider}\\\eqref{ax:Z-spider}}{\setlength{\fboxsep}{0pt}\colorbox{gray!15}{~\strut}}\\
				\eq{\eqref{ax:tick-through-W-spider}\\\eqref{ax:tick-through-Z-spider}}{\setlength{\fboxsep}{0pt}\colorbox{gray!15}{~\strut}}
				\eq{\ref{lem:pure-zw-lemmas}\\\eqref{ax:tick-through-Z-spider}\\\eqref{ax:Z-spider}}{\setlength{\fboxsep}{0pt}\colorbox{gray!15}{~\strut}}
				\eq{\eqref{ax:Z-spider}}{\setlength{\fboxsep}{0pt}\colorbox{gray!15}{~\strut\input{./figures/\fig/\fig_05.tikz}}}\\
				\eq{\ref{lem:pure-zw-lemmas}}{\setlength{\fboxsep}{0pt}\colorbox{gray!15}{~\strut\input{./figures/\fig/\fig_06.tikz}}}
				\eq{\eqref{ax:tick-through-Z-spider}\\\eqref{ax:Z-spider}}{\setlength{\fboxsep}{0pt}\colorbox{gray!15}{~\strut\input{./figures/\fig/\fig_07.tikz}}}
			\end{aligned}$\\
			Then:
			\def\fig{CPM-red-i}
			$\begin{aligned}[t]
				{\setlength{\fboxsep}{0pt}\colorbox{gray!15}{~\strut}}
				\eq{\ref{lem:pure-zw-lemmas}\\\eqref{ax:W-binary-diffusion}\\\eqref{ax:tick-through-W-spider}\\\eqref{ax:W-inv}}{\setlength{\fboxsep}{0pt}\colorbox{gray!15}{~\strut}}
				\eq{\ref{lem:pure-zw-lemmas}\\\eqref{ax:Z-spider}}{\setlength{\fboxsep}{0pt}\colorbox{gray!15}{~\strut}}\\
				\eq{\eqref{ax:tick-through-W-spider}}{\setlength{\fboxsep}{0pt}\colorbox{gray!15}{~\strut}}
				\eq{\ref{lem:pure-zw-lemmas}\\\eqref{ax:Z-spider}}{\setlength{\fboxsep}{0pt}\colorbox{gray!15}{~\strut}}
				\eq{\ref{lem:pure-zw-lemmas}}{\setlength{\fboxsep}{0pt}\colorbox{gray!15}{~\strut\input{./figures/\fig/\fig_05.tikz}}}\\
				\eq{\ref{lem:pure-zw-lemmas}\\\eqref{ax:tick-through-Z-spider}\\\eqref{ax:tick-through-fswap}}{\setlength{\fboxsep}{0pt}\colorbox{gray!15}{~\strut\input{./figures/\fig/\fig_06.tikz}}}
				\eq{\ref{lem:pure-zw-lemmas}}{\setlength{\fboxsep}{0pt}\colorbox{gray!15}{~\strut\input{./figures/\fig/\fig_07.tikz}}}
				\eq{}{\setlength{\fboxsep}{0pt}\colorbox{gray!15}{~\strut\input{./figures/\fig/\fig_08.tikz}}}\\
				\eq{\ref{lem:pure-zw-lemmas}}{\setlength{\fboxsep}{0pt}\colorbox{gray!15}{~\strut\input{./figures/\fig/\fig_09.tikz}}}
				\eq{\eqref{ax:W-bialgebra}\\\eqref{ax:W-spider}}{\setlength{\fboxsep}{0pt}\colorbox{gray!15}{~\strut\input{./figures/\fig/\fig_10.tikz}}}
				\eq{\eqref{ax:W-inv}}{\setlength{\fboxsep}{0pt}\colorbox{gray!15}{~\strut\input{./figures/\fig/\fig_11.tikz}}}
			\end{aligned}$\\[0.5em]
			Finally:
			\def\fig{CPM-axiom-H-derivation}
			$\begin{aligned}[t]
				{\setlength{\fboxsep}{0pt}\colorbox{gray!15}{~\strut}}
				\eq{\ref{lem:pure-zw-lemmas}}{\setlength{\fboxsep}{0pt}\colorbox{gray!15}{~\strut}}
				\eq{}{\setlength{\fboxsep}{0pt}\colorbox{gray!15}{~\strut}}
			\end{aligned}$
			\item First:
			\def\fig{CPM-CNot-derivation}
			$\begin{aligned}[t]
				{\setlength{\fboxsep}{0pt}\colorbox{gray!15}{~\strut}}
				\eq{\ref{lem:pure-zw-lemmas}\\\eqref{ax:Z-spider}}{\setlength{\fboxsep}{0pt}\colorbox{gray!15}{~\strut}}
				\eq{\eqref{ax:Z-spider}\\\eqref{ax:tick-through-W-spider}}{\setlength{\fboxsep}{0pt}\colorbox{gray!15}{~\strut}}\\
				\eq{\eqref{ax:tick-through-Z-spider}}{\setlength{\fboxsep}{0pt}\colorbox{gray!15}{~\strut}}
				\eq{\ref{lem:pure-zw-lemmas}\\\eqref{ax:Z-spider}}{\setlength{\fboxsep}{0pt}\colorbox{gray!15}{~\strut}}\\
				\eq{\ref{lem:pure-zw-lemmas}}{\setlength{\fboxsep}{0pt}\colorbox{gray!15}{~\strut\input{./figures/\fig/\fig_05.tikz}}}
				\eq{\eqref{ax:W-bialgebra}\\\eqref{ax:W-spider}\\\eqref{ax:W-inv}\\\eqref{ax:Z-id}}{\setlength{\fboxsep}{0pt}\colorbox{gray!15}{~\strut\input{./figures/\fig/\fig_06.tikz}}}
			\end{aligned}$\\[0.5em]
			Then:
			\def\fig{CPM-axiom-fswap-derivation}
			$\begin{aligned}[t]
				{\setlength{\fboxsep}{0pt}\colorbox{gray!15}{~\strut}}
				\eq[]{\ref{lem:pure-zw-lemmas}}{\setlength{\fboxsep}{0pt}\colorbox{gray!15}{~\strut}}
				\eq[]{}{\setlength{\fboxsep}{0pt}\colorbox{gray!15}{~\strut}}\\
				\eq[]{}{\setlength{\fboxsep}{0pt}\colorbox{gray!15}{~\strut}}
				\eq[]{}{\setlength{\fboxsep}{0pt}\colorbox{gray!15}{~\strut}}
			\end{aligned}$
		\end{itemize}
	\end{proof}
	
	\subsection{Proofs of \Cref{sec:completeness}}
	
	\begin{proof}[Proof of \Cref{lem:NF-negation}]
		First:
		\def\fig{NF-negation-1}
		\begin{align*}
			{\setlength{\fboxsep}{0pt}\colorbox{gray!15}{~\strut}}
			\eq{\eqref{ax:Z-spider}\\\eqref{ax:W-spider}\\\eqref{ax:tick-through-Z-spider}}{\setlength{\fboxsep}{0pt}\colorbox{gray!15}{~\strut}}
			\eq{\eqref{ax:Z-spider}\\\eqref{ax:Z-W-bialgebra}}{\setlength{\fboxsep}{0pt}\colorbox{gray!15}{~\strut}}\\
			\eq[\!]{\eqref{ax:tick-through-Z-spider}\\\eqref{ax:Z-id}}{\setlength{\fboxsep}{0pt}\colorbox{gray!15}{~\strut}}
			\eq[\!]{\eqref{ax:W-binary-diffusion}\\\eqref{ax:Z-W-bialgebra}\\\eqref{ax:W-spider}}{\setlength{\fboxsep}{0pt}\colorbox{gray!15}{~\strut}}
			\eq[\!]{\eqref{ax:tick-through-Z-spider}\\\eqref{ax:Z-spider}}{\setlength{\fboxsep}{0pt}\colorbox{gray!15}{~\strut\input{./figures/\fig/\fig_05.tikz}}}
		\end{align*}
		Then:
		\def\fig{NF-negation-2}
		\begin{align*}
			&{\setlength{\fboxsep}{0pt}\colorbox{gray!15}{~\strut}}\\
			&\eq{\eqref{ax:W-spider}\\\eqref{ax:Z-spider}\\\eqref{ax:tick-through-W-spider}\\\eqref{ax:Z-W-bialgebra}}{\setlength{\fboxsep}{0pt}\colorbox{gray!15}{~\strut}}\\
			&\eq{}{\setlength{\fboxsep}{0pt}\colorbox{gray!15}{~\strut}}\\
			&\eq{\eqref{ax:W-spider}\\\eqref{ax:Z-spider}\\\eqref{ax:tick-through-W-spider}\\\eqref{ax:Z-W-bialgebra}}{\setlength{\fboxsep}{0pt}\colorbox{gray!15}{~\strut}}
		\end{align*}
	\end{proof}
	
	\begin{lemma}
		\def\fig{NF-tensor-recomposing-branches-2}
		\label{lem:NF-tensor-branch-recomposition}
		\begin{align*}
			{\setlength{\fboxsep}{0pt}\colorbox{gray!15}{~\strut}}
			\eq{}{\setlength{\fboxsep}{0pt}\colorbox{gray!15}{~\strut}}
		\end{align*}
	\end{lemma}
	
	\begin{proof}[Proof of \Cref{lem:NF-tensor-branch-recomposition}]
		First:
		\def\fig{NF-tensor-recomposing-branches-1}
		\begin{align*}
			{\setlength{\fboxsep}{0pt}\colorbox{gray!15}{~\strut}}
			&\eq{\eqref{ax:W-binary-diffusion}\\\ref{lem:E3-through-Z}}{\setlength{\fboxsep}{0pt}\colorbox{gray!15}{~\strut}}
			\eq{\eqref{ax:Z-spider}\\\eqref{ax:nf-tensor-1}}{\setlength{\fboxsep}{0pt}\colorbox{gray!15}{~\strut}}
			\eq{\ref{lem:E3-through-Z}}{\setlength{\fboxsep}{0pt}\colorbox{gray!15}{~\strut}}\\
			&\eq{\eqref{ax:Z-spider}\\\eqref{ax:Z-id}\\\eqref{ax:Z-W-bialgebra}}{\setlength{\fboxsep}{0pt}\colorbox{gray!15}{~\strut}}
			\eq{\eqref{ax:tick-through-W-spider}\\\eqref{ax:W-spider}}{\setlength{\fboxsep}{0pt}\colorbox{gray!15}{~\strut\input{./figures/\fig/\fig_05.tikz}}}
		\end{align*}
		then:
		\def\fig{NF-tensor-recomposing-branches-2}
		\begin{align*}
			{\setlength{\fboxsep}{0pt}\colorbox{gray!15}{~\strut}}
			\eq{\eqref{ax:Z-spider}\\\eqref{ax:W-spider}}{\setlength{\fboxsep}{0pt}\colorbox{gray!15}{~\strut}}
			\eq{}{\setlength{\fboxsep}{0pt}\colorbox{gray!15}{~\strut}}
			\eq{\eqref{ax:Z-W-bialgebra}\\\eqref{ax:tick-through-Z-spider}\\\eqref{ax:Z-spider}\\\eqref{ax:W-spider}}{\setlength{\fboxsep}{0pt}\colorbox{gray!15}{~\strut}}
		\end{align*}
	\end{proof}
	
	\begin{lemma}
		\label{lem:NF-tensor-branch-decomposition}
		\def\fig{lemma-tensor}
		\begin{align*}
			{\setlength{\fboxsep}{0pt}\colorbox{gray!15}{~\strut}}
			\eq[]{}{\setlength{\fboxsep}{0pt}\colorbox{gray!15}{~\strut}}
		\end{align*}
	\end{lemma}
	
	\begin{proof}[Proof of \Cref{lem:NF-tensor-branch-decomposition}]
		\def\fig{lemma-tensor}
		\begin{align*}
			{\setlength{\fboxsep}{0pt}\colorbox{gray!15}{~\strut}}
			\eq{\eqref{ax:W-spider}\\\eqref{ax:Z-spider}\\\eqref{ax:Z-W-bialgebra}}{\setlength{\fboxsep}{0pt}\colorbox{gray!15}{~\strut}}\\
			\eq{\eqref{ax:nf-tensor-2}}{\setlength{\fboxsep}{0pt}\colorbox{gray!15}{~\strut}}
			\eq{\eqref{ax:Z-W-bialgebra}\\\eqref{ax:Z-spider}\\\eqref{ax:W-spider}}{\setlength{\fboxsep}{0pt}\colorbox{gray!15}{~\strut}}
		\end{align*}
	\end{proof}
	
	\begin{proof}[Proof of \Cref{prop:NF-tensor}]
		\def\fig{NF-tensor}
		\begin{align*}
			&{\setlength{\fboxsep}{0pt}\colorbox{gray!15}{~\strut}}
			\eq{\eqref{ax:W-bialgebra}}{\setlength{\fboxsep}{0pt}\colorbox{gray!15}{~\strut}}\\
			&\eq{\ref{lem:NF-negation}}{\setlength{\fboxsep}{0pt}\colorbox{gray!15}{~\strut}}
			\eq{\eqref{ax:Z-spider}\\\eqref{ax:W-spider}\\\eqref{ax:Z-W-bialgebra}}{\setlength{\fboxsep}{0pt}\colorbox{gray!15}{~\strut}}\\
			&\eq{\eqref{ax:W-bialgebra}}{\setlength{\fboxsep}{0pt}\colorbox{gray!15}{~\strut}}
			\eq{\eqref{ax:Z-W-bialgebra}\\\eqref{ax:W-spider}}{\setlength{\fboxsep}{0pt}\colorbox{gray!15}{~\strut\input{./figures/\fig/\fig_05.tikz}}}\\
			&\eq{\eqref{ax:fswap-through-Z}\\\eqref{ax:fswap-removal}\\\eqref{ax:Z-spider}}{\setlength{\fboxsep}{0pt}\colorbox{gray!15}{~\strut\input{./figures/\fig/\fig_06.tikz}}}
			\eq{\eqref{ax:Z-spider}\\\eqref{ax:W-spider}\\\eqref{ax:Z-W-bialgebra}}{\setlength{\fboxsep}{0pt}\colorbox{gray!15}{~\strut\input{./figures/\fig/\fig_07.tikz}}}\\
			&\eq{\eqref{ax:Z-spider}\\\eqref{ax:W-inv}}{\setlength{\fboxsep}{0pt}\colorbox{gray!15}{~\strut\input{./figures/\fig/\fig_08.tikz}}}
			\eq{\ref{lem:NF-tensor-branch-decomposition}}{\setlength{\fboxsep}{0pt}\colorbox{gray!15}{~\strut\input{./figures/\fig/\fig_09.tikz}}}\\
			&\eq{\eqref{ax:Z-W-bialgebra}\\\eqref{ax:tick-through-W-spider}\\\eqref{ax:W-spider}}{\setlength{\fboxsep}{0pt}\colorbox{gray!15}{~\strut\input{./figures/\fig/\fig_10.tikz}}}\\
			&\eq{\ref{lem:NF-tensor-branch-recomposition}\\\eqref{ax:tick-through-W-spider}\\\eqref{ax:W-spider}}{\setlength{\fboxsep}{0pt}\colorbox{gray!15}{~\strut\input{./figures/\fig/\fig_11.tikz}}}
		\end{align*}
	\end{proof}
	
	\begin{proof}[Proof of \Cref{lem:NF-bra-0}]
		Notice first that by the Negation Lemma, if the result is true for $\tikzfig{bra-0}$ then it is true for $\tikzfig{bra-1}$.  If we look at the case $\tikzfig{bra-0}$, we get:
		\def\fig{NF-bra-0}
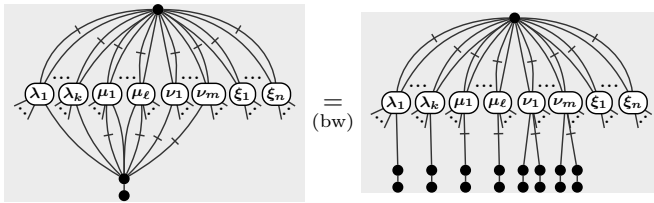
		\begin{align*}
			\scalebox{0.9}{${\setlength{\fboxsep}{0pt}\colorbox{gray!15}{~\strut}}$}
			\eq{\eqref{ax:W-bialgebra}}\scalebox{0.9}{${\setlength{\fboxsep}{0pt}\colorbox{gray!15}{~\strut}}$}
		\end{align*}
		It then remains to apply the copy rule (i.e.~\eqref{ax:Z-W-bialgebra} with $m=0$) and the W-spider rule \eqref{ax:W-spider} as many times as possible. This will remove the white nodes that were initially connected to the output (through $\tikzfig{id}$, $\tikzfig{tickedge}$, or both) and all the $\tikzfig{ket-0}$ produced will be absorbed thanks to the spider rule to either the top W-spider or the ones at the outputs.
	\end{proof}
	
	\begin{proof}[Proof of \Cref{lem:NF-W-2-1}]
		We have, starting from an arbitrary diagram in normal form:
		\def\fig{NF-W-2-1}
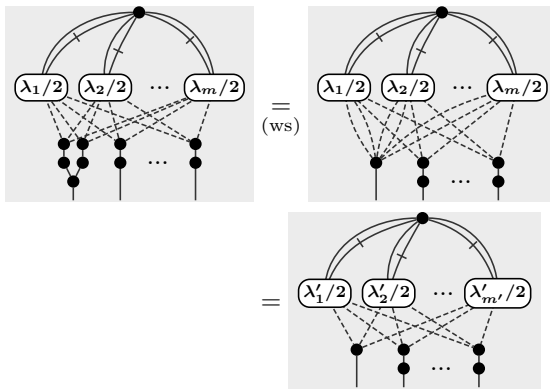
		\begin{align*}
			{\setlength{\fboxsep}{0pt}\colorbox{gray!15}{~\strut}}
			&\eq{\eqref{ax:W-spider}}{\setlength{\fboxsep}{0pt}\colorbox{gray!15}{~\strut}}\\
			&\eq{}{\setlength{\fboxsep}{0pt}\colorbox{gray!15}{~\strut}}
		\end{align*}
		where in the last equality we use equations \def\fig{Hopf}${\setlength{\fboxsep}{0pt}\colorbox{gray!15}{~\strut}}\eq{\eqref{ax:Z-spider}\\\eqref{ax:W-spider}\\\eqref{ax:Z-W-Hopf}\\\eqref{ax:Z-W-bialgebra}}{\setlength{\fboxsep}{0pt}\colorbox{gray!15}{~\strut}}$ and \def\fig{Hopf-tick}${\setlength{\fboxsep}{0pt}\colorbox{gray!15}{~\strut}}\eq{}{\setlength{\fboxsep}{0pt}\colorbox{gray!15}{~\strut}}$ (derived similarly with the addition of \eqref{ax:tick-through-W-spider} for instance) to remove parallel edges of the same type. In doing so, the white node is deleted and again it suffices to merge all occurrences of ~$\tikzfig{ket-0}$~ and apply \Cref{lem:NF-negation} to arrive to the normal form.
	\end{proof}
	
	\begin{lemma}
		\label{lem:NF-Z}
		\def\fig{lemma-NF-Z-aux}
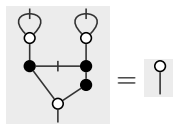
		\begin{align*}
			{\setlength{\fboxsep}{0pt}\colorbox{gray!15}{~\!\!\!\!\strut\!\!\!\!}}
			\eq{}{\setlength{\fboxsep}{0pt}\colorbox{gray!15}{~\strut\input{./figures/\fig/\fig_07.tikz}}}
		\end{align*}
	\end{lemma}
	
	\begin{proof}[Proof of \Cref{lem:NF-Z}]
		\def\fig{lemma-NF-Z-aux}
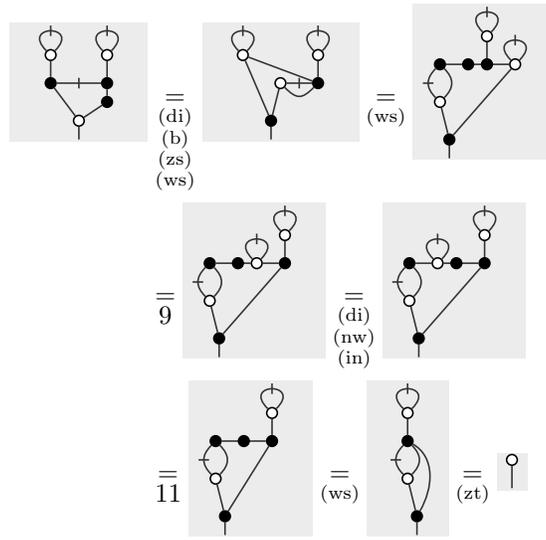
		\begin{align*}
			{\setlength{\fboxsep}{0pt}\colorbox{gray!15}{~\strut}}
			&\eq{\eqref{ax:W-binary-diffusion}\\\eqref{ax:Z-W-bialgebra}\\\eqref{ax:Z-spider}\\\eqref{ax:W-spider}}{\setlength{\fboxsep}{0pt}\colorbox{gray!15}{~\strut}}
			\eq{\eqref{ax:W-spider}}{\setlength{\fboxsep}{0pt}\colorbox{gray!15}{~\strut}}\\
			&\eq{\ref{lem:untitled}}{\setlength{\fboxsep}{0pt}\colorbox{gray!15}{~\strut}}
			\eq{\eqref{ax:W-binary-diffusion}\\\eqref{ax:tick-through-W-spider}\\\eqref{ax:W-inv}}{\setlength{\fboxsep}{0pt}\colorbox{gray!15}{~\strut}}\\
			&\eq{\ref{lem:ground-removed-by-E3}}{\setlength{\fboxsep}{0pt}\colorbox{gray!15}{~\strut\input{./figures/\fig/\fig_05.tikz}}}
			\eq{\eqref{ax:W-spider}}{\setlength{\fboxsep}{0pt}\colorbox{gray!15}{~\strut\input{./figures/\fig/\fig_06.tikz}}}
			\eq{\eqref{ax:NF-Z}}{\setlength{\fboxsep}{0pt}\colorbox{gray!15}{~\strut\input{./figures/\fig/\fig_07.tikz}}}
		\end{align*}
	\end{proof}
	
	\begin{proof}[Proof of \Cref{prop:NF-generators}]~
		\begin{itemize}
			\setlength{\itemsep}{0.5em}
			\item The normal form of the Z-spider can be obtained as:
			\def\fig{NF-Z}
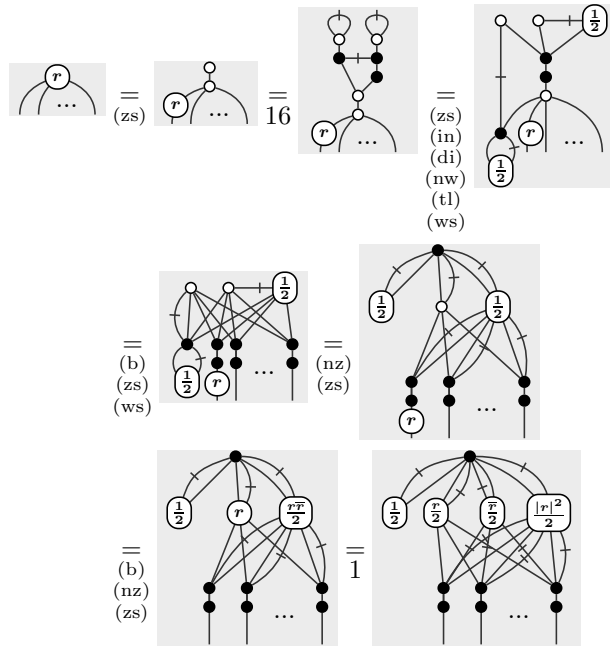
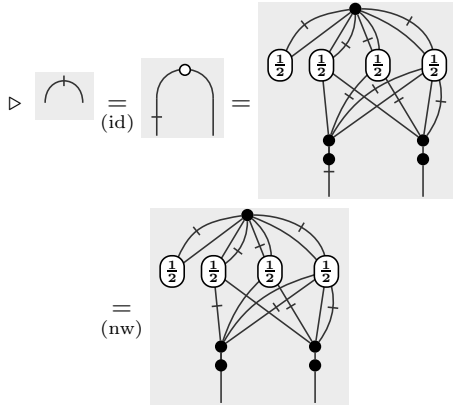
			\begin{align*}
				{\setlength{\fboxsep}{0pt}\colorbox{gray!15}{~\strut}}
				&\eq{\eqref{ax:Z-spider}}{\setlength{\fboxsep}{0pt}\colorbox{gray!15}{~\strut}}
				\eq{\ref{lem:NF-Z}}{\setlength{\fboxsep}{0pt}\colorbox{gray!15}{~\strut}}
				\eq{\eqref{ax:Z-spider}\\\eqref{ax:W-inv}\\\eqref{ax:W-binary-diffusion}\\\eqref{ax:tick-through-W-spider}\\\eqref{ax:tick-loop}\\\eqref{ax:W-spider}}{\setlength{\fboxsep}{0pt}\colorbox{gray!15}{~\strut}}\\
				&\eq{\eqref{ax:Z-W-bialgebra}\\\eqref{ax:Z-spider}\\\eqref{ax:W-spider}}{\setlength{\fboxsep}{0pt}\colorbox{gray!15}{~\strut}}
				\eq{\eqref{ax:tick-through-Z-spider}\\\eqref{ax:Z-spider}}{\setlength{\fboxsep}{0pt}\colorbox{gray!15}{~\strut\input{./figures/\fig/\fig_05.tikz}}}\\
				&\eq{\eqref{ax:Z-W-bialgebra}\\\eqref{ax:tick-through-Z-spider}\\\eqref{ax:Z-spider}}{\setlength{\fboxsep}{0pt}\colorbox{gray!15}{~\strut\input{./figures/\fig/\fig_06.tikz}}}
				\eq{\ref{lem:sum-branch-NF}}{\setlength{\fboxsep}{0pt}\colorbox{gray!15}{~\strut\input{./figures/\fig/\fig_07.tikz}}}
			\end{align*}
			\item \def\fig{NF-tick}$
			\begin{aligned}[t]
				{\setlength{\fboxsep}{0pt}\colorbox{gray!15}{~\strut}}
				&\eq{\eqref{ax:Z-id}}{\setlength{\fboxsep}{0pt}\colorbox{gray!15}{~\strut}}
				\eq{}{\setlength{\fboxsep}{0pt}\colorbox{gray!15}{~\strut}}\\
				&\eq{\eqref{ax:tick-through-W-spider}}{\setlength{\fboxsep}{0pt}\colorbox{gray!15}{~\strut}}
			\end{aligned}$
			\item For the W-spider, let us first look at the first 4 possible degrees: a W-spider of degree 0 is already in normal form. For the degree 1: \def\fig{NF-W-1}${\setlength{\fboxsep}{0pt}\colorbox{gray!15}{~\strut}}\eq{}{\setlength{\fboxsep}{0pt}\colorbox{gray!15}{~\strut}}$, since the cap can be put in normal form, and thanks to \Cref{lem:NF-bra-0}, this generator can be put in normal form. For the degree 2: \def\fig{NF-W-2}${\setlength{\fboxsep}{0pt}\colorbox{gray!15}{~\strut}}\eq{}{\setlength{\fboxsep}{0pt}\colorbox{gray!15}{~\strut}}$, and this time it is \Cref{lem:NF-negation} that puts the diagram in normal form. Finally for degree 3: \def\fig{NF-W-simp}${\setlength{\fboxsep}{0pt}\colorbox{gray!15}{~\strut}}\eq{}{\setlength{\fboxsep}{0pt}\colorbox{gray!15}{~\strut}}$, that is, the $0\to3$ W-spider can be obtained by applying the $2\to 1$ W-spider on the tensor product of a cap with itself. Since the cap can be put in normal form, thanks to \Cref{prop:NF-tensor} its tensor product with itself can be put in normal form, and by \Cref{lem:NF-W-2-1} the application of $\tikzfig{W-2-1}$ yields a diagram that can be put in normal form. W-spiders with a larger degree can then be obtained by compositions of the smaller ones, as: \def\fig{W-spider-decomposition}${\setlength{\fboxsep}{0pt}\colorbox{gray!15}{~\strut}}\eq{}{\setlength{\fboxsep}{0pt}\colorbox{gray!15}{~\strut}}$.
			\item By completeness of the $\cat{ZW}$-Calculus:
			\def\fig{fswap-decomposition}
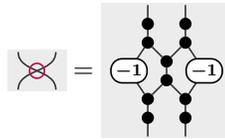
			\[{\setlength{\fboxsep}{0pt}\colorbox{gray!15}{~\strut}}\eq{}{\setlength{\fboxsep}{0pt}\colorbox{gray!15}{~\strut}}\]i.e.~it is a composition of generators we already showed how to put in normal form. It can hence be put in normal form.
		\end{itemize}
	\end{proof}
	
	\subsection{Pure ZW-Calculus}
	
	\label{sec:pure-zw}
	
	The equational theory for the pure part of the ZW-calculus (given in \Cref{fig:ZW_rules}) is not technically the one that was shown to be complete in \cite{Hadzihasanovic2018complete}. Indeed, while most rules are the same, some are given in a different manner. We show in this section that we can recover all of those from \cite{Hadzihasanovic2018complete} from the ones in \Cref{fig:ZW_rules}. We obviously only focus on the axioms that differ:
	
	\begin{itemize}
		\setlength\itemsep{0.5em}
		\item \def\fig{orig-fswap-involution}
		$\begin{aligned}[t]
			{\setlength{\fboxsep}{0pt}\colorbox{gray!15}{~\strut}}
			\eq{}{\setlength{\fboxsep}{0pt}\colorbox{gray!15}{~\strut}}
			\eq{\eqref{ax:fswap-rotated}}{\setlength{\fboxsep}{0pt}\colorbox{gray!15}{~\strut}}
			\eq{\eqref{ax:fswap-involution}}{\setlength{\fboxsep}{0pt}\colorbox{gray!15}{~\strut}}
		\end{aligned}$
		\item Similarly: \def\fig{orig-fswap-involution}
		$
		\rotatebox{180}{{\setlength{\fboxsep}{0pt}\colorbox{gray!15}{~\strut}}}
		\eq{}\rotatebox{180}{{\setlength{\fboxsep}{0pt}\colorbox{gray!15}{~\strut}}}
		$
		\item \def\fig{orig-floop-sym}
		$\begin{aligned}[t]
			{\setlength{\fboxsep}{0pt}\colorbox{gray!15}{~\strut}}
			\eq{}{\setlength{\fboxsep}{0pt}\colorbox{gray!15}{~\strut}}
			\eq{\eqref{ax:fswap-swapped}}{\setlength{\fboxsep}{0pt}\colorbox{gray!15}{~\strut}}
			\eq{}{\setlength{\fboxsep}{0pt}\colorbox{gray!15}{~\strut}}
			\eq{}{\setlength{\fboxsep}{0pt}\colorbox{gray!15}{~\strut}}
		\end{aligned}$
		\item 
		\def\fig{orig-Z-loop}
		$\begin{aligned}[t]
			{\setlength{\fboxsep}{0pt}\colorbox{gray!15}{~\strut}}
			\eq{\eqref{ax:W-inv}}{\setlength{\fboxsep}{0pt}\colorbox{gray!15}{~\strut}}
			\eq{\eqref{ax:Z-id}\\\eqref{ax:Z-spider}\\\eqref{ax:W-spider}}{\setlength{\fboxsep}{0pt}\colorbox{gray!15}{~\strut}}
			\eq{\eqref{ax:Z-W-bialgebra}}{\setlength{\fboxsep}{0pt}\colorbox{gray!15}{~\strut}}\\
			\eq{\eqref{ax:W-binary-diffusion}\\\eqref{ax:Z-W-bialgebra}}{\setlength{\fboxsep}{0pt}\colorbox{gray!15}{~\strut}}
			\eq{\eqref{ax:W-spider}\\\eqref{ax:W-bialgebra}}{\setlength{\fboxsep}{0pt}\colorbox{gray!15}{~\strut\input{./figures/\fig/\fig_05.tikz}}}
			\eq{\eqref{ax:W-binary-diffusion}}{\setlength{\fboxsep}{0pt}\colorbox{gray!15}{~\strut\input{./figures/\fig/\fig_06.tikz}}}
		\end{aligned}$
		\item
		\def\fig{orig-W-through-fswap}
		$\begin{aligned}[t]
			{\setlength{\fboxsep}{0pt}\colorbox{gray!15}{~\strut}}
			\eq{\eqref{ax:fswaps-through-W}}{\setlength{\fboxsep}{0pt}\colorbox{gray!15}{~\strut}}
			\eq{\eqref{ax:fswaps-through-W}}{\setlength{\fboxsep}{0pt}\colorbox{gray!15}{~\strut}}
		\end{aligned}$
		\item
		\def\fig{orig-floop-through-Z}
		$\begin{aligned}[t]
			{\setlength{\fboxsep}{0pt}\colorbox{gray!15}{~\strut}}
			\eq{\eqref{ax:f-loop}}{\setlength{\fboxsep}{0pt}\colorbox{gray!15}{~\strut}}
			\eq{\eqref{ax:Z-spider}}{\setlength{\fboxsep}{0pt}\colorbox{gray!15}{~\strut}}
			\eq{\eqref{ax:f-loop}}{\setlength{\fboxsep}{0pt}\colorbox{gray!15}{~\strut}}
		\end{aligned}$
		\item
		\def\fig{orig-fswap-removal}
		$\begin{aligned}[t]
			{\setlength{\fboxsep}{0pt}\colorbox{gray!15}{~\strut}}
			&\eq{\eqref{ax:fswap-through-Z}}{\setlength{\fboxsep}{0pt}\colorbox{gray!15}{~\strut}}
			\eq{\eqref{ax:fswap-through-Z}}{\setlength{\fboxsep}{0pt}\colorbox{gray!15}{~\strut}}\\
			&\eq{\eqref{ax:fswap-removal}}{\setlength{\fboxsep}{0pt}\colorbox{gray!15}{~\strut}}
			\eq{}{\setlength{\fboxsep}{0pt}\colorbox{gray!15}{~\strut}}
		\end{aligned}$
		\item
		\def\fig{orig-sum}
		$\begin{aligned}[t]
			{\setlength{\fboxsep}{0pt}\colorbox{gray!15}{~\strut}}
			&\eq{\eqref{ax:Z-spider}}{\setlength{\fboxsep}{0pt}\colorbox{gray!15}{~\strut}}
			\eq{\eqref{ax:Z-W-bialgebra}}{\setlength{\fboxsep}{0pt}\colorbox{gray!15}{~\strut}}\\
			&\eq{\eqref{ax:sum}}{\setlength{\fboxsep}{0pt}\colorbox{gray!15}{~\strut}}
			\eq{\eqref{ax:Z-spider}}{\setlength{\fboxsep}{0pt}\colorbox{gray!15}{~\strut}}
		\end{aligned}$
	\end{itemize}
	All the other axioms of \cite{Hadzihasanovic2018complete} are either found in \Cref{fig:ZW_rules} directly, or are particular cases of those found there.
	
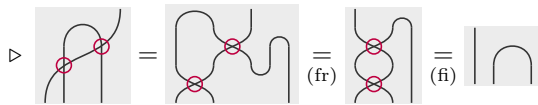
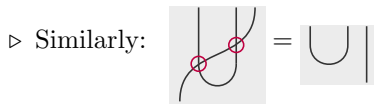
\end{document}